\documentclass[12pt]{article}

\usepackage{bbm}
\usepackage{setspace,graphicx,epstopdf,amsmath,amsfonts,amssymb,amsthm,versionPO}
\usepackage{thmtools}

\usepackage{multicol}
\usepackage{marginnote,datetime,enumitem,subfigure,rotating,fancyvrb}
\usepackage[longnamesfirst]{natbib}
\usepackage{relsize}
\usepackage{lipsum}
\usepackage[toc,page]{appendix}
\usepackage{tikz}
\usetikzlibrary{shapes,arrows}
\usepackage{booktabs}
\usepackage{tabularx}
\usepackage{physics}

\usepackage{multibib}
\newcites{Ap}{References}
\newcites{Sm}{References}

\usepackage[labelfont=bf]{caption}
\newcommand\fnote[1]{\vspace{2ex}\captionsetup{font=small}\caption*{\textit{Note}: #1}}

\usetikzlibrary{snakes}
\usetikzlibrary{positioning}
\usepackage{enumitem}

\newcommand\blfootnote[1]{%
  \begingroup
  \renewcommand\thefootnote{}\footnote{#1}%
  \addtocounter{footnote}{-1}%
  \endgroup
}
\usdate

\makeatother

\usepackage[unicode=true,
 bookmarks=false,
 breaklinks=false,pdfborder={0 0 1},
 colorlinks=false]
 {hyperref}
\hypersetup{
,colorlinks,linkcolor={red!50!black},citecolor={blue!50!black},urlcolor={blue!80!black}}
\makeatletter

\excludeversion{notes}		
\includeversion{links}          

\iflinks{}{\hypersetup{draft=true}}

\ifnotes{%
\usepackage[margin=1in,paperwidth=10in,right=2.5in]{geometry}%
\usepackage[textwidth=1.4in,shadow,colorinlistoftodos]{todonotes}%
}{%
\usepackage[margin=1in]{geometry}%
\usepackage[disable]{todonotes}%
}

\makeatletter\let\chapter\@undefined\makeatother 

\usepackage{titlesec}
\titlespacing*{\subsection}{0pt}{2ex plus 1ex minus .2ex}{2ex plus .2ex}
\makeatletter
\renewcommand{\paragraph}{%
  \@startsection{paragraph}{4}%
  {\z@}{1.25ex \@plus 1ex \@minus .2ex}{-1em}%
  {\normalfont\normalsize\bfseries}%
}

\let\oldparagraph=\paragraph
\renewcommand\paragraph[1]{\oldparagraph{#1.}}

\makeatother

\usepackage[toc,page]{appendix}
\newtheorem{thm}{Theorem}
\newtheorem{claim}{Claim}
\newtheorem{assumption}{Assumption}
\newtheorem{prop}{Proposition}

\newtheorem{lem}{Lemma}
\newtheorem{defn}{Definition}
\newtheorem{cor}{Corollary}

\newtheorem{alg}{Algorithm}

\newtheoremstyle{definition_space} 
  {\topsep} 
  {\topsep} 
  {} 
  {} 
  {\bfseries} 
  {.} 
  {.5em} 
  {} 

\theoremstyle{definition_space}

\newtheorem{examplex}{Example}
\newenvironment{example}
  {\pushQED{\qed}\examplex}
  {\popQED\endexamplex}

\begin{document}

\setlist{noitemsep}  

\title{Adaptive Priority Mechanisms}
\author{
O\u{g}uzhan \c{C}elebi\footnote{Stanford University Department of Economics, 579 Jane Stanford Way, Stanford, CA 94305. Email: \href{mailto:ocelebi@stanford.edu}{ocelebi@stanford.edu}}\\
Stanford\\
\and
Joel P. Flynn\footnote{Yale University Department of Economics, 30 Hillhouse Avenue, New Haven, CT, 06511. Email: \href{mailto:joel.flynn@yale.edu}{joel.flynn@yale.edu}} \\
Yale\\
}
\vspace{-2ex}
\date{\today}

\onehalfspacing
\maketitle
\thispagestyle{empty}
\vspace{-3ex}
\begin{abstract} 
\noindent
How should authorities that care about match quality and diversity allocate resources when they are uncertain about the market? We introduce \textit{adaptive priority mechanisms} (APM) that prioritize agents based on both their scores and characteristics. We derive an APM that is optimal and show that the ubiquitous priority and quota mechanisms are optimal if and only if the authority is risk-neutral or extremely risk-averse over diversity, respectively. With many authorities, each authority using the optimal APM is dominant and implements the unique stable matching. Using Chicago Public Schools data, we find that the gains from adopting APM may be considerable.

\end{abstract}

\blfootnote{We are grateful to Daron Acemoglu, Mohammad Akbarpour, George-Marios Angeletos, Nick Arnosti, Jonathan Cohen, Viola Corradini, Roberto Corrao, Mert Demirer, Glenn Ellison, Aytek Erdil, Arda Gitmez, \"{O}mer Karaduman, Stephen Morris, Anh Nguyen, Parag Pathak, Charlie Rafkin, Karthik Sastry, Tayfun S\"{o}nmez, Bertan Turhan, Alexander Wolitzky, Bumin Yenmez and seminar participants at the 2022 INFORMS Workshop on Market Design, the 2022 Society for the Advancement of Economic Theory Conference, Iowa State University and the MIT Theory Lunch for helpful comments. We thank Chicago Public Schools for graciously sharing their data and Eryn Heying, Talia Gerstle and Jim Shen for invaluable administrative support. The views expressed here are those of the authors and do not reflect the views of Chicago Public Schools.}

\clearpage
\maketitle
\addtocontents{toc}{\protect\setcounter{tocdepth}{0}}
\setcounter{page}{1}
\section{Introduction}
\setlength{\abovedisplayskip}{5pt}
\setlength{\belowdisplayskip}{5pt}
Authorities that allocate resources such as school seats, university places, and medical supplies often face conflicting objectives. On the one hand, they want to maximize match quality or appear fair by allocating resources to the highest-scoring agents according to various criteria such as academic attainment, mortality risk, or distance. On the other hand, they want to achieve diversity across a range of socioeconomic attributes including race, religion, and gender. Resolving this conflict is complicated, especially in new markets, due to uncertainty regarding the distribution of individuals' scores, characteristics, and preferences.

To balance these trade-offs, when the use of prices is seen as infeasible or unethical, authorities have broadly used two classes of policies: \textit{quotas},\footnote{We use {\it quota} as a general term that includes the widely used reserve policies (see Definition \ref{quotadef}).} where a certain portion of the resource is set aside for given groups; and \textit{priorities}, where individuals in given groups receive higher scores. These policies have been applied across many different markets in many different countries, for example: the Indian government reserves some government jobs for disadvantaged groups; Chicago Public Schools employs quotas for students from different socioeconomic groups at its competitive exam schools; the University of California, Davis instituted a quota system for minority students; many countries gave differential priority to healthcare workers in the receipt of Covid-19 vaccines; church-run schools in the UK give explicit priority points to students from various religious groups; and the University of Michigan and the University of Texas have used different priority scales for minority students.

But what mechanism \textit{should} such an authority use?  Despite the practical importance of this question, we do not know if (and under what circumstances) an authority should use a priority mechanism, a quota mechanism, or something else entirely.

In this paper, we formulate and solve the optimal mechanism design problem of an authority that allocates a resource to agents who are heterogeneous in their individual scores and belong to different groups. The authority cares about individuals' \textit{scores}, through some aggregate index, and \textit{diversity}, through the numbers of agents from different groups who are allocated the resource.\footnote{This diversity preference can be interpreted more generally as encoding a preference of the authority over the composition of assigned agents across a range of attributes, \textit{e.g.,} when allocating medical resources, the authority may care about ensuring that frontline medical workers are treated. Moreover, when scores represent individuals' property rights over objects (\textit{e.g.,} higher-scoring students \textit{deserve} better schools), we can interpret the preference for higher scores as a preference for procedural fairness. Whenever a lower-scoring agent obtains the resource while a higher-scoring agent does not, the latter agent has \textit{justified envy} towards the former. In the two-sided matching literature, justified envy is often seen as inimical to fairness \citep[see \textit{e.g.,}][]{balinski/sonmez:99}.} Moreover, they are uncertain about the market they face and have some beliefs about the joint distribution of scores and groups in the population.

We propose a new class of \textit{adaptive priority mechanisms} (APM) that adjust agents' scores as a function of the number of assigned agents with the same characteristics and that allocate the resource to the set of agents with the highest adjusted scores. With a single authority, we derive an APM that is optimal, implements a unique outcome, and can be specified solely in terms of the \textit{preferences} of the authority (\textit{i.e.,} it is optimal regardless of their beliefs). By contrast, we show that priorities and quotas are optimal if and only if risk aversion over diversity is extremely low or high, respectively. Moreover, optimally set priority and quota policies depend on both the preferences and beliefs of the authority. Thus, the optimal APM improves outcomes, is robust to uncertainty, and requires less information. With many authorities, it is dominant for each of them to implement this APM and this leads to the unique stable outcome.

\paragraph{Single-Authority Model} We begin our analysis by studying a setting with a single authority that has some amount of a homogeneous resource (\textit{e.g.,} seats at a school, medical resources) that it can allocate to a continuum of agents.\footnote{In Appendix \ref{discapp}, we generalize our analysis and results to a setting with discrete agents.} Agents differ in their scores (\textit{e.g.,} exam score, clinical need) and discrete attributes (\textit{e.g.,} socioeconomic status, whether they are a frontline health worker). The authority cares separably about some index of the score distribution (\textit{e.g.,} the average score) of those to whom it allocates the resource and the numbers of agents from different groups. Thus, the authority's preferences over agents depend on the joint distribution of agents' scores and groups. We assume that this distribution is potentially unknown and varies arbitrarily across states of the world. The authority's problem is to design a \textit{first-best optimal} mechanism: a mechanism that is optimal regardless of their beliefs and implements an \textit{ex post} optimal allocation in all states.

\paragraph{Adaptive Priority Mechanisms} To this end, we introduce the class of adaptive priority mechanisms (APM), which proceed in two steps. First, each agent is given an \textit{adaptive priority} that is a function of their own score and the number of agents from the same group to whom the resource is assigned. Second, APM allocate the resource to agents in order of adaptive priorities, subject to fully allocating the available amount. This class of mechanisms allows the implicit preference for agents from different groups to depend upon the ultimate allocation. The allocation under an APM is defined as the fixed point of the above operation: an allocation is implemented by APM if the adaptive priority of all agents who are allocated the resource (evaluated at the allocation) is higher than those who are not allocated the resource. When an agent's adaptive priority is increasing in their own score and decreasing in the number of agents with the same attributes that are assigned the resource -- a property we call \textit{monotonicity} -- the APM implements a unique allocation. Moreover, this allocation can be computed greedily by prioritizing agents according to their adaptive priority, evaluated at the number of higher-scoring agents in their group.

Most importantly, we derive a particular, monotone APM that is first-best optimal. Under this optimal APM, an agent's priority is equal to the contribution of their own score plus their marginal contribution to diversity utility. Intuitively, this mechanism equates the benefits and costs of allocating to the marginal agent, regardless of the ultimate joint distribution of agents' scores and groups. Moreover, this APM can be described \textit{ex ante} as a function of the authority's preferences, without any reference to its beliefs or hypothetical states of the world.

\paragraph{(Sub)Optimality of Priorities and Quotas} We next establish that priority and quota mechanisms are generally dominated by APM. We do so by fully characterizing the conditions on the preferences of the authority such that priorities and quotas attain first-best optimality. Concretely, we find that priorities and quotas are first-best optimal if and only if (i) the authority is risk-neutral over diversity, in which case priorities are optimal, or (ii) the authority is extremely risk-averse over diversity, in which case quotas are optimal. Hence, outside of extreme cases, APM deliver strict improvements relative to the \textit{status quo}.

 \paragraph{A Price-Theoretic Intuition} To both illustrate and develop the intuition behind these results, we study a detailed example that allows for a closed-form comparison of priorities, quotas, and the optimal adaptive priority mechanism. We do this in the spirit of the seminal analysis of \cite{weitzman1974prices}, who compares price and quantity regulation in product markets. In the example, the resource corresponds to seats at a school and there are two groups of students (minority and majority students). The authority is uncertain over the relative scores of minority and majority students, and has preferences over the scores of admitted students and the number of minority students admitted to the school.

 The preference of the authority between priority and quota mechanisms is governed by its risk aversion over the number of admitted minority students: there is a cutoff value such that quotas are preferred when risk aversion exceeds this threshold and priorities are otherwise preferred. On the one hand, by mandating a minimal level of minority admissions, quotas \textit{guarantee} a level of diversity. On the other hand, as relatively more minority students receive the resource in the states in which they have relatively higher scores, priorities \textit{positively select} minority students. Adaptive priority mechanisms optimally exploit the guarantee effects of quotas and the positive selection effects of priorities, and are always optimal.

\paragraph{Dominance, Stability, and Efficiency with Multiple Authorities} While the single-authority model is relevant for studying settings with a single resource, in many markets there are multiple authorities who control heterogeneous resources (\textit{e.g.,} school seats) over which agents have heterogeneous preferences. We generalize our analysis to this setting and show that APM arise under both cooperative (stability) and non-cooperative (dominant-strategy equilibrium) solution concepts. We first show that there is a unique stable allocation. This constitutes a methodological contribution as we establish the uniqueness of stable allocations in continuum economies in which there are multiple socioeconomic groups and authorities have non-linear preferences over the composition of the agents that they admit. Importantly, this means that authorities have \textit{endogenous} preferences over various agents: how an authority ranks one agent relative to another depends on the representation of their groups in the ultimate allocation. Moreover, we characterize this unique stable allocation and show that a mechanism is consistent with stability if and only if it coincides with the single-authority-optimal APM. Furthermore, when authorities sequentially admit agents, each authority using its single-authority-optimal APM is a dominant strategy and implements the unique stable matching. These results imply that one could advise authorities to use APMs with confidence that outcomes will be stable and that they could do no better under any alternative mechanism. This notwithstanding, the decentralized allocation under single-authority optimal APM is not utilitarian efficient for the authorities. To remedy this, we propose a modification of APM for centralized markets that yields utilitarian efficient allocations for the authorities.

\paragraph{Benchmarking the Gains from APM} To obtain a sense of the benefit of using APM, we benchmark the improvements from APM using application and admission data from 2013-2017 on the selective exam schools of Chicago Public Schools (CPS), a setting also empirically studied by \cite{angrist2019choice} and \cite{ellison2021efficiency}. CPS uses a reserve system to increase the admissions of underrepresented groups. In this system, as we later detail, academic scores and the socioeconomic characteristics of the census tracts in which students live determine the schools that students can attend. Moreover, there is substantial variation in the joint distribution of student characteristics over time. This justifies our focus on the importance of uncertainty and implies that APM could generate gains for the authority. Estimating preference parameters to best rationalize the pursued reserve policy, we find that the gains from using the optimal APM are equivalent to eliminating 37.5\% of the loss to CPS' payoffs from failing to admit a diverse class of students. This gain is 2.3 times larger than the estimated gain from a 2012 policy change that increased the size of all reserves. This proof-of-concept exercise shows both that APM could be practically implemented and that the gains from so doing may be considerable.

\paragraph{Related Literature}
The market design literature has largely studied the comparative statics and axiomatic foundations of mechanisms. In this context, our paper relates to the literature on matching with affirmative action concerns initiated by \cite{abdulkadiroglu/sonmez:03a} and \cite{abdulkadiroglu:05}.  For example, in the study of quotas, \cite{kojima:12} shows how affirmative action policies that place an upper bound on the enrollment of non-minority students may hurt all students, \cite{hafalir/yenmez/yildirim:13} introduce the alternative and more efficient minority reserve policies, \cite{ehlers/hafalir/yenmez/yildirim:14} generalize reserves to accommodate policies that have floors and ceilings for minority admissions, and \cite{dougan2016responsive} shows that stronger affirmative action can (weakly) harm all minority students under reserve policies and proposes a new rule that fixes this issue. The quota policies studied in this paper are a special case of the slot-specific priorities introduced in \cite{kominers2016matching}. Further related papers study quota policies in university admissions in India
\citep{aygun2020dynamic,sonmez2022affirmative,sonmez2022constitutional}, in Germany \citep{westkamp2013analysis} and in Brazil \citep{aygun2021college}. \cite{kamada2017stability,kamada2018stability} and \cite{goto2017designing} study stability and efficiency in more general matching-with-constraints models. \cite{echenique2015control} characterize a class of substitutable choice rules under diversity preferences, \cite{erdil2019efficiency} study tie-breaking rules under substitutable priorities under stable matching mechanisms and distributional constraints, and \cite{imamura2020meritocracy} presents axioms to compare the meritocracy and diversity of different choice rules and characterizes reserves and quotas. \cite{celebi2022diversity} studies when affirmative action policies can be rationalized by diversity preferences.

In this paper, we instead pursue the methodological approach of mechanism design and welfare economics by analyzing optimal mechanisms from the perspective of an authority with some given preferences over allocations. \cite{chan2003does} share this perspective in their analysis of the costs and benefits of banning affirmative action.\footnote{Other analyses of this issue include \cite{epple2008diversity} and \cite{temnyalov2021information}.} In this vein, we have previously analyzed the narrower problem of how to optimally coarsen agents' scores into priorities \citep{celebi2020priority} in a continuum matching market framework in the style of \cite{abdulkadirouglu2015expanding} and \cite{azevedo2016supply}. This analysis nevertheless restricted authorities to use a priority mechanism that does not consider agents' characteristics and implement only allocations that are stable with respect to these priorities. Thus, our focus on comparing priorities, quotas, and optimal mechanisms distinguishes our analysis from our prior work and the previous literature, which study the properties of each policy in isolation and without an explicit treatment of uncertainty.

\paragraph{Outline} Section \ref{weitzmansection} exemplifies our main results. Section \ref{sec:single} studies optimal mechanisms with a single authority. Section \ref{sec:eq} studies equilibrium mechanisms with many authorities. Section \ref{sec:cent} studies authority-efficient mechanisms. Section \ref{sec:quant} quantifies the gains from APM using Chicago Public Schools' data. Section \ref{sec:conc} concludes. The proofs are in Appendix \ref{ap:proofs}.

\section{Comparing Mechanisms: An Example}
\label{weitzmansection}

\paragraph{The Setting} A single school has capacity $q$. Students are of unit total measure, have scores in $[0,1]$, and are either minority or majority students. The authority has linear-quadratic preferences $\xi:\mathbb{R}^2\rightarrow\mathbb{R}$ over students' total scores $\bar{s}$ and the measure of admitted minority students $x$:
\begin{equation}
\label{ufn}
   \xi(\bar{s},x)=\bar{s}+\gamma\left(x-\frac{\beta}{2}x^2\right)
\end{equation}
where $\gamma\ge0$ indexes their general preference for admitting minority students and $\beta\ge 0$ indexes the degree of risk aversion regarding the measure of admitted minority students.

The minority students are of measure $\kappa$ and have scores that are uniform over $[0,1]$. The majority students are of measure $1-\kappa$ and all have common underlying score $\omega\in[\underline{\omega},\overline{\omega}]\subseteq[0,1]$ with distribution $\Lambda$. The score of the majority students, $\omega$, parameterizes how well the majority students score relative to the minority students. Finally, we assume that the affirmative action preference is neither too small nor too large with the following: $\min\{\kappa,q\}>\frac{1+\gamma-\underline{\omega}}{\frac{1}{\kappa}+\gamma\beta}+\kappa(\overline{\omega}-\underline{\omega})$,    $\kappa(1-\underline{\omega})<\frac{1+\gamma-\overline{\omega}}{\frac{1}{\kappa}+\gamma\beta}$. These conditions ensure that optimal affirmative action policies will neither be so large as to award all slots to minority students in some states nor so small that there is no affirmative action in some states.

The authority can implement an APM, a priority mechanism, or a quota mechanism. An APM increases the scores of minority students by $A(y)$ when $y$ other minority students are admitted, does not change the scores of majority students, and allocates seats to the students with highest transformed scores.\footnote{Formally, this mechanism allocates seats to $x(\omega)$ minority students and $q-x(\omega)$ majority students, where $s(x(\omega)) + A(x(\omega)) = \omega$, and $s(x(\omega))$ denotes the score of the marginal minority student when the highest-scoring $x(\omega)$ minority students are admitted.} An (additive) priority mechanism $\alpha\in\mathbb{R}_{+}$ increases uniformly the scores of minority students by $\alpha$. The authority then admits the highest-scoring measure $q$ students. A quota policy $Q\in[0,\min\{\kappa,q\}]$ sets aside measure $Q$ of the capacity for the minority students. The highest-scoring minority students of measure $Q$ are first allocated to quota slots, and all other agents are then admitted to the residual $q-Q$ places according to the underlying score.\footnote{This corresponds to a precedence order that processes quota slots first. We discuss the importance of precedence orders in Section \ref{sec:discussion} and in Appendix \ref{precedencesection}.} 

\begin{figure}
    \centering
    \caption{How Priorities, Quotas, and APM Prioritize Minority Students}
    \includegraphics[width=0.6\textwidth]{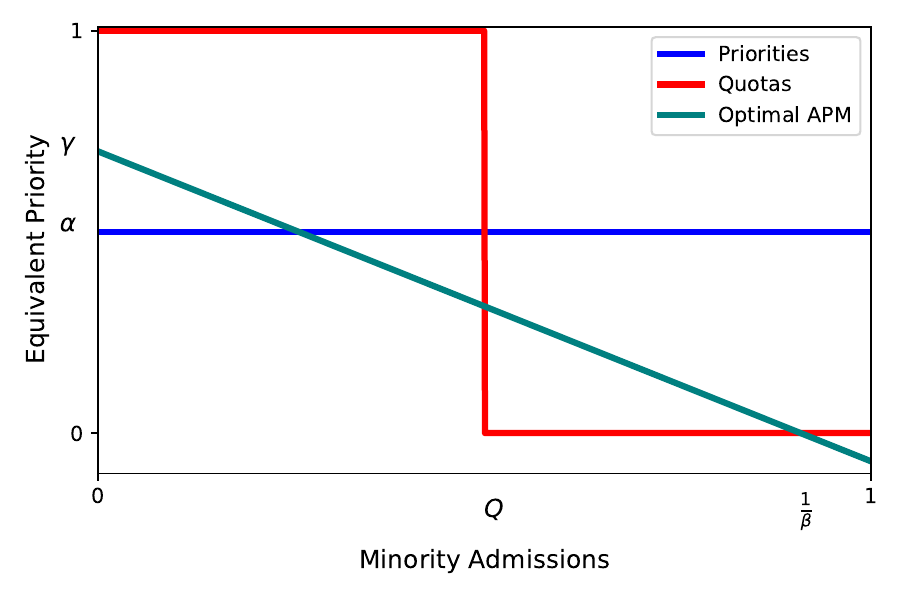}
    \fnote{Illustration of the equivalent priority given to a minority student as a function of the measure of admitted minority students under: the optimal APM (see Proposition \ref{mechcomp}), a priority mechanism $\alpha$, and a quota mechanism $Q$.}
    \label{fig:policyfig}
\end{figure}

We illustrate how these three policies prioritize minority students in Figure \ref{fig:policyfig}. Priority mechanisms award a constant score boost of $\alpha$. Quota mechanisms give enough points to always ensure admission until measure $Q$ is reached and then give no advantage. APM allow any pattern of prioritization as a function of minority admissions (we plot only the optimal APM, which turns out to be linear in this context).

\paragraph{Comparing Mechanisms} Let the authority's expected utility be $V^*$ under any optimal (expected utility maximizing) mechanism, $V_A$ under an optimal adaptive priority mechanism, $V_P$ under an optimal priority mechanism, and $V_Q$ under an optimal quota mechanism. The following proposition characterizes the relationships between these mechanisms:

\begin{prop}
\label{mechcomp}
The following statements are true:
\begin{enumerate}
\item The APM $A(y)=\gamma(1-\beta y)$ is optimal, $V^*=V_A$
\item The comparative advantage of priorities over quotas is given by:
\begin{equation}
\label{delta}
    \Delta \equiv V_P-V_Q=\frac{\kappa}{2}\left(1-\kappa\gamma\beta \right)\text{\normalfont Var}[\omega]
\end{equation}
\item The comparative advantage of APM over priorities and quotas is given by:
\begin{equation}
\label{eq:loss}
    \Delta^*\equiv\min\{V^*-V_P,V^*-V_Q\}=    \begin{cases}
        \frac{1}{2}\left(\kappa\gamma\beta \right)^2\frac{\kappa\text{\normalfont Var}[\omega]}{1+\kappa\gamma\beta }, &\kappa\gamma\beta \leq1, \\
        \frac{1}{2}\frac{\kappa\text{\normalfont Var}[\omega]}{1+\kappa\gamma\beta }, &\kappa\gamma\beta >1.
    \end{cases}
\end{equation}
\end{enumerate}
\end{prop}
We now develop intuition for the comparative advantage of priorities over quotas. First, observe that a quota of $Q$ admits measure $Q$ minority students in all states of the world under our assumptions. However, a priority policy induces variability in the measure of admitted minority students across states of the world. We call the gain to quota policies in eliminating this variation the \textit{guarantee effect} and find mathematically that it is equal to $\frac{\kappa}{2}\left(1+\kappa\gamma\beta \right)\text{\normalfont Var}[\omega]$ in payoff terms.

\begin{figure}
    \centering
    \caption{Comparative Statics for the Positive Selection and Guarantee Effects}
    \includegraphics[width=0.6\textwidth]{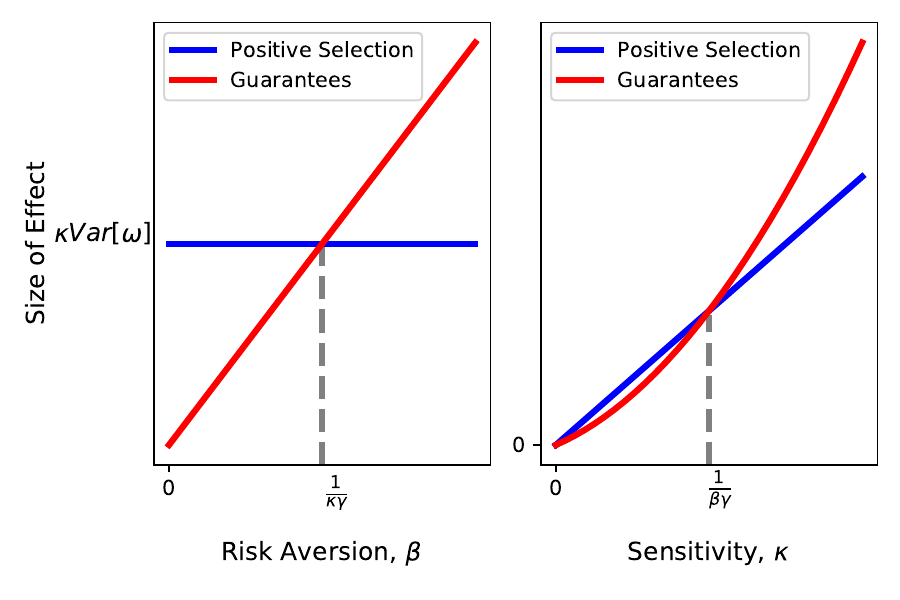}
    \fnote{Illustration of the comparative statics for the trade-offs between priority and quota mechanisms. Positive Selection plots the positive selection effect, $\kappa\text{\normalfont Var}[\omega]$, and Guarantee plots the guarantee effect, $\frac{\kappa}{2}\left(1+\kappa\gamma\beta \right)\text{\normalfont Var}[\omega]$. As per Equation \ref{delta} in Proposition \ref{mechcomp}, priorities dominate quotas if and only if $1\ge\kappa\gamma\beta $, where the point of indifference is denoted by the dashed grey line. }
    \label{fig:PvsQfig}
\end{figure}

Second, the optimal priority policy admits more minority students when minority students score relatively well and fewer when minority students score relatively poorly. To demonstrate this, we show that minority admissions in state $\omega$ under the optimal priority policy are $x(\alpha,\omega)=\bar{x}(\alpha)+\varepsilon(\omega)$ where $\bar{x}(\alpha)=\kappa(1+\alpha-\mathbb{E}[\omega])$ and $\varepsilon(\omega)=\kappa\left(\mathbb{E}[\omega]-\omega\right)$. Thus, in the states where minority students score relatively better ($\omega<\mathbb{E}[\omega]$), we have that $\varepsilon(\omega)>0$ and $x(\alpha,\omega)>\bar{x}(\alpha)$. We call this effect the \textit{positive selection} effect and find that this benefits a priority policy by $-\text{\normalfont Cov}[\omega,\varepsilon(\omega)]=\kappa\text{\normalfont Var}[\omega]$ in payoff terms.

The ultimate preference between priority and quota mechanisms is determined by which of the guarantee and positive selection effects dominates. This is itself determined by the extent to which the authority values diversity $\gamma$, the risk preferences of the authority $\beta$, and the measure of minority students $\kappa$. We illustrate how risk aversion and the measure of minority students affect the sizes of the positive selection and guarantee effects in Figure \ref{fig:PvsQfig}. If the authority is close enough to risk-neutral (\textit{i.e.,} $\frac{1}{\kappa\gamma}>\beta $), then priorities are strictly preferred as positive selection dominates guarantees. If the authority is sufficiently risk-averse (\textit{i.e.,} $\frac{1}{\kappa\gamma}<\beta $), then quotas are strictly preferred as the guarantee effects dominate positive selection. The threshold for risk aversion scales inversely with the measure of minority students $\kappa$. Because minority students' scores are uniform, $\kappa$ corresponds to the density of minority students' scores. Hence, the change in minority admissions from a small change in their priority equals $\kappa$. Thus, $\kappa$ indexes the \textit{sensitivity} of minority admissions to the state under priority policies. As a result, higher $\kappa$ favors quota policies by increasing the magnitude of the guarantee effect relative to the positive selection effect. Finally, the extent of uncertainty $\text{\normalfont Var}[\omega]$ may intensify an underlying preference but never determines which regime is preferred.

\begin{figure}
    \centering
    \caption{Comparative Statics for the Losses from Priorities and Quotas}
    \includegraphics[width=0.6\textwidth]{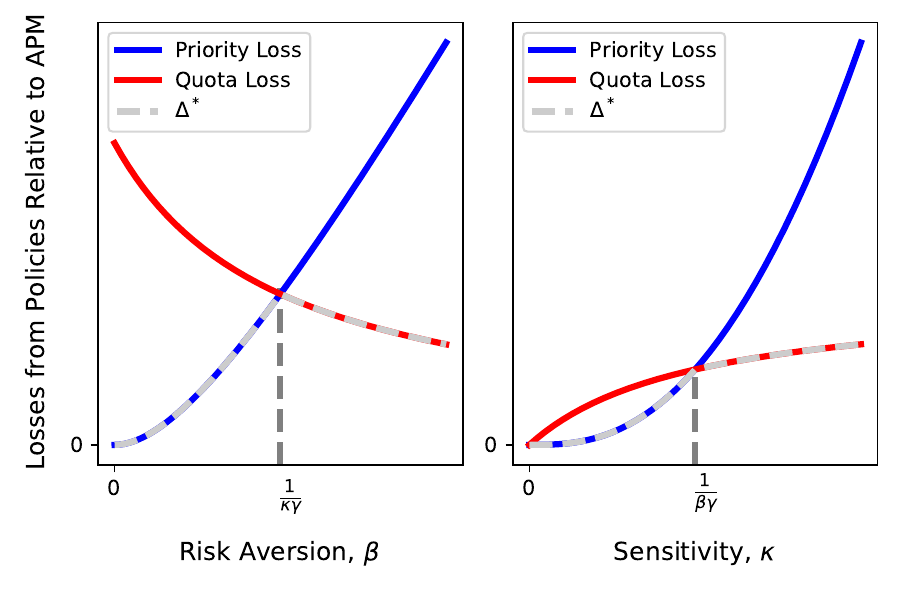}
    \fnote{Illustration of the comparative statics for the losses from optimal priority and quota policies relative to the optimal APM (as presented in Equation \ref{eq:loss} in Proposition \ref{mechcomp}). The lower envelope of the losses, $\Delta^*$, corresponds to the comparative advantage of the optimal APM over priorities and quotas. The point of indifference between priorities and quotas is denoted by the dashed grey line.}
    \label{fig:LossFig}
\end{figure}

An APM is optimal and overcomes the limitations posed by both priorities and quotas. In this case, the optimal APM is linear in the measure of admitted minority students, with slope given by the authority's risk aversion over minority admissions, awarding each minority student a subsidy equivalent to their marginal contribution to the diversity preferences of the authority. This allows the adaptive priorities to optimally balance the positive selection and guarantee effects, and implement the first-best allocation in every state. In Figure \ref{fig:LossFig}, we illustrate how the losses from priority mechanisms and quota mechanisms vary with risk aversion and sensitivity. As risk aversion moves, the loss from priority and quota policies relative to the optimum is greatest when the authority is indifferent between the two regimes. The loss from restricting to priority or quota policies is zero when the authority is risk-neutral or there is no uncertainty regarding relative scores, and decreases as the authority becomes extremely risk-averse. As sensitivity increases, the scope for affirmative action increases and so the gains from APM also increase. Thus, we should expect there to be large gains from switching to APM precisely when authorities have intermediate levels of risk aversion and/or the scope for implementing affirmative action is significant.

Finally, optimal APM have a further advantage that we have not yet highlighted: they depend only on the authority's preferences, $\gamma$ and $\beta$, and not their beliefs, $\Lambda$. This contrasts with the optimal priority and quota policies, which depend on $\Lambda$.\footnote{The optimal quota policy is given by $Q^*=\frac{1+\gamma-\mathbb{E}[\omega]}{\frac{1}{\kappa}+\gamma\beta }$, while the optimal priority policy sets the expected measure of minorities to $Q^*$. The policies depend on $\Lambda$ through $\mathbb{E}[\omega]$.} As a result, APM improve outcomes while using \textit{less} information and are robust to changes in beliefs.

\subsection{Discussion}\label{sec:discussion}
Before moving to the general analysis, we discuss three additional findings that emphasize the broader economics and scope of these results.

\paragraph{A Price-Theoretic Intuition} This comparison of \textit{priorities vs. quotas} echoes the comparison of \textit{prices vs. quantitities} by \cite{weitzman1974prices}. We show in Appendix \ref{weitzmancomp} that there is a formal mapping between the two. Intuitively, the positive selection effect is equivalent to the effect that price regulation gives rise to the greatest production in states where the firm's marginal cost is lowest. Moreover, the guarantee effect is equivalent to the ability of quantity regulation to stabilize the level of production. An APM corresponds in the \cite{weitzman1974prices} setting to a regulator setting neither a price nor a quantity, but completely specifying the optimal demand curve. Thus, the comparison of mechanisms for allocating goods without prices boils down to similar trade-offs between well-understood price-based mechanisms for goods allocation.

\paragraph{Medical Resource Allocation}  In Appendix \ref{sec:Medical}, we apply this model to understand the trade-offs between priority and quotas in the context of medical resource allocation. This topic received enormous attention during the Covid-19 pandemic \citep[see \textit{e.g.,}][]{pathak2020leaving}. Our analysis provides a formal justification for the idea that priorities may lose out relative to quotas from ignoring some groups or ethical values in the allocation of scarce resources (the guarantee effect). However, we also uncover a benefit of priorities that was not previously understood: they induce positive selection. Thus, if we care mostly about treating the neediest ($\beta $ is low), priorities may yet be optimal.

\paragraph{Optimal Precedence Orders}
 We have modelled quotas by first allocating minority students to quota slots and then allocating all remaining students according to the underlying score. However, we could have done the opposite. The orders in which quotas are processed are called \textit{precedence orders} in the matching literature and their importance has been the subject of a growing literature \citep[see \textit{e.g.,}][]{dur2018reserve,dur2020explicit,pathak2020immigration}. In Corollary \ref{prec} in Appendix \ref{precedencesection}, we show that processing quotas second is equivalent to using a priority policy in this setting. Thus, processing quotas first is better than processing them second if and only if $1\leq \kappa\gamma\beta $. The main aspect of this conclusion is robust in the general theory: in Theorem \ref{rationalization}, we show that for any quota policy to be optimal in the presence of uncertainty, it must process quotas first. In the absence of uncertainty, in Appendix \ref{ap:implementation}, we show that priority and quota mechanisms are equivalent and use this to quantify the impact of changes in precedence orders for US H1-B visa allocation.

\section{Optimal Mechanisms with a Single Authority}
\label{sec:single}
We begin our general analysis by studying the resource allocation problem of a single authority. In this context, we define APM and derive an optimal APM that attains the first-best. We moreover provide necessary and sufficient conditions for the optimality of the ubiquitous priority and quota mechanisms and find that they are extremely restrictive, implying that there are likely gains from switching to APM.

\subsection{Model}
An authority allocates a single resource of measure $q\in(0,1)$. Agents differ in their type $\theta\in\Theta=[0,1]\times\mathcal{M}$ comprising their scores $s\in[0,1]$ and the group to which they belong, $m\in\mathcal{M}$, where their score denotes their suitability for the resource and $\mathcal{M}$ is a finite set comprising potential attributes such as race, gender, or socioeconomic status. We denote the score and group of any type $\theta$ by $s(\theta)$ and $m(\theta)$, respectively. The true distribution of types is unknown to the authority. The authority's uncertainty is parameterized by $\omega\in\Omega$, where $\Omega$ is the set of all distributions over $\Theta$ that admit a density. The authority believes that $\omega$ has distribution $\Lambda\in\Delta(\Omega)$. In state of the world $\omega$, we denote the measure of types by $F_{\omega}$ with density $f_{\omega}$.\footnote{Formally, we mean that $f_{\omega}(s,m)=\frac{\partial}{\partial s}F_{\omega}(s,m)$ exists for all $s\in[0,1]$ and $m\in\mathcal{M}$.} In Appendix \ref{discapp}, we translate our analysis and results to the discrete context.\footnote{Concretely, we establish the optimality of APM, characterize the (sub)-optimality of priorities and quotas, and demonstrate the dominance of APM in discrete economies.}

An allocation $\mu:\Theta\rightarrow\{0,1\}$ specifies for any type $\theta\in\Theta$ whether they are assigned to the resource.\footnote{Formally, $\mu$ is a measurable function with respect to the Borel $\sigma-$algebra of the product topology in $\Theta$.} Two allocations $\mu$ and $\mu'$ are \textit{essentially the same} if they coincide up to a measure zero set. The set of possible allocations is $\mathcal{U}$. An allocation is feasible if it allocates no more than measure $q$ of the resource. A mechanism is a function $\phi:\Omega\rightarrow \mathcal{U}$ that returns a feasible allocation for any possible measure of types.

As motivated, authorities often have preferences over scores and diversity. To model this, we define the aggregate score index of any allocation as:
\begin{equation}
    \bar{s}_{h}(\mu,\omega)=\int_{\Theta}\mu(s,m)h(s)\dd F_{\omega}(s,m)
\end{equation}
for some continuous, strictly increasing function $h:[0,1]\rightarrow\mathbb{R}_{+}$, which determines the extent to which the authority values agents with higher scores. To capture diversity, we compute the measure of agents of each group allocated the resource $x(\mu,\omega)=\{x_m(\mu,\omega)\}_{m\in\mathcal{M}}$ as:
\begin{equation}
    x_m(\mu,\omega)=\int_{[0,1]}\mu(s,m)f_{\omega}(s,m)\dd s
\end{equation}

To separate the roles of scores and diversity, we impose that their utility function over these dimensions $\xi:\mathbb{R}^{|\mathcal{M}|+1}\rightarrow\mathbb{R}$ satisfies the following assumption:

\begin{assumption}
\label{sep}
The authority's utility function can be represented as:
\begin{equation}
    \xi\left(\bar{s}_{h},x\right)\equiv g\left(\bar{s}_{h}+\sum_{m\in\mathcal{M}}u_{m}(x_m)\right)
\end{equation}
for some continuous, strictly increasing function $g:\mathbb{R}\rightarrow\mathbb{R}$ and differentiable and concave functions $u_m:\mathbb{R}\rightarrow\mathbb{R}$ for all $m\in\mathcal{M}$.
\end{assumption}

We also assume that the authority always prefers to allocate the entire resource.\footnote{A necessary and sufficient condition for this is: $h(0) + u'_m(q) \geq 0$ for all $m \in \mathcal M$. This condition is sufficient as the lowest utility the authority can get from allocating the resource is always positive. It is also necessary: if $h(0) + u_m'(q) < 0$ for some $m$, in the state of the world where there are only measure $q$ of group $m$ agents with uniform score distribution, the authority would prefer not to allocate a portion of the resource to the lowest-scoring agents.} The preference of the authority is a monotone transformation of a quasi-linear utility index comprised of scores and a diversity preference. Intuitively, $u_m$ determines the preference for assigned agents of group $m$, with its concavity following from a preference for diversity.\footnote{Note that $u_m$ depends on $m$, so our specification allows the designer to have different preferences for allocating the resource to agents from different groups. For example, this allows for a designer with affirmative action motives who prefers to assign the resource to some particular group $m$: $u'_m(x) > u'_{m'}(x)$ for all $x$ or a designer who prefers a balanced composition of allocated agents: $u'_m(x) = u'_{m'}(x)$ for all $m \in \mathcal M$.} The function $g$ determines their risk preferences as well as the complementarity/substitutability of scores and diversity (if $g$ is convex (concave) at a point, then scores and diversity are complements (substitutes) at that point).

As we later show, this assumption allows for particularly simple functional forms for optimal mechanisms. We explore the robustness of our results to relaxing the separability, differentiability, and concavity embedded in Assumption \ref{sep} in Appendix \ref{ap:preferences}. Most importantly, we show that our results are essentially unchanged when preferences are non-separable over diversity, \textit{i.e.,} when $\sum_{m\in\mathcal{M}}u_m(x_m)$ is replaced with $u(x)$. Among other things, this allows our model to capture preferences in situations with overlapping group membership \citep{aygun2021college,sonmez2022affirmative}, \textit{e.g,} when people have different genders and belong to different socioeconomic groups.\footnote{As a simple example, if people can be men $m$ or women $w$ and rich $r$ or poor $p$ (so the groups are $\mathcal{M}=\{wp,wr,mp,mr\}$) and the authority cares about increasing the representation of women and poor people, then the utility function could be given by $u(x)=\hat{u}(x_{wp}+x_{wr},x_{mp}+x_{wp})$ and our analysis would apply so long as $u$ is concave.} The essential assumption for our results is the weak separability of diversity and score preferences.\footnote{The assumption of separable preferences over scores and diversity is common in the literature on affirmative action concerns \citep[see e.g.,][]{athey2000mentoring,chan2003does,ellison2021efficiency}.} When this fails, it is no longer possible to specify optimal mechanisms without explicitly conditioning the allocation on the realized distribution of agents. While separability does impose structure, we re-emphasize that the separability embodied in Assumption \ref{sep} does not rule out complementarity or substitutability between scores and diversity and does not even rule out that scores and diversity can be substitutes local to some allocations and complements local to other allocations.

We define the value of a mechanism $\phi$ under distribution $\Lambda$ as the authority's expected utility of the allocations induced by that mechanism:
\begin{equation}
\label{eq:expectedu}
    \Xi(\phi,\Lambda)=\int_{\Omega}\xi(\bar{s}_h(\phi(\omega),\omega),x(\phi(\omega),\omega))\dd\Lambda(\omega)
\end{equation}
We say that a mechanism is first-best optimal if it maximizes the authority's expected utility for all possible \textit{distributions of} measures of agents' characteristics.

\begin{defn}[First-Best Optimality]
\label{def:opt}
A mechanism $\phi^*$ is \textit{first-best optimal} if:
\begin{equation}
    \Xi(\phi^*,\Lambda)=\sup_{\phi}\Xi(\phi,\Lambda)
\end{equation}
for all $\Lambda\in\Delta(\Omega)$.
\end{defn}

This is a demanding property for a mechanism to possess as it requires a mechanism to implement an \textit{ex post} optimal allocation in all states of the world. Moreover, as the example from Section \ref{weitzmansection} shows, priority and quota mechanisms can fail to be first-best optimal while APM can attain first-best optimality. This is despite the fact that the optimal APM can be described without reference to the state of the world. This allows the optimal APM to be defined \textit{ex ante} and communicated to stakeholders (for example, students' families) in a simple way, without any reference to the beliefs of the authority or the states of the world. In the remainder of this section, we formally define APM, show that (when suitably designed) they are first-best optimal, and characterize the conditions under which priorities and quotas are first-best optimal.

\subsection{Adaptive Priority Mechanisms}
Toward deriving a first-best optimal mechanism, we introduce APMs. To this end, we first introduce an \textit{adaptive priority policy} $A=\{A_m\}_{m\in\mathcal{M}}$, where $A_m:\mathbb{R}\times[0,1]\rightarrow\mathbb{R}$. The adaptive priority policy assigns priority $A_m(y_m,s)$ to an agent with score $s$ in group $m$ when measure $y_m$ of agents of the same group is allocated the object. Given an adaptive priority policy, an APM implements allocations in the following way: 

\begin{defn}[Adaptive Priority Mechanism]\label{def1}
An adaptive priority mechanism, induced by an adaptive priority $A$, implements an allocation $\mu$ in state $\omega$ if the following are satisfied:
\begin{enumerate}
    \item Allocations are in order of priorities: $\mu(\theta) = 1$  if and only if for all $\theta'$ with $\mu(\theta') = 0$, we have that:
    \begin{equation}
        A_{m(\theta)}(x_{m(\theta)}(\mu,\omega),s(\theta)) > A_{m(\theta')}(x_{m(\theta')}(\mu,\omega),s(\theta'))
    \end{equation}
    \item The resource is fully allocated:
    \begin{equation}
        \sum_{m \in \mathcal M} x_m(\mu,\omega) = q
    \end{equation}
\end{enumerate}
\end{defn}

With some abuse of terminology, we will often refer to an APM as the adaptive priority $A$ that induces it. By way of illustration, we provide a simple example of the flexibility of APM to act like a hybrid of priority and quota policies.
\begin{example}
Let $\mathcal M = \{m,n\}$ and the capacity be $q = 0.5$. We consider the adaptive priority policy $A=\{A_m,A_n\}$ given by:
\begin{equation}
A_m(x,s)=
     s,   
 \hspace{4ex}
A_n(x,s) = \begin{cases}
            s+1 &\text{ if } x \leq 0.1\\
            s+0.1 &\text{ if } x \in (0.1,0.25)\\
            s &\text{ if } x \geq 0.25
    \end{cases}
\end{equation}
This leaves the score of group $m$ agents unchanged and gives agents of group $n$ a score boost of: $1$ if less than measure $0.1$ group $n$ agents is assigned, $0.1$ if between measure $0.1$ and $0.25$ group $n$ agents is assigned, and no score boost at all if measure greater than $0.25$ group $n$ agents is assigned. 

To understand the properties of this adaptive priority policy, observe that the highest-scoring measure $0.1$ group $n$ agents is guaranteed the resource, even in states where they score badly. Therefore, $A_n$ practically embeds a quota of $0.1$. For admissions levels between $0.1$ and $0.25$, the APM acts like a priority policy and boosts the scores of group $n$ agents by $0.1$, increasing the admissions of group $n$ when group $n$ agents score moderately well. For admissions levels beyond $0.25$, group $n$ agents are given no further advantage. Thus, when diversity is attained, this APM ``phases out'' and no longer advantages any group.
\end{example}

At this point, we have not established that a given APM implements any allocation at all, or that it implements a unique allocation (\textit{i.e.,} it may not even be a mechanism). However, there is a natural subclass of APM that do implement a unique allocation: those that are monotone. An APM $A$ is \textit{monotone} when (i) $A_m(\cdot,s)$ is a decreasing function for all $m\in\mathcal{M}, s\in[0,1]$ and (ii) $A_m(y_m,\cdot)$ is a strictly increasing function for all $m\in\mathcal{M}, y_m\in\mathbb{R}$.\footnote{Observe that monotone adaptive priority mechanisms are fair in the sense that they preserve the ranking of agents within any group and assign higher priority to an agent whenever there are fewer agents from her group who are allocated the resource.}

\begin{prop}
\label{lem:APMprop}
Any Monotone APM $A$ implements an essentially unique allocation.
\end{prop}

Moreover, the unique outcome of a monotone APM can be implemented ``greedily:''\footnote{Formally, when we consider an Adaptive Priority \textit{Mechanism}, we are studying any selection from the set of allocations that the APM implements. As monotone APMs implement an essentially unique allocation, this is without loss of optimality. When we refer to the ``unique'' allocation, we refer to the cutoff allocation defined in the proof of Proposition \ref{lem:APMprop} and which is implemented by Algorithm \ref{alg1}.}

\begin{alg}[Greedy Algorithm for Implementation of APM]
\label{alg1}
The greedy APM algorithm proceeds in the following four steps:
\begin{enumerate}
    \item For each $\theta$, define
    \begin{equation}
        \overline{x}(\theta) = \int_{s(\theta)}^1 f_{\omega}(s,m(\theta))\dd s
    \end{equation}
    as the measure of agents who have higher scores than $\theta$ and belong to same group.
    \item Construct a ranking of the agents as 
    \begin{equation}
        R(\theta) = A_{m(\theta)}(\overline{x}(\theta),s(\theta))
    \end{equation}
    \item Define the cutoff ranking for the agents as $\overline{R}_{\omega}$ by
    \begin{equation}
        \int_{\Theta} \mathbb{I}\{R(\theta) \geq \overline R_{\omega} \} \dd F_{\omega}(\theta) = q
    \end{equation}
    \item Allocate the resource to all $\theta$ with $R(\theta) \geq \overline{R}_{\omega}$.
\end{enumerate}
\end{alg}

Intuitively, this algorithm works by ranking all agents by their score within each group $m$ and assigning agents in order of their transformed scores evaluated at the measure of \textit{already assigned} agents of the same group, conditional on their admission. Informally, the algorithm greedily moves down the ranking of agents until the resource is exhausted.

\subsection{Adaptive Priority Mechanisms Achieve the First-Best}
Having shown that monotone APM implement a unique allocation and provided an algorithm to compute this allocation, we now show that a certain, monotone APM is first-best optimal:

\begin{thm}
\label{subsched}
The following APM is monotone and first-best-optimal:
\begin{equation}
    A_m^*(y_m,s)\equiv h^{-1}(h(s)+u_m'(y_m))
\end{equation}
Moreover, if a mechanism is first-best-optimal, then it implements essentially the same allocations as $A^*$.
\end{thm}

Observe that $A^*$ is not only uniquely first-best optimal, it also requires only that the authority knows its preferences over scores $h$ and diversity $u_m$. Importantly, it need have no knowledge of the underlying distribution of agents and can be fully specified even without any knowledge of the nature or extent of uncertainty, $\Lambda$. Moreover, this mechanism does not depend at all on the authority's across-state risk and complementarity preferences, $g$. This is because it achieves the \textit{ex post} optimal allocation in all states and so there is no need to trade off gains and losses across states.

To gain intuition for the form of this mechanism, suppose that the authority has linear utility over scores $h(s)\equiv s$. In this case, $A_m^*(y_m,s)=s+u_m'(y_m)$, so an agent in group $m$ is awarded a boost of $u_m'(y_m)$ when there are $y_m$ higher-scoring agents of the same group, their direct marginal contribution to the diversity preferences of the authority. This is optimal because this boost precisely trades off the marginal benefit of additional diversity with the marginal costs of reduced scores. Moreover, failing to award this precise level of boost would result in a suboptimal allocation. Thus, any optimal mechanism must be essentially identical to the optimal APM we have characterized. To generalize this beyond linear utility of scores, consider the following observation: we can map agents' scores from $s$ to $h(s)$, and consider the optimal boost in this space. As $h$ is monotone, this preserves the ordinal structure of the optimal allocation, and the authority has linear preferences over $h(s)$. Thus, in this transformed space, the optimal boost remains additive and given by $u_m'(y_m)$. To find the optimal transformed score in the original space, we simply invert the transformation $h$ and apply it to the optimal score in the transformed space, yielding the formula for the optimal mechanism in Theorem \ref{subsched}.

\subsection{(Sub)Optimality of Priorities and Quotas}
We have shown that APM are optimal. However, the primary classes of mechanisms that have been used in practice are priority and quota mechanisms. Therefore, it is important to understand whether (and when) these mechanisms are also optimal. We now establish that APM generally provide a strict improvement over priority and quota mechanisms and characterize when priority and quota mechanisms attain optimality.

We first formally define priority and quota mechanisms. A \textit{priority policy} $P:\Theta \to [0,1]$ awards an agent of type $(s,m)\in\Theta$ a priority $P(s,m)$, that depends on both their score and group.

\begin{defn}[Priority Mechanisms]
A priority mechanism, induced by a priority policy $P$, allocates the resource in order of priorities until measure $q$ has been allocated, with ties broken uniformly and at random. 
\end{defn}

We define a \textit{quota policy} as $(Q,D)$, where $Q=\{Q_m\}_{m \in \mathcal M}$ and $D:\mathcal{M}\cup\{R\}\rightarrow \{1,2,\ldots,|\mathcal{M}|+1\}$ is a bijection. The vector $Q$ reserves measure of the capacity $Q_m$ for agents in group $m$, with residual capacity $Q_R=q-\sum_{m\in\mathcal{M}}Q_m$ open to agents of all types. The bijection $D$ (the precedence order) gives the order in which the groups are processed.

\begin{defn}[Quota Mechanisms]
\label{quotadef}
A quota mechanism, induced by a quota policy $(Q,D)$, proceeds by allocating the measure $Q_{D^{-1}(k)}$ agents from group $D^{-1}(k)$ (if there are sufficient agents from this group) to the resource in ascending order of $k$, and in descending order of score within each $k$. If there are insufficiently many agents of any group to fill the quota, the residual capacity is allocated to a final round in which all agents are eligible.
\end{defn}

We now characterize when priority and quota mechanisms are (sub)optimal. To do this, we first provide some definitions. Authority preferences are \textit{non-trivial} if for all $m,n\in\mathcal{M}$, we have that:
\begin{equation}
     h(1)+ u'_{n}(0) > h(0)+ u'_{m}(q)
\end{equation}
Intuitively, the authority's preferences are non-trivial when their concerns for representation of certain groups do not always outweigh the consideration of scores.\footnote{Note that failure of non-triviality means there exists $m$ and $n$ such that $ h(1) +  u'_{n}(0) \leq +  h(0)+u'_{m}(q)$, \textit{i.e.,} a group $n$ agent with the maximum score is less preferred than a group $m$ agent with the minimum score even when all of the entire capacity is allocated to group $m$ agents.} The authority is \textit{risk-neutral} over diversity if for all $m \in \mathcal M$, $u'_{m}: [0,q] \to \mathbb{R}$ is constant, \textit{i.e.,} there are constant marginal returns to admitting more agents from all groups. If there are decreasing marginal returns, then the authority's preferences feature risk aversion. We define extremely risk-averse preferences as follows. Let $\tilde u$ and $\tilde h$ be functions describing diversity and score preferences, and let $\{x_m^{\text{tar}}\}_{m \in \mathcal M}$ be a vector of target allocation levels. Moreover, assume that these satisfy: (i) $\tilde u_m'(x_m)= 0$ for all $x_m>x_m^{\text{tar}}$ (ii) $\tilde u_m'(x_m) \geq \tilde h(1)-\tilde h(0)$ for $x_m\leq x_m^{\text{tar}}$ and (iii) $\sum_{m\in\mathcal{M}}x_m^{\text{tar}} \leq q$. Intuitively, an authority whose preferences are represented by $\tilde u$ and $\tilde h$ is very risk-averse as the condition that $\tilde u_m'(x_m) \geq \tilde h(1)-\tilde h(0)$ implies that the loss from being below the target level for a group $x_m^{\text{tar}}$ dominates any benefit from increased scores. Thus, they are infinitely risk-averse to failing to meet this target. We say that the authority is \textit{extremely risk-averse} if the authority's preferences over the optimal allocations can be represented by $(\tilde u, \tilde h)$.\footnote{More formally, this means that there exists $(\tilde u,\tilde h)$ such that the optimal allocation under $(u,h)$ is also optimal under $(\tilde u,\tilde h)$ for all $\omega\in\Omega$.}

\begin{thm}
\label{rationalization}
Suppose that the authority has non-trivial preferences. The following statements are true:
\begin{enumerate}
    \item There exists a first-best optimal priority mechanism if and only if the authority is risk-neutral. Moreover, this mechanism is given by $P(s,m)=h^{-1}(h(s)+u_m')$.
    \item There exists a first-best optimal quota mechanism if and only if the authority is extremely risk-averse. Moreover, this mechanism is given by $Q_m=x_{m}^{\text{tar}}$ and $D(R) = \vert \mathcal M \vert + 1$.
\end{enumerate}
\end{thm}

Theorem \ref{rationalization} provides precise conditions on preferences such that the inability of priorities and quotas to adapt to the state is not problematic. That risk-neutrality and high risk aversion are sufficient for the optimality of priority and quota mechanisms is intuitive. On the one hand, if the authority is risk-neutral over the measure of agents from different groups, then they can perfectly balance their score and diversity goals without regard for the state of the world. This is because, under risk-neutrality, there is a constant ``exchange rate'' between the two: how the authority compares any two agents does not depend on the final allocation and thus can be specified \textit{ex ante} by a priority policy. On the other hand, if the authority is extremely risk-averse as to the prospect of failing to assign $x_{m}^{\text{tar}}$ agents from group $m$, then a quota allows them to always achieve this target level of allocation in all states of the world while minimally sacrificing scores. It is less obvious that risk-neutrality and high risk aversion are necessary. We prove this result by constructing certain adversarial measures of agents that render any priority or quota mechanism suboptimal unless the authority is risk-neutral or extremely risk-averse, respectively. Importantly, this result also shows that the only optimal quota mechanisms are those that process open slots last.

This result highlights the fragility of priority mechanisms to uncertainty absent the strong assumption of risk-neutrality over diversity. Intuitively, this is because they feature no guarantees as to how many agents of different groups will be assigned. Indeed, the unfortunate interaction between priority mechanisms and unforeseen market realizations has led to public backlash against priority mechanisms. For example, in the Vietnamese university admissions system, which combines exam scores with priority boosts for students from disadvantaged groups, a year of unexpectedly easy exams led to ``top-scoring students missing out on the opportunity to attend their university of choice'' and generated backlash against the system \citep{vietnam}. 

Moreover, our result highlights that quota mechanisms similarly fail to achieve the first-best away from high levels of risk aversion as they do not take advantage of the potential for positive selection. Our quantitative analysis in Section \ref{sec:quant} in the context of quota mechanisms in Chicago Public Schools suggests that the variation in the distribution of characteristics across years generates meaningful welfare gains from switching to APM.

To formalize the connection between uncertainty and the importance of the adaptability of APM, we consider a setting with \textit{no uncertainty}, where $\Lambda$ is a Dirac measure. In this context, we say that a mechanism is optimal without uncertainty if it is a utility maximizer. 

\begin{prop}\label{prop:nouncertainty}
If there is no uncertainty, then there exist optimal priority and quota mechanisms.
\end{prop}
This result shows that if an authority is certain about the market, then appropriately constructed priority and quota mechanisms would be optimal. This formalizes the idea that the suboptimality of priority and quota mechanisms stems from their inability to adapt to the state. Of course, in practice, an authority is always somewhat uncertain of the market they face. We will later show empirically that this is the case for CPS as we find substantial variation in the joint distribution of student scores and groups across years. Thus, absent the strong conditions on authority preferences that we have characterized in Theorem \ref{rationalization}, APM dominate priority and quota mechanisms.

\section{Equilibrium Mechanisms with Multiple Authorities}
\label{sec:eq}
The single-authority model is relevant for many resource allocation contexts, such as the medical resource allocation problem of a hospital and the allocation of (homogeneous) government jobs to candidates. However, in other settings such as school or university admissions, multiple authorities must decide upon their admissions policies and rules. In this section, we generalize our single authority model to a setting with multiple authorities. We define \textit{stability} in this setting and show that there is a unique stable allocation. Moreover, we show that a mechanism is consistent with stability if and only if it coincides with the \textit{single-authority-optimal} APM from Theorem \ref{subsched}. We then consider a model where agents sequentially apply to the authorities, who decide which agents to admit. We show that the optimal APM is a dominant strategy. Moreover, we show that in any equilibrium in which authorities use the optimal APM, the resulting allocation corresponds to the unique stable matching of the economy. Taken together, our results provide cooperative (stability) and non-cooperative (dominance) foundations for recommending the use of APM in multi-authority settings.

\subsection{The Multi-Authority Model}
There are authorities $c\in\mathcal{C}= c_0 \cup \bar{\mathcal C}=\{c_0,c_1,\ldots,c_{|\mathcal{C}|-1}\}$ with capacities $q_c$, where $c_0$ is a dummy authority that corresponds to an agent going unmatched. The agents differ in their authority-specific scores, the group to which they belong, and their preferences over the authorities, $\succ$. We index agents by their type $\theta=(s,m,\succ)\in[0,1]^{|\mathcal C|}\times\mathcal{M}\times\mathcal{R}=\Theta$, where $\mathcal{R}$ is set of all complete, transitive, and strict preference relations over $\mathcal{C}$. For each type $\theta$, $s_c(\theta)$ denotes the score of $\theta$ at authority $c$ and $m(\theta)$ denotes the group of $\theta$. From now, to economize on notation, we suppress indexing states by $\omega\in\Omega$ and let the measure of types be $F$, with density $f$.\footnote{Formally, this density is given by $f(s,m,\succ)=\frac{\partial}{\partial s} F(s,m,\succ)$.} We assume that $f$ has full support over $\Theta$ (\textit{i.e.,} $f>0$) and that $F(\Theta)$ is less than the capacity of $c_0$ and greater than the capacity of $\bar{\mathcal C}$.

Each authority has preferences over the agents they are assigned of the form introduced in the previous section:
    \begin{equation}\label{schoolutility}
        \xi_{c}(\bar{s}_{h_c},x_c)=g_c\left(\bar{s}_{h_c}+\sum_{m\in\mathcal{M}}u_{m,c}(x_{m,c})\right)
    \end{equation}
where the extent to which they care about risk $g_c$, scores $h_c$, and diversity $\{u_{m,c}\}_{m\in\mathcal{M}}$ are potentially specific to each authority. 

A matching is a function $\mu: \mathcal C \cup \Theta \to 2^{\Theta} \cup \mathcal C$ where $\mu(\theta) \in \mathcal C$ is the authority that any type $\theta$ is assigned and $\mu(c) \subseteq \Theta$ is the set of agents that is assigned to authority $c$.\footnote{The mathematical definition of a matching for the continuum economy we study follows \cite{azevedo2016supply} and requires that $\mu$ satisfies the following four properties: (i) for all $\theta \in \Theta$, $\mu(\theta) \in \mathcal C$; (ii) for all $c \in \mathcal C$, $\mu(c) \subseteq \Theta$ is measurable and $F(\mu(c)) \leq q_c$; (iii) $c = \mu(\theta)$ iff $\theta \in \mu(c)$; (iv) (open on the right) for any $c \in \mathcal C$, the set $\{\theta \in \Theta: c \succ_{\theta} \mu(\theta)\}$ is open.} Given a matching $\mu$, $\bar{s}_{h_c,c}(\mu)$ and $x_c(\mu) = \{x_{m,c}(\mu)\}_{m \in \mathcal M}$ denote the score indices and measures of agents from different groups matched to $c$ at $\mu$. We say that $c$ \textit{prefers} $\mu$ to $\mu'$, which we denote by $\mu\succ_c \mu'$, if $   \xi_c(\bar{s}_{h_c,c}(\mu), x_c(\mu)) > \xi_c(\bar s_{h_c,c}(\mu'),x_c(\mu'))$. Toward representing a matching as a lower-dimensional object, we moreover define a cutoff matching as one in which agents are assigned to the authority that they most prefer among the set of authorities in which their score clears a group-specific threshold:

\begin{defn}\label{defn:cutoffmatching}
A matching $\mu$ is a cutoff matching if there exist cutoffs $S =\{S_{m,c}\}_{m \in \mathcal M, c \in \mathcal C}$ such that  $\mu(\theta) = c$ if (i) $s_c(\theta) \geq S_{m(\theta),c}$ and (ii) for all $c'$ with $c' \succ_{\theta} c$, $s_{c'}(\theta) <  S_{m(\theta),c'}$.
\end{defn}

Given $S$, the \textit{demand} of an agent $\theta$ is their favorite authority among those for which they clear the cutoff:
\begin{equation}
    D^{\theta}(S) = \{c: s_c(\theta) \geq  S_{m(\theta),c} \text{ and } c \succeq_{\theta} c' \text{ for all } c' \text{ with } s_{c'}(\theta) \geq  S_{m(\theta),c')} \}
\end{equation}
The \textit{aggregate demand} for authority $c$ is the set of agents who demand it $D_c(S) = \{\theta: D^{\theta}(S) = c\}$, while $\tilde D_c(S_{-c}) = D_c((0,\ldots,0),S_{-c})$ returns the set of all agents who would demand $c$ if offered admission when other authorities' cutoffs are $S_{-c}$.

\subsection{Characterization of Stable Allocations}\label{sec:characterizationofstability}
We first characterize the set of stable allocations. Our context presents two challenges in this regard. First, the priorities which are typically used to define stability are not primitives of our model. Therefore, to define stability, we will use the preferences of the authorities induced by Equation \ref{schoolutility}. Second, unlike discrete models, a single agent does not affect the preferences of an authority. Therefore, we need to consider a positive mass of agents to define blocking.

For each matching $\mu$, authority $c \neq c_0$, and two sets of agents $\tilde \Theta$ and $\hat \Theta$, we let $\hat\mu_{(\hat \Theta,\tilde \Theta, c, \mu)}$ denote the matching that maps $\hat \Theta$ to $c$ and $\tilde \Theta$ to $c_0$ and otherwise coincides with $\mu$.\footnote{Formally, \begin{equation}
    \hat \mu_{(\hat \Theta, \tilde \Theta, c,  \mu)}(\theta) = \begin{cases}
            c_0 &\text{ if } \theta \in \tilde \Theta\\
            c & \text{ if } \theta \in \hat \Theta\\
            \mu(\theta) & \text{otherwise}
    \end{cases}
\end{equation}} A set of agents $\hat \Theta$ \textit{blocks} matching $\mu$ at authority $c$ by $\tilde \Theta$ if (i) for all $\theta \in \hat \Theta$, $c \succ_{\theta} \mu(\theta)$, (ii) $\tilde \Theta \subseteq \mu(c)$, (iii) $F(\tilde \Theta) = F(\hat \Theta)$, and (iv) $\hat\mu_{(\hat \Theta, \tilde \Theta,c, \mu)}\succ_c\mu$. A matching $\mu$ is \textit{not blocked} if there does not exist such a $(\hat\Theta,\tilde\Theta,c)$. A matching $\mu$ satisfies \textit{within-group fairness} if for all $\theta, \theta'\in\Theta$ such that $m(\theta') = m(\theta)$ and $s_{\mu(\theta)}(\theta') > s_{\mu(\theta)}(\theta)$, it holds that $\mu(\theta') \succeq_{\theta'} \mu(\theta)$.\footnote{Within-group fairness requires an authority to not reject an agent if it is admitting an agent from the same group with a lower score. Under our assumption that authorities prefer higher scores ($h_c$ is strictly increasing), within-group fairness is satisfied if there is no blocking in discrete models.} A matching $\mu$ is \textit{stable} if it satisfies within-group fairness, is not blocked, and all non-dummy authorities fill their capacity. The following result establishes that there exists a unique stable matching and that this is a cutoff matching.

\begin{thm}\label{prop:multischoolstable}
There is a unique stable matching. This matching is a cutoff matching.
\end{thm}

This result extends existing results on the uniqueness of stable allocations in continuum economies \citep[see Theorem 1.1 of][]{azevedo2016supply} to our setting, in which agents belong to different socioeconomic groups and the preferences of authorities depend non-linearly on their admissions of various groups. This results in the authority having preferences over various sets of agents that are endogenous to the composition of the admitted agents and necessitates new arguments.

To gain intuition for this result, first imagine that there is only one group of agents $|\mathcal M| = 1$, so that authorities' preferences are determined by the scores of the agents. Given a set of cutoffs $S_{-c}$, a cutoff $t_c$ clears the market for $c$ if $F(D_c(t_c,S_{-c}))=q_c$. When $|\mathcal M| = 1$, for a given $S_{-c}$, there is a unique $t_c$ that clears the market since a smaller cutoff will exceed the capacity while a larger one will leave a positive measure of the capacity empty. Define $T = \{T_c\}_{c\in\mathcal{C}}$, where $T_c(S)$ is the function that maps each $S$ to the market-clearing cutoff $t_c$ under $S_{-c}$. The result then follows from (i) showing the fixed points of $T$ correspond to market-clearing cutoffs of stable matchings, (ii) establishing that $T$ is monotone, (iii) applying Tarski's fixed point theorem to show that the set of market-clearing cutoffs is a lattice, and (iv) showing that there can only be one market-clearing cutoff as, if there were two, one would strictly exceed the capacities of at least one authority.

When $|\mathcal M| > 1$, there is a potential continuum of cutoffs that would clear the market for authority $c$. A selection from this set is provided by the cutoffs induced by the optimal APM, $A^*_{m,c}(y_m,s)\equiv h_c^{-1}(h_c(s)+u_{m,c}'(y_m))$. We show that the APM cutoffs are unique among the market-clearing cutoffs in being compatible with stability. This is because, for any other $t_c'$, there is a set $\hat \Theta$ of agents (with positive measure) who have scores lower than the cutoff for their group and a set $\tilde \Theta$ of agents (with positive measure) who have scores higher than the cutoff for their group, but the authority is strictly better off by admitting $\hat \Theta$ and rejecting $\tilde \Theta$. We define $T_c(S)$ as the market-clearing cutoffs induced by the optimal APM, show that the fixed points of $T_c$ correspond to market-clearing cutoffs of stable matchings, and follow the same steps as above to demonstrate uniqueness.

This hints at a connection between the stable allocation and the allocation induced by all authorities pursuing the optimal APM, which we now make explicit. The demand set of $c$ at $\mu$, $D_c(\mu)$, is the set of agents who prefer $c$ to their allocation under $\mu$. A mechanism is \textit{consistent with stability} if for all $F$ with stable matching $\mu_F$, it chooses $\mu_F(c)$ from $D_c(\mu_F)$. In other words, evaluated at the set of agents who demand an authority, this mechanism chooses the set of agents with which the authority is already matched. Moreover, we say that a mechanism $\phi$ is \textit{equivalent} to $\phi'$ if it chooses the same agents under all full support measures. We now establish that single-authority-optimal APMs (and equivalent mechanisms) comprise the full set of mechanisms that are consistent with stability.

\begin{prop}\label{thm:consistencycharacterization}
A mechanism $\phi$ is consistent with stability if and only if it is equivalent to $A^*_c$.
\end{prop}

Thus, not only is the optimal APM $A^*$ inherent to the structure of stable allocations, but it also characterizes stability in this setting in the sense that any deviation from $A^*$ would result in a violation of stability.

\subsection{APM Are Dominant Under Decentralized Admissions}\label{section:decentralized}
To study which mechanisms are optimal under decentralized admissions, we now consider a setting in which the agents apply sequentially to the authorities, who then decide which agents to admit. We index the stage of the game by $t\in \mathcal{T}=\{1,\ldots,|\mathcal{C}|-1\}$. Each stage corresponds to a (non-dummy) authority $I(t)$, where $I:\mathcal{T}\rightarrow \mathcal{T}$. At each stage $t$, any unmatched agents choose whether to apply to authority $I(t)$. Given the set of applicants, authority $I(t)$ chooses to admit a subset of these agents, who are then matched to the authority. Given this, histories are indexed by the path of the measure of agents who have not yet matched, $h^{t-1}=(F,F_1,\ldots,F_{t-1})\in\mathcal{H}^{t-1}$. Given each history $h^{t-1}$ and set of applicants $\Theta^{A}_c\subseteq\Theta$, a strategy for an authority returns a set of agents $\Theta^{G}_c\subseteq \Theta$ whom they will admit such that $\Theta^G_c\subseteq\Theta^{A}_c$ and $F_t(\Theta^G_c)\leq q_c$ for each time at which they could move $t\in\mathcal{T}$, $a_{c,t}:\mathcal{H}^{t-1}\times\mathcal{P}(\Theta)\rightarrow\mathcal{P}(\Theta)$, where $\mathcal{P}(\Theta)$ is the power set over $\Theta$.\footnote{Formally, so that $F_t(\Theta^G_c)$ is well defined, we require that authorities' strategies be measurable in the Borel sigma algebra over $\Theta$.} A strategy for an agent returns a choice of whether to apply to authorities at each history and time $t\in\mathcal{T}$ for all agent types $\theta\in\Theta$, $\sigma_{\theta,t}:\mathcal{H}^{t-1}\rightarrow[0,1]$.

Within this context, our notion of equilibrium is that of subgame perfect equilibrium:

\begin{defn}[Equilibrium]
A strategy profile $\Sigma=\{\{a_{c,t}\}_{c\in\bar{\mathcal{C}}},\{\sigma_{\theta,t}\}_{\theta\in\Theta}\}_{t\in\mathcal{T}}$ is a subgame perfect equilibrium if $a_{c,t}$ maximizes authority utility given $\Sigma$ for all $c
\in\bar{\mathcal{C}}$ and $t\in\mathcal{T}$ and $\sigma_{\theta,t}$ is maximal according to agent preferences for all $\theta\in\Theta$ and $t\in\mathcal{T}$.
\end{defn}

We moreover say that a strategy $a_{\tilde c,t}$ for an authority $\tilde c$ at time $t$ is \textit{dominant} if it maximizes authority utility regardless of the strategies of all other authorities and agents, $\{\{a_{c,t}\}_{c\in \bar{\mathcal{C}}\setminus\{\tilde c\}},\{\sigma_{\theta,t}\}_{\theta\in\Theta}\}_{t\in\mathcal{T}}$, and the order in which authorities admit agents, $I$. Moreover, an equilibrium $\Sigma$ is in \textit{dominant strategies} if $a_{ c,t}$ is dominant for all $c\in\bar{\mathcal{C}}$ and $t\in\mathcal{T}$. We denote the unique probabilistic allocation of agents to authorities induced by $\Sigma$ as $\mu_{\Sigma}:\Theta\rightarrow\Delta(\mathcal C)$. A probabilistic allocation $\mu_{\Sigma}$ is deterministic if $\mu_{\Sigma}(\theta)$ is a Dirac measure on some authority $c\in\mathcal{C}$ for all $\theta\in\Theta$. A deterministic allocation $\mu_{\Sigma}$ corresponds to a matching $\mu$ if $\mu_{\Sigma}(\theta)$ is a Dirac measure on $\mu(\theta)$ for all $\theta\in\Theta$.

We now establish that the single-authority optimal APM characterizes dominance.

\begin{thm}
\label{thm:eq}
A mechanism implements a dominant strategy for an authority if and only if it implements essentially the same allocations as $A^*_c$.
\end{thm}

The intuition behind this result is that each authority takes as given the set of agents that will accept it. Thus, given this measure of agents, they can do no better than to employ the same APM that a single authority would, which is $A^*_c$ by Theorem \ref{subsched}.

Theorem \ref{thm:eq} provides a powerful rationale for focusing on APMs in decentralized markets at both positive and normative levels.  Normatively, this result allows an analyst to advise an authority regarding how it should conduct its admissions. This is important because any policy that does not coincide with the APM we derive --- such as the popular priority and quota mechanisms outside of the cases delimited by Theorem \ref{rationalization} --- will disadvantage an authority. Positively, this result allows a sharp prediction that the equilibrium matching between agents and authorities will be the unique stable matching (as per Theorem \ref{prop:multischoolstable}):

\begin{prop}\label{prop:apmallocationstable}
For all equilibria $\Sigma^*$ where authorities use $A^*$, the allocation $\mu_{\Sigma^*}$ is deterministic and corresponds to the unique stable matching of this economy.
\end{prop}

The intuition is that if an equilibrium matching under $A^*$ was not the unique stable matching, then it must be that some agents are applying suboptimally and failing to select the most preferred authority that they can attend, which contradicts that the outcome is consistent with equilibrium. Proposition \ref{prop:apmallocationstable} also shows that the allocation implemented in the dominant strategy equilibrium is the same for all possible orderings of authorities, $I$.

\section{Efficient Mechanisms with Multiple Authorities}
\label{sec:cent}
We have so far characterized the decentralized outcome, but two natural questions remain. First, is the decentralized outcome efficient for the authorities?\footnote{We have shown that the decentralized outcome corresponds to the unique stable matching. Naturally, stable allocations need not be efficient for the students. Here, we will see if they are efficient for the authorities.} Second, if not, what kind of centralized solution can remedy any inefficiency? We show that the decentralized outcome is generally inefficient and that a modified, centralized APM mechanism restores efficiency.

\subsection{Inefficiency of the Decentralized Outcome}
The notion of efficiency that we will consider is utilitarian efficiency over authorities. A mechanism in the multi-authority setting is a function $\phi:\Omega\rightarrow\mathcal{U}$, where $\mathcal{U}$ is the set of matchings (which, by definition, encodes the feasibility requirement imposed in the single authority setting). We define the total authority value $\Xi_T$ of a mechanism $\phi$ under distribution $\Lambda\in\Delta(\Omega)$ as the total expected utility of the allocations induced by that mechanism:
\begin{equation}
    \Xi_T(\phi,\Lambda)=\sum_{c\in\bar{\mathcal{C}}}\Xi_c(\phi,\Lambda)
\end{equation}
A mechanism is efficient if it maximizes total authority value for all possible distributions:

\begin{defn}[Efficiency]
A mechanism $\phi^*$ is \textit{efficient} if:
\begin{equation}
    \Xi_T(\phi^*,\Lambda)=\sup_{\phi}\Xi_T(\phi,\Lambda)
\end{equation}
for all $\Lambda\in\Delta(\Omega)$.
\end{defn}

For this section, so that scores are directly comparable across authorities and allocations are interior, we impose the following assumption:

\begin{assumption}
\label{ass2}
Scores and preferences are such that $s_c(\theta) = s_{c'}(\theta)$, $h_c=h$ and $g_c=\text{Id}$, where $\text{Id}$ is the identity function, for all $c,c'\in\bar{\mathcal{C}}$ and $\theta\in\Theta$. Moreover, $\lim_{x\rightarrow^{+}0}u_{m,c}'(x)=\infty$, $u_{m,c}$ is strictly concave for all $m\in\mathcal{M}$ and $c\in\bar{\mathcal{C}}$, and for all $\theta\in\Theta$, $c_0$ is less preferred than $c$ for all $c\in\bar{\mathcal{C}}$.
\end{assumption}

Assumption \ref{ass2} makes scores a common numeraire across authorities and is akin to the standard quasi-linearity assumption in mechanism design. For example, it may be suitable in settings where the score is derived from a common index of academic attainment, such as in Chicago Public Schools. This assumption does not impose that all authorities have common marginal rates of substitution between scores and diversity, as they are allowed unrestricted heterogeneity in diversity preferences. We add the Inada condition for analytical tractability. We argue that it is also reasonable to assume that failing to admit any individuals from a given group is intolerable for authorities. We add that the outside option is ranked lower than all authorities so that any agent is willing to be assigned to any of the authorities. When authorities control highly desirable resources, such as elite school or university seats or essential medical supplies, we argue that this is a reasonable assumption.

With the efficiency benchmark defined, we now demonstrate that the decentralized equilibrium outcome can fail to be efficient. We prove this result via an explicit example with two authorities, $c$ and $c'$ of capacity $\frac{1}{4}$, and two groups of agents, $m$ and $m'$ of measure $\frac{1}{2}$. All agents in group $m$ prefer $c'$ to $c$ and all agents in group $m'$ prefer $c$ to $c'$. Authority $c$ values admitting group $m$ agents more on the margin than group $m'$ agents, and authority $c'$ values admitting group $m'$ agents more on the margin than group $m$ agents. Using the optimal APMs, both authorities admit more agents of the group whose admissions they value relatively less than the efficient benchmark. The intuition for this is that both authorities ``steal'' the high-scoring agents of the group whom they relatively less value from the other authority, an externality that they do not internalize.

\begin{prop}[Equilibrium Inefficiency]
\label{prop:ineff}
All authorities using the privately optimal APMs $\{A_c^*\}_{c\in\mathcal{C}}$ is not necessarily efficient.
\end{prop}

\subsection{An Efficient Centralized Mechanism}
The inefficiency of each authority using a decentralized APM stems from the implicit incompleteness of markets: if we added the ability for authorities to pay each other for agents, then they would have willingness-to-pay to do so at the equilibrium allocation. A centralized mechanism can remedy this issue by ensuring the cross-sectional allocation of agents to authorities is optimal. 

We propose the following augmentation of an APM to solve this problem, an \textit{adaptive priority mechanism with quotas} (APM-Q). The idea behind this hybrid mechanism is to use aggregate, market-level priorities with authority-specific quotas. To this end, an APM-Q comprises an aggregate non-separable APM $\tilde A=\{\tilde A_m\}_{m\in\mathcal{M}}$ with $\tilde A_m:\mathbb{R}^{|\mathcal{M}|}\times[0,1]\rightarrow\mathbb{R}$ and a profile of quota functions $Q=\{Q_{m,c}\}_{m,c\in\mathcal{M}}$ with with $Q_{m,c}:\mathbb{R}^{|\mathcal{M}|}\rightarrow\mathbb{R}_{+}$. Intuitively, the aggregate APM pins down the aggregate measures of allocations of each group to \textit{any} authority $\{ x_m\}_{m\in\mathcal{M}}$, where $x_m = \sum_{c \in \bar{\mathcal C}} x_{m,c}$. The non-separability of this APM simply means that the measures of all groups matter for the adaptive priority of any agent. Given the aggregate measure of allocation for group $m$, the quota function for authority $c$ assigns $Q_{m,c}(\{ x_m\}_{m\in\mathcal{M}})$ agents of type $m$ to authority $c$.

\begin{defn}[Adaptive Priority Mechanism with Quotas] An adaptive priority mechanism with quotas $(\tilde A,Q)$ comprises a non-separable APM $\tilde A$ and a quota function $Q$. An APM-Q implements allocation $\mu$ if the following are satisfied:
\begin{enumerate}
    \item Aggregate allocations are in order or priorities: $\mu(\theta) \in \bar{\mathcal C}$  if and only if for all $\theta'$ with $\mu(\theta') = c_0$, we have that:
    \begin{equation}
        \tilde A_{m(\theta)}(\{ x_m(\mu)\}_{m\in\mathcal{M}},s(\theta)) > \tilde A_{m(\theta')}(\{ x_m(\mu)\}_{m\in\mathcal{M}},s(\theta'))
    \end{equation}
    \item Authority-level allocations are given by the corresponding quota functions:
    \begin{equation}
        x_{m,c}(\mu) = Q_{m,c}(\{ x_m(\mu)\}_{m\in\mathcal{M}})
    \end{equation}
    \item The resources are fully allocated:
    \begin{equation}
        \sum_{m \in \mathcal M} x_{m,c}(\mu) = q_c
    \end{equation}
\end{enumerate}
\end{defn}

By appropriate choice of the APM and quota functions, we can derive an APM-Q that is efficient. To this end, define the optimally-allocated aggregate utility from diversity:
\begin{equation}
\begin{split}
    &\tilde u(\{ x_m\}_{m\in\mathcal{M}}) = \max_{\{x_{m,c}\}_{c\in\mathcal{C}}}\sum_{c\in\mathcal{C}}\sum_{m\in\mathcal{M}}u_{m,c}(x_{m,c})\\
    & \quad \text{s.t.}\, \sum_{c\in\mathcal{C}}x_{m,c}\leq x_m, \, \sum_{m\in\mathcal{M}}x_{m,c}\leq q_c, \, \forall m\in\mathcal{M}, c\in\mathcal{C}
\end{split}
\end{equation}
Moreover, define the marginal value of aggregate group $m$ admissions $\tilde u^{(m)}(y)=\frac{\partial}{\partial y_m}\tilde u(y)$ and the marginal value of authority capacity $\tilde u_{q_c}(y)=\frac{\partial}{\partial q_c}\tilde u(y)$. Using these marginal values, we can design an efficient APM-Q that combines market-level APMs with authority-level quotas: 

\begin{prop}[Efficient APM-Q]
\label{thm:apmq}
Every allocation induced by the following APM-Q $(\tilde A,Q)$ is efficient:
\begin{enumerate}
    \item The non-separable APM is given by $\tilde A_m(y,s)=h^{-1}\left(h(s)+\tilde u^{(m)}(y)\right)$
    \item The quota functions are given by $Q_{m,c}(y)=\left(u_{m,c}^{'}\right)^{-1}\left(\tilde u^{(m)}(y)+\tilde u_{q_c}(y)\right)$
\end{enumerate}
\end{prop}

The proof of this result constructs a fictitious aggregate authority in our single object setting. The claimed APM is optimal for this aggregate authority by a non-separable adaptation of Theorem \ref{subsched}. The substantial step in this proof establishes that $\tilde u$ is increasing, concave, and differentiable by employing the restrictions provided by Assumption \ref{ass2}. Then, given the allocation induced by this APM, we construct the quota function to optimally allocate the level of aggregate admissions induced by the APM.

Intuitively, this mechanism remedies inefficiency by ``completing markets.'' There is a common ``market price'' for each group given by $\mathcal{P}_m=\tilde u^{(m)}(x)$ and an authority-level ``shadow price of admissions'' $\mathcal{P}_c=\tilde u_{q_c}(x)$. Authorities are allocated agents so that the marginal benefit of additional agents equals the sum of the market price and shadow price of admissions $u_{m,c}'(x_{m,c})=\mathcal{P}_m+\mathcal{P}_c$. Hence, through the completion of markets, a centralized planner can allocate agents efficiently and internalize the externalities that prevented efficiency under the decentralized outcome. Notice that this market involves relatively few prices as it involves only $|\mathcal{M}|+|\mathcal{C}|$ shadow prices rather than the full set of $|\mathcal{M}|\times|\mathcal{C}|$ marginal values.

\section{Benchmarking the Quantitative Gains from APM}
\label{sec:quant}
We have so far shown theoretically that APM outperform conventional priority and quota mechanisms. In this section, we attempt to benchmark the magnitude of the gains from implementing APM relative to the reserve system employed by Chicago Public Schools (CPS). To do this, we use application and admission data from CPS for the 2013-2017 academic years. Estimating preference parameters to best rationalize the pursued reserve policy, we find that the gains from using the optimal APM are equivalent to removing 37.5\% of the loss to CPS' payoffs from failing to admit a diverse class of students. Our analysis therefore suggests that the gains from APM are considerable.

\subsection{Institutional Detail on Chicago Public Schools}

We first describe the institutional context of CPS. Under current policy, CPS admits students to its selective exam schools based on two criteria. First, CPS ranks students according to a score which combines the results of a specialized entrance exam, prior standardized test scores, and grades in prior coursework. This composite score ranges from $0$ to $900$ and higher-scoring students are admitted before lower-scoring ones, reflecting CPS's desire to allocate seats in exam schools to the students with the best academic standing. In our model, these are the students' scores. Second, CPS divides the census tracts in the city into four \textit{tiers} based on socioeconomic characteristics.\footnote{Concretely, 800 census tracts are divided into four tiers based on six characteristics of each census tract: (i) median family income, (ii) percentage of single-parent households, (iii) percentage of households where English is not the first language, (iv) percentage of homes occupied by the homeowner, (v) adult educational attainment, and (vi) average Illinois Standards Achievement Test scores for attendance-area schools. These characteristics are then combined to construct the socioeconomic score for the tract. Finally, the tracts are ranked according to socioeconomic scores and partitioned into 4 tiers with approximately the same number of school-age children. See \cite{ellison2021efficiency} for a more detailed account of the CPS system.} Tier 1 tracts are the most disadvantaged, while Tier 4 tracts are the most advantaged. This is reflected in the composite scores of students from Tier 1, who represent $25\%$ of the city's population but comprise relatively few of the high-scoring students \citep{ellison2021efficiency}. As a result, Tier 1 students would have a very small share in the city's top exam schools without affirmative action. To ensure more equal representation across socioeconomic status in these schools, between 2013-2017 CPS implemented a quota policy that reserves $17.5$\% of the seats for each tier, yielding a total of $70$\% reserve seats and $30$\% merit slots that are open to students from all tiers. CPS allocates the seats by first assigning the highest-scoring students (regardless of their tier) to the merit slots and then the highest-scoring students from each tier to the $17.5$\% reserve seats.

We focus on the most selective CPS school, Walter Payton College Preparatory High School (Payton), which has the highest cutoffs for each tier in each year in our data and would have very few tier $1$ students without affirmative action.\footnote{This approach follows the analysis in \cite{ellison2021efficiency}, who focus on the two most competitive schools, Northside College Preparatory High School (Northside) alongside Payton. In the years we study, the cutoff scores for Northside are below some other schools frequently, which is why we restrict attention to Payton.} Table \ref{table:cutoffscores} presents the cutoff scores of each tier (\textit{i.e.,} the composite score of the last admitted student from each tier). 

We make two observations. First, the cutoff students from less advantageous tiers face is lower than the cutoff for students from more advantageous tiers. Therefore, CPS has a \textit{revealed} preference (and not merely a stated preference) for a diverse student body. Second, cutoff scores vary across years. This implies that the distribution of applicant characteristics varies from year to year. Given this uncertainty and the fact that CPS uses a policy that processes quotas after open slots, we know by Theorem \ref{rationalization} that CPS' baseline policy cannot be rationalized as optimal (even if they are extremely risk-averse). Nevertheless, it is always possible that the gains from APM could be small.

\begin{table}[]
\caption{Admissions Cutoff Scores for Payton}
\vspace{-3ex}
\begin{center}
\begin{tabular}{@{}cccccc@{}}
\toprule
Cutoff Score & 2013 & 2014 & 2015 & 2016 & 2017 \\ \midrule
Tier 1        & 801  & 838  & 784  & 769  & 771  \\
Tier 2        & 845  & 840  & 831  & 826  & 846  \\
Tier 3        & 871  & 883  & 877  & 853  & 875  \\
Tier 4        & 892  & 896  & 891  & 890  & 894  \\ \bottomrule
\end{tabular}
\end{center}
\fnote{The table reports the score of the lowest-scoring student that was admitted to Payton in each of the four tiers and five years.}
\label{table:cutoffscores}
\end{table}

\subsection{Preferences and Estimation Methodology}
We perform our analysis in two steps: establishing a parametric framework for evaluating welfare gains and losses and then estimating its parameters. 

\paragraph{Preferences} We assume a parsimonious, parametric form for CPS's preferences to evaluate the gains from APM. In particular, we impose that the preferences of CPS over the scores and diversity of the student body are represented by the following parametric utility function:
\begin{equation}\label{CPSutility}
  \xi(\bar{s},x;\beta,\gamma) = \bar{s} + \sum_{t=1}^4 \beta |x_t - 0.25|^{\gamma}
\end{equation}
where $\bar{s}$ is the average score of admitted students, $x_t$ is the percentage of tier $t$ students. Motivated by CPS' desire to allocate the highest-scoring students, $\xi$ is increasing in $\bar{s}$. To model the diversity preferences of CPS, we assume that CPS loses as the gap from equal representation in each tier increases. We do this through the functional form $\beta |x_t - 0.25|^{\gamma}$. The parameters $\beta$ and $\gamma$ index the slope and curvature of utility in losses from unequal representation and are the two free parameters of our framework. The preferences that we assume follow \cite{ellison2021efficiency}, but also allow for a score-diversity trade-off that depends on the level of diversity. Our goal in assuming these preferences is to arrive at some sense of the gains from APM while acknowledging that it is, as in all contexts, impossible to know the parametric class in which the authority's preferences lie. To probe robustness to this functional form assumption, in Appendix \ref{sec:alternativeutilityfunctions} we consider two other parametric utility functions that: i) estimate separate coefficients for underrepresented and overrepresented tiers and ii) only consider losses from underrepresented tiers.

\paragraph{Estimation} We estimate $\beta$ and $\gamma$ to best rationalize the choice of 17.5\% reserves for each tier as optimal. We believe this to be a reasonable approach, as the size of the reserves is an important issue that is decided only after much deliberation.\footnote{These points are emphasized in \cite{dur2020explicit}: ``This change was made at the urging of a Blue Ribbon Commission (BRC, 2011), which examined the racial makeup of schools under the $60$\% reservation compared to the old Chicago’s old system of racial quotas. They advocated for the increase in tier reservations on the basis it would be ``improving the chances for students in neighborhoods with low performing schools, increasing diversity, and complementing the other variables.''} Moreover, CPS has used the size of the reserves as a policy tool, increasing them from $15 \%$ to $17.5 \%$ in 2012 and is currently deliberating another change that would further boost the representation of tier 1 and tier 2 students \citep{cpspolicy}.

Given our functional form, the optimality of the chosen reserve sizes yields moment conditions that we use to estimate the parameters $\beta$ and $\gamma$. Formally, we index reserve mechanisms by the reserve sizes of the four socioeconomic tiers $r=(r_1,r_2,r_3,r_4)$. We let $\bar{s}(r,y)$ and $x(r,y)$ denote the average scores and tier percentages that would be obtained in year $y$, with distribution $F_y$, under reserve policy $r$. The payoff of the policymaker under reserve policy $r$ is given by $\Xi(r,\Lambda;\beta,\gamma)$, as per Equation \ref{eq:expectedu}:
\begin{equation}
      \Xi(r,\Lambda;\beta,\gamma) =  \mathbb E_{\Lambda}[\xi(\bar{s}(r,y),x(r,y);\beta,\gamma)]
\end{equation}
where the expectation is taken over distributions of agents' characteristics $F_y$ under the subjective probability measure $\Lambda$. Define the expected marginal benefit of increasing reserve $i$ and decreasing reserve $j$ as:
\begin{equation}
    G_{ij}(r,\Lambda;\beta,\gamma) = \frac{\partial}{\partial r_i}\Xi(r,\Lambda;\beta,\gamma)-\frac{\partial}{\partial r_j}\Xi(r,\Lambda;\beta,\gamma)
\end{equation}
Any (interior) optimal reserve policy $r^*$ must equate the expected marginal benefit of increasing reserve $i$ and decreasing reserve $j$ at $r^*$ to zero for all $(i,j)$ pairs, \textit{i.e.,} $G_{ij}(r^*,\Lambda;\beta,\gamma)=0$ for all $\{i,j\} \subset \{1,2,3,4\}$ such that $j>i$. These six first-order conditions yield six moments.

We take empirical analogs of the theoretical moments and estimate preference parameters by minimizing the sum of squared deviations of these moments from zero. We take CPS' pursued reserve policy from 2012 to 2017 as optimal, $\hat{r}^*=(0.175,0.175,0.175,0.175)$. We estimate the empirical joint distribution of students' scores and tiers in CPS in each year $\hat{F}_y$ for $y\in\{2013,2014,2015,2016,2017\}$ and estimate $\hat{\Lambda}$ as a distribution that places equal probability on each of these five measured distributions. We plug these sample estimates into the theoretical moment functions. This yields six empirical moment functions that depend only on the preference parameters, $G_{ij}(\hat{r}^*,\hat{\Lambda};\beta,\gamma)$. Motivated by the theoretical necessity of $G_{ij}(r^*,\Lambda;\beta,\gamma)=0$, we estimate the preference parameters by minimizing the sum of squared deviations of the empirical moments from zero:
\begin{equation}
   (\beta^*,\gamma^*)\in \arg\min_{\beta,\gamma} \sum_{i=1}^4\sum_{j>i} G_{ij}(\hat{r}^*,\hat{\Lambda};\beta,\gamma)^2
\end{equation}
Performing this estimation yields estimated parameter values of $\beta^* = -209.5$ and $\gamma^* = 2.11$.

\subsection{The Estimated Gains from APM}
\begin{figure}
    \centering
        \caption{The Estimated Optimal APM}
    \includegraphics[width=0.6\textwidth]{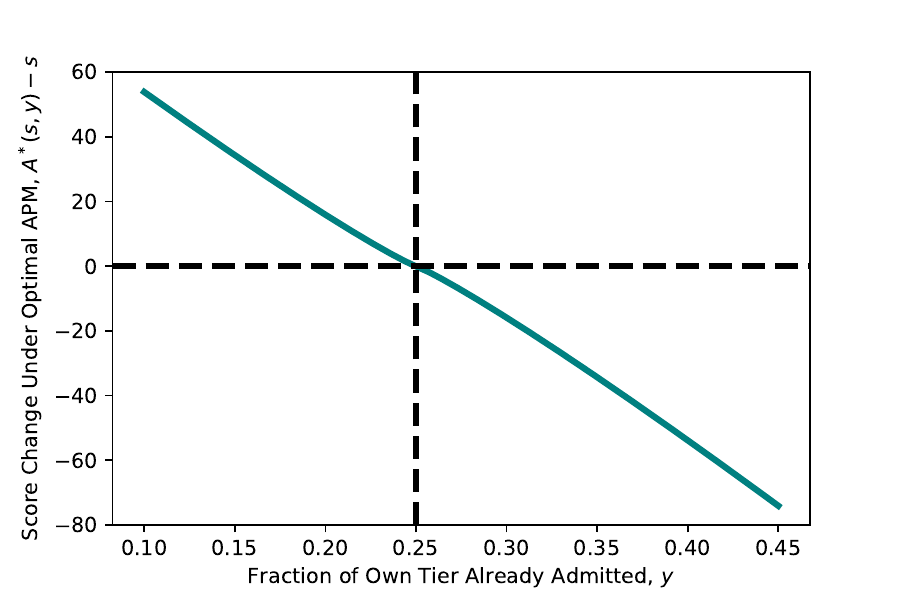}
    \fnote{This figure plots the change in a student's score when fraction $y$ of students in their own tier has already been admitted under the estimated optimal APM, $A^*$. At $y=0.25$ (the vertical dashed black line), the score in unchanged. For $y<0.25$, students receive a score boost. For $y>0.25$, students receive a score penalty. The range of the x-axis, $[0.1,0.45]$, is chosen to cover the full range of fractions of admitted students under both the optimal APM and the CPS reserve policy from all tiers in all years of our sample (see Figure \ref{fig:comparison}).}
    \label{fig:apm_cps}
\end{figure}

We now use our estimated model to quantify the welfare gains from using APM. To do this, we compare the empirical payoff $\Xi(\phi,\hat \Lambda,\beta^*,\gamma^*)$ under two mechanisms: the pursued quota policy, $r^*$, and the optimal APM from Theorem \ref{subsched}, $A^*$. In Figure \ref{fig:apm_cps}, we illustrate how the estimated optimal APM changes students' scores to arrive at their ultimate priorities. In accordance with the preferences we have assumed, students receive a score boost when their tier is underrepresented and a score penalty when their tier is overrepresented. As we found $\gamma^*=2.11$, the estimated diversity preference is very close to quadratic. Thus, the optimal APM is very close to linear. From Theorems \ref{subsched} and \ref{rationalization}, we know that this APM achieves the first-best allocation in each year while the implemented quota policy does not. However, our theorems do not guarantee that the gains from APM are economically meaningful.

The empirical payoff under APM is $876.9$, while it is $874.8$ under the CPS reserve policy. Thus, the gains from APM are equivalent to increasing average scores by $2.1$, holding diversity fixed. To benchmark the size of the gains, we require units in which they can be meaningfully expressed. To this end, we define the \textit{loss from underrepresentation} as the payoff lost by CPS under its baseline policy from not admitting a fully balanced class, while holding fixed the average score of the class. This is equal to $5.6$ points under our estimated parameters. Thus, the gains from APM are equal to $37.5\%$ of the loss from underrepresentation incurred under the CPS policy.\footnote{This is equivalent to increasing the percentage of students from tier 1 from $0.179$ to $0.21$ and decreasing the percentage of students from tier $4$ from $0.407$ to $0.378$. This corresponds to swapping $8.7$ students from the most overrepresented group (tier 4) for the most underrepresented group (tier 1) each year.}

To contextualize the magnitude of this improvement, we can compare the gains from the optimal APM to the gains from the 2012 CPS reform that gave rise to the CPS policy from 2012-2017 and increased the size of all reserves from $15\%$ to $17.5\%$. Under the estimated preferences, the empirical payoff under the $15\%$ reserve rule is $873.9$, and so the gains from the reform are equivalent to increasing average scores by $0.9$. Thus, the gains from switching to the optimal APM are $2.3$ times larger than the gains from this recent reform.

\begin{figure}
    \centering
    \caption{Comparing Admissions under the Optimal APM and the CPS Policy}
    \includegraphics[width=0.6\textwidth]{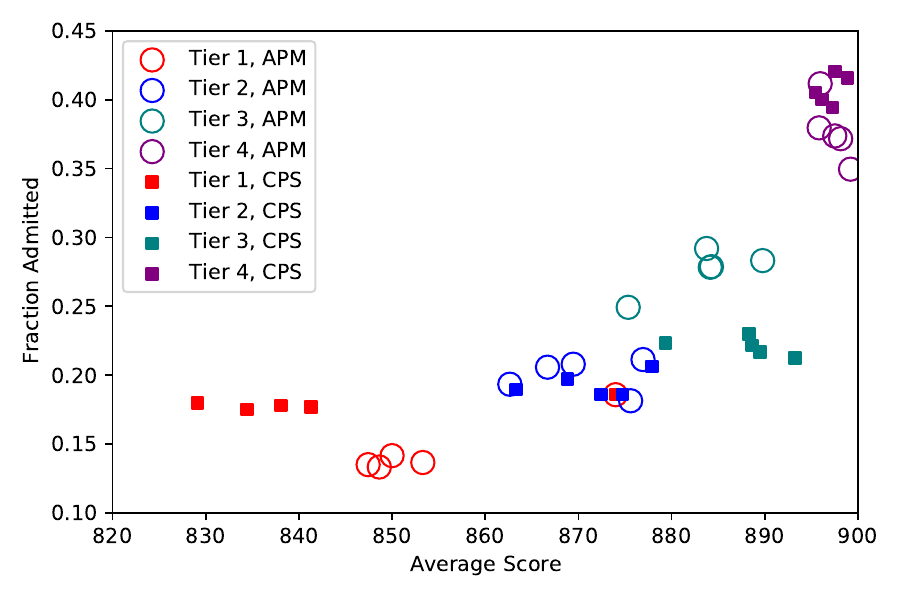}
    \fnote{Each point corresponds to one of the four tiers of students in one of the five years under either the optimal APM or the CPS policy. The x-axis corresponds to the average score of those admitted from that tier in that year under that policy. The y-axis corresponds to the fraction of admitted students from that tier in that year under that policy.}
    \label{fig:comparison}
\end{figure}

These estimates suggest that the gains from APM are economically meaningful. These gains stem from the variation across years in the joint distribution of student scores and tiers. This can be seen in Table \ref{table:cutoffscores}, which shows the variability in the scores of the marginally admitted students from tiers 1, 2 and 3. More systematically, we visualize the difference in outcomes under CPS' reserves and the optimal APM by plotting the average scores and fraction admitted for each tier for each year under both mechanisms in Figure \ref{fig:comparison}. There are two main differences between the allocations. First, the APM allocates systematically fewer tier 1 and tier 4 students and more tier 3 students. These level effects are a consequence of the second difference: the APM admits a greater fraction of students from each tier (especially tiers 1 and 3) in the years in which that tier scores well. The fact that tier 3 students score well relative to their admissions level is what leads the authority to admit more tier 3 students and fewer tier 1 students. These positive selection and level effects generate the welfare gains.

\paragraph{Robustness} We now explore the robustness of APM to the three core assumptions of our analysis: (i) that CPS has the correct beliefs about the distribution of distributions of students, (ii) that CPS has preferences that lie in the assumed parametric family, and (iii) that CPS optimizes the sizes of all four tiers.

\begin{figure}
    \centering
    \caption{Robustness of the Gains from APM}
    \includegraphics[width=0.6\textwidth]{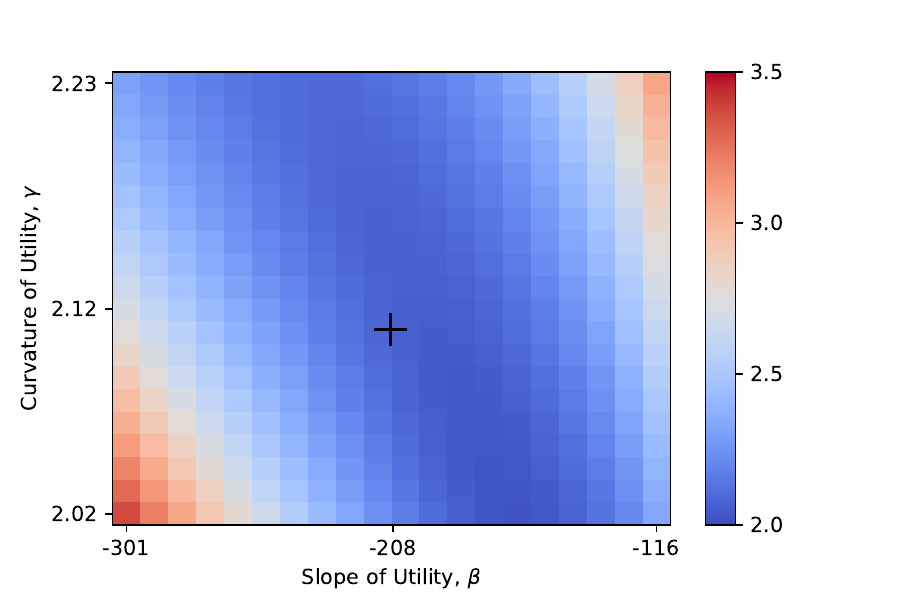}
    \fnote{This chart plots the difference in empirical payoffs from the optimal APM and CPS reserve policy under alternative parameter values, with the shaded colors corresponding to the numerical value of the gains from APM, ranging from 2.0 to 3.5. The black `+' indicates our baseline parameter values. The ranges for the axes are obtained by estimating $\beta$ and $\gamma$ separately for each year of our data and separately taking the minimum and maximum estimated values of each set of estimated parameters.}
    \label{fig:difference}
\end{figure}

Our baseline analysis took the beliefs of CPS to be the true empirical distribution of student distributions over years. To test robustness to this assumption, we take $\hat{\Lambda}$ as a Dirac distribution on the realized distribution for each of the five years of our data and re-estimate the preference parameters. In Figure \ref{fig:difference}, we plot the difference in welfare under the optimal APM and CPS reserve policies over the full range of these re-estimated parameters (\textit{i.e.,} we take the minimum and maximum of the estimated parameters across years as the ranges for the axes). We find that the gains from APM range from 2.0 to 3.5, while our baseline estimate was 2.1. Thus, the point estimate of our welfare gains from APM appears to be conservative by this metric.

To gauge robustness to the functional form we have assumed for CPS' preferences, in Appendix \ref{sec:alternativeutilityfunctions}, we estimate two different parametric specifications of utility. First, we consider a utility function that includes a loss term only for underrepresented tiers (and does not penalize overrepresentation of any tier). Second, we allow for CPS to care differentially about underrepresentation and overrepresentation by considering a utility function with separate coefficients for underrepresented and overrepresented tiers. We find that, under these specifications, the improvement from APM corresponds to $9.7 \%$ and $8.7 \%$ of the loss from underrepresentation, which is attenuated relative to our baseline, but remains considerable.

To study the robustness of our findings to the assumption that CPS optimizes the size of all four tiers, in Appendix \ref{sec:homogeneousreserves} we consider a setting where CPS sets a \textit{single} reserve size for all tiers. As we now have only one moment condition, we vary $\gamma$ over the interval [1,10], estimate $\beta^*(\gamma)$ as the exact solution to the moment condition, and compute the gains from APM as a function of $\gamma$. The \textit{minimum} gain from APM over the estimated range is $1.98$ points, which corresponds to $26.2\%$ of the loss from underrepresentation under that parameterization. This is slightly smaller than our baseline estimate but still considerable.

\paragraph{Limitations} We conclude our quantitative analysis by stating some limitations of our analysis. First, even though we argue that our functional form assumptions are reasonable and parsimonious in this setting, there are possibly many other parametric utility functions that might represent the preferences of CPS and these alternative preference structures may give different estimates of the gains from APM. This notwithstanding, we have documented the robustness of our conclusion that APM deliver considerable gains relative to the \textit{status quo} under several alternative estimation methods and preference assumptions. Moreover, if a researcher prefers an alternative structure for preferences, they could estimate the optimal APM and the gains from the optimal APM using precisely the same approach that we developed. Furthermore, given that CPS does face uncertainty, we have shown theoretically that the pursued quota policy (with merit slots processed first) cannot be rationalized as optimal \textit{even} if CPS' preferences are extremely risk averse (by Theorem \ref{rationalization}). Thus, while the numerical gains from APM will be sensitive to parametric assumptions on utility and should be treated as a benchmarking exercise, the fundamental conclusion that APM deliver strict gains is not sensitive to such assumptions.

Second, one of the main aims of the tier system employed by CPS is to increase racial diversity in the prestigious exam schools. Indeed, the pursued tier system is a race-neutral alternative that replaced the previous race-based system following two Supreme Court Rulings in 2003 and 2007 \citep[see][for a summary]{ellison2021efficiency}. Because of this, CPS uses tiers based on socioeconomic status instead of race and so we estimate their preferences over tiers. Of course, if admission rules could depend on race, then one could perform a similar analysis in which APM simply prioritize based on race rather than tier.

Finally, we also note that one key benefit of APM is to yield optimal allocations even under rare circumstances. By assuming that the designer is risk-neutral (rather than risk-averse) over their score-diversity index (\textit{i.e,} $g$ is taken to be linear rather than concave), we are likely to understate the gains from APM in dealing with such situations.
\vspace{-3ex}
\section{Conclusion}
\label{sec:conc}
Motivated by the use of priority and quota policies in resource allocation settings with diversity concerns, we consider an authority that has preferences over scores and diversity. We introduce Adaptive Priority Mechanisms (APM) and characterize an APM that is both optimal and can be specified solely in terms of the preferences of the authority. We study the priority and quota policies that are used in practice and show that they are optimal if and only if the authority is either risk-neutral or extremely risk-averse over diversity. Analyzing a setting with multiple authorities that dynamically admit agents, we show that the optimal APM is a dominant strategy. Thus, one could potentially advise authorities to follow an optimal APM with confidence (under our maintained assumptions on preferences) that they could do no better. Moreover, all authorities following the optimal APM implements the unique stable allocation. Finally, in centralized settings, we propose a modified APM that delivers authority-efficient allocations.

Our analysis has potential implications for improving the design of real-world allocation mechanisms. First, we show that while both priorities and quotas can be better than one another (depending on the risk preferences of the authority), they are generally suboptimal. Second, we show how to improve upon these mechanisms using APM that harness the strengths of these policies: APM benefit both from the guarantee effect of quotas in ensuring certain levels of admissions from various groups and the positive selection effect of priorities in expanding affirmative action when it is least costly. Our quantitative analysis using CPS data suggests that the use of APM can yield considerable welfare gains over the \textit{status quo}. We conclude that APM may have practical utility in delivering more desirable allocations of resources. Moreover, APM could be applied in many settings, including the allocation of seats at schools, places at universities, and medical resources to patients.

\bibliographystyle{econometrica.bst}
\bibliography{bib-matching}

\clearpage

\begin{appendices}
\vspace{-2ex}
\section{Omitted Proofs}
\label{ap:proofs}
\vspace{-2ex}
\subsection{Proof of Proposition \ref{mechcomp}}
\label{mechcompproof}
\begin{proof}
Part (i): In state $\omega$ the payoff from admitting the highest-scoring minority students of measure $x(\omega)$ is:
\begin{equation}
    q\omega+(1+\gamma-\omega)x(\omega)-\frac{1}{2}\left(\frac{1}{\kappa}+\gamma\beta \right)x(\omega)^2
\end{equation}
Thus, the $x(\omega)$ that solves the FOC is given by:
\begin{equation}\label{eq:weitzmanoptimal}
    x(\omega)=\frac{\kappa(1+\gamma-\omega)}{1+\kappa\gamma\beta }
\end{equation} 
Under our maintained assumptions, we have that:
\begin{equation}
   x(\omega)=\frac{\kappa(1+\gamma-\omega)}{1+\kappa\gamma\beta }\leq  \frac{1+\gamma-\underline{\omega}}{\frac{1}{\kappa}+\gamma\beta }+\kappa(\overline{\omega}-\underline{\omega}) < \min\{\kappa,q\}
\end{equation}
and:
\begin{equation}
    x(\omega)=\frac{\kappa(1+\gamma-\omega)}{1+\kappa\gamma\beta }\ge \frac{1+\gamma-\overline{\omega}}{\frac{1}{\kappa}+\gamma\beta }>\kappa(1-\underline{\omega})\ge 0
\end{equation}
Thus, this level of minority admissions is feasible. Substituting, we have that:
\begin{equation}
\begin{split}
    V^* 
    &=q\mathbb{E}[\omega]+\frac{1}{2}\frac{\mathbb{E}[\kappa(1+\gamma-\omega)^2]}{1+\kappa\gamma\beta }
\end{split}
\end{equation}
Consider now the APM $A(y)=\gamma(1-\beta  y)$. Agents are allocated the resource if their modified scores exceed $\omega$, with a uniform lottery over students with score exactly $\omega$. Thus, in state $\omega$, this policy admits measure $y(\omega)$ minorities that solve the fixed point equation:
\begin{equation}\label{eq:weitzmanfixedpoint}
    y(\omega)=\min\left\{\kappa\int_0^1\mathbb{I}[s+A(y(\omega))\ge\omega]\dd s,q\right\} = \min\{\kappa\left(1-\max\{\omega-A(y(\omega)),0\}\right),q\}
\end{equation}
Denote the RHS of this fixed point equation by the function $\text{RHS}(y,\omega)$, which is continuous and decreasing in $y$. Moreover, $\text{RHS}(0,\omega)=\min\{\kappa(1-\max\{\omega-\gamma,0\}),q\}>0$ and $\text{RHS}(\min\{\kappa,q\},\omega)<\min\{\kappa,q\}$. The second of these inequalities is true because the condition $\underline{\omega}>\gamma(1-\beta \min\{q,\kappa\})$ follows from our assumption that $\min\{\kappa,q\}>\frac{1+\gamma-\underline{\omega}}{\frac{1}{\kappa}+\gamma\beta }+\kappa(\overline{\omega}-\underline{\omega})$. Thus, there exists a unique $y(\omega)$ implemented by the APM. Moreover, let $y_A(\omega)$ denote the unique solution to the equation \ref{eq:weitzmanfixedpoint}, which gives the measure of admitted minority students under APM $A$ at state $\omega$.
\begin{equation}
\begin{split}
    y_A(\omega) &= \kappa (1 - (\omega - \gamma(1-\beta  y_A(\omega))))\\
                &= \kappa(1-\omega+\gamma) - \kappa \gamma \beta  y_A(\omega)\\
                &= \frac{ \kappa(1-\omega+\gamma)}{1+\kappa \gamma \beta }
\end{split}
\end{equation}
Thus, $A$ implements the optimal level of minority admissions characterized in equation \ref{eq:weitzmanoptimal} and $V_A=V^*$.

Part (ii): First, if we admit all minority students over some threshold $\hat s$, the total score of admitted minority students is $\kappa\int_{\hat s}^1 s\dd s$. Moreover, when we admit measure $x$ minority students where $x\leq\min\{\kappa,q\}$, this admissions threshold is defined by $x=\kappa\int_{\hat s}^1 \dd s=\kappa(1-\hat s)$. Thus, we have that $\hat s =1-\frac{x}{\kappa}$. Finally, the residual measure $q-x$ admitted majority students all score $\omega$. Thus, the total score is given by $\bar{s}= q\omega+(1-\omega)x-\frac{1}{2\kappa}x^2$  for $0\leq x\leq\min\{\kappa,q\}$. As both quota and priority policies always admit the highest-scoring minority students, the authority's utility is given by:
\begin{equation}
\begin{split}
    \mathcal{U}
    &= q\mathbb{E}[\omega]+\mathbb{E}[(1+\gamma-\omega)x]-\frac{1}{2}\left(\frac{1}{\kappa}+\gamma\beta \right)\mathbb{E}[x^2]
\end{split}
\end{equation}

We now derive the admitted measure of minority students. In the absence of a priority or quota policy, $\alpha=0$ or $Q=0$, we have that $x=\kappa(1-\omega)$ measure minority students is admitted. Thus, under a quota policy $Q$, measure $x=\max\{Q,\kappa(1-\omega)\}$ minority students are admitted. Under a priority policy, the measure of admitted minority students is $x=\kappa\int_{\omega-\alpha}^1 \dd x=\kappa(1+\alpha-\omega)$. In each case $x$ is capped by $\min\{\kappa,q\}$ and floored by 0.

The expected utility function over quotas is given by one of four cases. First, $Q>\min\{\kappa,q\}$ and:
\begin{equation}
    \mathcal{U}_Q(Q)=q\mathbb{E}[\omega]+(1+\gamma-\mathbb{E}[\omega])\min\{\kappa,q\}-\frac{1}{2}\left(\frac{1}{\kappa}+\gamma\beta \right)\min\{\kappa,q\}^2
\end{equation}
Second, $Q\in[\kappa(1-\underline{\omega}),\min\{\kappa,q\})$ and:\footnote{By our maintained assumptions we have that this interval has non-empty interior.}
\begin{equation}
    \mathcal{U}_Q(Q)=q\mathbb{E}[\omega]+(1+\gamma-\mathbb{E}[\omega])Q-\frac{1}{2}\left(\frac{1}{\kappa}+\gamma\beta \right)Q^2
\end{equation}
Third, $Q\in(\kappa(1-\overline{\omega}),\kappa(1-\underline{\omega}))$ and:
\begin{equation}
\begin{split}
    \mathcal{U}_Q(Q)&=q\mathbb{E}[\omega]+\int_{1-\frac{Q}{\kappa}}^{\overline{\omega}}\left((1+\gamma-\omega)Q-\frac{1}{2}\left(\frac{1}{\kappa}+\gamma\beta \right)Q^2\right) \dd \Lambda(\omega) \\
    &\quad +\int_{\underline{\omega}}^{1-\frac{Q}{\kappa}}\left((1+\gamma-\omega)\kappa(1-\omega)-\frac{1}{2}\left(\frac{1}{\kappa}+\gamma\beta \right)\left(\kappa(1-\omega)\right)^2\right) \dd \Lambda(\omega)
\end{split}
\end{equation}
Finally, $Q\leq\kappa(1-\overline{\omega})$ and:
\begin{equation}
    \mathcal{U}_Q(Q)=q\mathbb{E}[\omega]+\mathbb{E}\left[(1+\gamma-\omega)\kappa(1-\omega)\right]-\frac{1}{2}\left(\frac{1}{\kappa}+\gamma\beta \right)\mathbb{E}\left[\left(\kappa(1-\omega)\right)^2\right]
\end{equation}
We claim that the optimum lies in the second case. See that in case two the strict maximum is attained at $Q^*=\frac{1+\gamma-\mathbb{E}[\omega]}{\frac{1}{\kappa}+\gamma\beta }\in(\kappa(1-\underline{\omega}),\min\{\kappa,q\})$, by our assumptions that $ \min\{\kappa,q\}>\frac{1+\gamma-\underline{\omega}}{\frac{1}{\kappa}+\gamma\beta }+\kappa(\overline{\omega}-\underline{\omega})$ and    $\kappa(1-\underline{\omega})<\frac{1+\gamma-\overline{\omega}}{\frac{1}{\kappa}+\gamma\beta }$. Moreover, in case three, the first derivative of the payoff is given by:
\begin{equation}
    \mathcal{U}_Q'(Q)=\int_{1-\frac{Q}{\kappa}}^{\overline{\omega}}\left((1+\gamma-\omega)-\left(\frac{1}{\kappa}+\gamma\beta \right)Q\right) \dd \Lambda(\omega)
\end{equation}
Thus, checking that the sign of this is positive amounts to verifying that for all $Q\in(\kappa(1-\overline{\omega}),\kappa(1-\underline{\omega}))$, we have that:
\begin{equation}
    Q<\frac{1+\gamma-\mathbb{E}[\omega|\omega\ge 1-\frac{Q}{\kappa}]}{\frac{1}{\kappa}+\gamma\beta }
\end{equation}
As the RHS is an increasing function of $Q$, it suffices to show that:
\begin{equation}
    \kappa(1-\underline{\omega})<\frac{1+\gamma-\overline{\omega}}{\frac{1}{\kappa}+\gamma\beta }
\end{equation}
which we have assumed. Moreover, the expected utility in the first case equals  $\mathcal U_{Q}(\kappa(1-\underline{\omega}))$, thus is lower than the optimum of the second case. The expected utility in the fourth case equals $\mathcal U_{Q}(\kappa(1-\overline{\omega}))$, thus is lower than the optimum of the third case. We therefore have that:
\begin{equation}
    V_Q=q\mathbb{E}[\omega]+(1+\gamma-\mathbb{E}[\omega])Q^*-\frac{1}{2}\left(\frac{1}{\kappa}+\gamma\beta \right){Q^*}^2
\end{equation}
We now turn to characterizing the value of priorities. There are three cases to consider. First, when $\kappa(1+\alpha-\overline{\omega})\ge\min\{\kappa,q\}$ we have that $x=\min\{\kappa,q\}$ and:
\begin{equation}
    \mathcal{U}_P(\alpha)=q\mathbb{E}[\omega]+(1+\gamma-\mathbb{E}[\omega])\min\{\kappa,q\}-\frac{1}{2}\left(\frac{1}{\kappa}+\gamma\beta \right)\min\{\kappa,q\}^2
\end{equation}
Second, when $\kappa(1+\alpha-\underline{\omega})\ge\min\{\kappa,q\}\ge\kappa(1+\alpha-\overline{\omega})$ we have that:
\begin{equation}
\begin{split}
    &\mathcal{U}_P(\alpha)=q\mathbb{E}[\omega]+\int_{\underline{\omega}}^{1+\alpha-\min\{\frac{q}{\kappa},1\}}\left((1+\gamma-\omega)\min\{\kappa,q\}-\frac{1}{2}\left(\frac{1}{\kappa}+\gamma\beta \right)\min\{\kappa,q\}^2\right)\dd \Lambda(\omega) \\
    &\quad + \int_{1+\alpha-\min\{\frac{q}{\kappa},1\}}^{\overline{\omega}}\left((1+\gamma-\omega)\kappa(1+\alpha-\omega)-\frac{1}{2}\left(\frac{1}{\kappa}+\gamma\beta \right)\left[\kappa(1+\alpha-\omega)\right]^2\right) \dd \Lambda(\omega)
\end{split}
\end{equation}
Finally, when $\min\{\kappa,q\}\ge \kappa(1+\alpha-\underline{\omega})$, we have that:
\begin{equation}\label{eq:weitzmanoptimalpriority}
    \mathcal{U}_P(\alpha)=q\mathbb{E}[\omega]+\mathbb{E}[(1+\gamma-\omega)\kappa(1+\alpha-\omega)]-\frac{1}{2}\left(\frac{1}{\kappa}+\gamma\beta \right)\mathbb{E}[\left(\kappa(1+\alpha-\omega)\right)^2]
\end{equation}
We claim that the optimum under our assumptions lies only in the third case. First, we argue that there is a unique local maximum in the third case. Second, we show the value in the second case is decreasing in $\alpha$. By continuity, the unique optimum then lies in the third case.

First, it is helpful to write $\bar{x}(\alpha)=\kappa(1+\alpha-\mathbb{E}[\omega])$ and $\varepsilon=\kappa\left(\mathbb{E}[\omega]-\omega\right)$. The value in the third case can then be re-expressed as:
\begin{equation}
\begin{split}
        \mathcal{U}_P(\alpha)&=q\mathbb{E}[\omega]+\mathbb{E}[(1+\gamma-\omega)\left(\bar x(\alpha)+\varepsilon\right)]-\frac{1}{2}\left(\frac{1}{\kappa}+\gamma\beta \right)\mathbb{E}[\left(\bar x(\alpha)+\varepsilon\right)^2] \\
        &= q\mathbb{E}[\omega]+(1+\gamma-\mathbb{E}[\omega])\bar x(\alpha)-\mathbb{E}[\omega\varepsilon]-\frac{1}{2}\left(\frac{1}{\kappa}+\gamma\beta \right)\bar x(\alpha)^2-\frac{1}{2}\left(\frac{1}{\kappa}+\gamma\beta \right)\mathbb{E}[\varepsilon^2]
\end{split}
\end{equation}
Finally, we have that $\mathbb{E}[\varepsilon^2]=\kappa^2\text{\normalfont Var}[\omega] $ and $\mathbb{E}[\omega\varepsilon]=\text{\normalfont Cov}[\omega,\varepsilon]=-\kappa\text{\normalfont Var}[\omega]$. Thus:
\begin{equation}
    \mathcal{U}_P(\alpha)=q\mathbb{E}[\omega]+(1+\gamma-\mathbb{E}[\omega])\bar x(\alpha)-\frac{1}{2}\left(\frac{1}{\kappa}+\gamma\beta \right)\bar x(\alpha)^2+\frac{\kappa}{2}\left(1-\kappa\gamma\beta \right)\text{\normalfont Var}[\omega]
\end{equation}
We then see that the optimal $\alpha^*$ in this range sets $ \bar x(\alpha^*)=Q^*<\min\{\kappa,q\}$. It remains only to check that this optimal $\alpha^*$ indeed lies within this case, or equivalently that $\kappa(1+\alpha^*-\underline{\omega})\leq\min\{\kappa,q\}$. To this end, see that $\kappa(1+\alpha^*-\mathbb{E}[\omega])=Q^*$, and:
\begin{equation}
\begin{split}
    \kappa(1+\alpha^*-\underline{\omega})&=Q^*+\kappa(\mathbb{E}[\omega]-\underline{\omega})\leq Q^*+\kappa(\overline{\omega}-\underline{\omega})\\
    &\leq \frac{1+\gamma-\underline{\omega}}{\frac{1}{\kappa}+\gamma\beta }+\kappa(\overline{\omega}-\underline{\omega}) <\min\{\kappa,q\}
\end{split}
\end{equation}
where the final inequality follows by our assumption that $\min\{\kappa,q\}>\frac{1+\gamma-\underline{\omega}}{\frac{1}{\kappa}+\gamma\beta }+\kappa(\overline{\omega}-\underline{\omega})$.

Second, in the second case we have that the first derivative of the payoff in $\alpha$ is given by:
\begin{equation}
\begin{split}
    \mathcal{U}_P'(\alpha)&=\int_{1+\alpha-\min\{\frac{q}{\kappa},1\}}^{\overline{\omega}}\frac{d}{d\alpha}\left((1+\gamma-\omega)\kappa(1+\alpha-\omega)-\frac{1}{2}\left(\frac{1}{\kappa}+\gamma\beta \right)\left[\kappa(1+\alpha-\omega)\right]^2\right) \dd \Lambda(\omega) \\
    &=\kappa\int_{1+\alpha-\min\{\frac{q}{\kappa},1\}}^{\overline{\omega}}\left((1+\gamma-\omega)-\left(\frac{1}{\kappa}+\gamma\beta \right)(\bar x(\alpha)+\varepsilon(\omega))\right) \dd \Lambda(\omega)
\end{split}
\end{equation}
Checking that the sign of this is negative for all $\alpha$ such that $\kappa(1+\alpha-\underline{\omega})\ge\min\{\kappa,q\}\ge\kappa(1+\alpha-\overline{\omega})$ then amounts to checking that:
\begin{equation}
    \bar x(\alpha)>\frac{1+\gamma-\mathbb{E}[\omega|\omega\ge 1+\alpha-\min\{\frac{q}{\kappa},1\}]}{\frac{1}{\kappa}+\gamma\beta }-\mathbb{E}\left[\varepsilon(\omega)|\omega\ge 1+\alpha-\min\{\frac{q}{\kappa},1\}\right]
\end{equation}
for all $\bar x(\alpha)\in[\min\{\kappa,q\}-\kappa(\mathbb{E}[\omega]-\underline{\omega}),\min\{\kappa,q\}-\kappa(\mathbb{E}[\omega]-\overline{\omega})]$. So it suffices to check that the minimal possible value of the LHS exceeds the maximal possible value of the RHS. A sufficient condition for this is that:
\begin{equation}
    \min\{\kappa,q\}-\kappa(\mathbb{E}[\omega]-\underline{\omega})>\frac{1+\gamma-\underline{\omega}}{\frac{1}{\kappa}+\gamma\beta }-\kappa(\mathbb{E}[\omega]-\overline{\omega})
\end{equation}
Which holds as we assumed that $\min\{\kappa,q\}>\frac{1+\gamma-\underline{\omega}}{\frac{1}{\kappa}+\gamma\beta }+\kappa(\overline{\omega}-\underline{\omega})$. Substituting the optimal priority policy $\bar x(\alpha) = Q^*$ in equation \ref{eq:weitzmanoptimalpriority}, we obtain
\begin{equation}
        V_P=q\mathbb{E}[\omega]+(1+\gamma-\mathbb{E}[\omega])Q^*-\frac{1}{2}\left(\frac{1}{\kappa}+\gamma\beta \right) {Q^*}^2+\frac{\kappa}{2}\left(1-\kappa\gamma\beta \right)\text{\normalfont Var}[\omega]
\end{equation}
We have now established that:
\begin{equation}
\begin{split}
    \Delta = V_P-V_Q
    &=\frac{\kappa}{2}\left(1-\kappa\gamma\beta \right)\text{\normalfont Var}[\omega]
\end{split}
\end{equation}

Part (iii): We have $V^*,V_Q,V_P$. Thus, we can compute the loss from restricting to quota policies:
\begin{equation}
\begin{split}
    \mathcal{L}_Q
    &=\frac{1}{2}\frac{\kappa\text{\normalfont Var}[\omega]}{1+\kappa\gamma\beta }
\end{split}
\end{equation}
To find the loss from restricting to priority policies, we compute:
\begin{equation}
\begin{split}
    \mathcal{L}_P&=\mathcal{L}_Q-\Delta=
    \frac{1}{2}\left(\kappa\gamma\beta \right)^2\frac{\kappa\text{\normalfont Var}[\omega]}{1+\kappa\gamma\beta }
\end{split}
\end{equation}
Enveloping over these losses yields the claimed formula.
\end{proof}

\subsection{Proof of Proposition \ref{lem:APMprop}}
\label{lem:APMpropproof}
\begin{proof}
Adapting Definition \ref{defn:cutoffmatching} to single object setting, we say that a matching $\mu$ admits a cutoff structure if there exists $S(\omega) = \{S_m(\omega)\}_{m \in\mathcal M}$ such that $\mu(s,m;\omega)=1$ if and only if $s\ge S_m(\omega)$. A mechanism admits a cutoff structure if it admits a cutoff structure at every $\omega$. We will first prove that any monotone APM admits a cutoff structure. 

\begin{lem}
\label{lem:cutoff}
A monotone APM admits a cutoff structure.
\end{lem}
\begin{proof}
For a contradiction, assume it does not. Then there exists $\omega$ and matching $\mu$ implemented by the monotone APM such that for some $m \in \mathcal M$, $s > s'$,  $\mu(s,m;\omega)=0$ but  $\mu(s',m;\omega)=1$. Let $x_m$ denote the measure of group $m$ agents allocated the resource at $\mu$. Since $A$ is a monotone APM and $s > s'$, we have that $A(x_m,s) > A(x_m,s')$, which contradicts that $A$ implements $\mu$.
\end{proof}

We now use Lemma \ref{lem:cutoff} to show that a monotone APM implements a unique allocation. Assume for a contradiction that $A_m(y_m,s)$ is monotone and implements two different allocations, $\mu$ and $\mu'$. Let $x_m$ and $x'_m$ denote the measure of type $m$ students assigned the resource at $\mu$ and $\mu'$. First, we prove that if $\mu$ and $\mu'$ admit the same measure of students from each group, \textit{i.e.,} $x_m = x_m'$ for all $m$, then the average score of admitted students are the same. Let $s_{m}$ and $s'_{m}$ denote the score of the lowest-scoring type $m$ students assigned the resource at $\mu$ and $\mu'$.

\begin{claim}
If $x_m = x_m'$ for all $m \in \mathcal M$, then $\bar{s}_h(\mu,\omega) = \bar{s}_h(\mu',\omega)$
\end{claim}
\begin{proof}
Fix an $m$. Without loss of generality, let $s_m \geq s_m'$. First, since APM has cutoff structure and $x_m = x_m'$, we have that
\begin{equation}\label{zeromeasure}
    \int_{\Theta} \mathbb I\{s(\theta) \in [s_{m}',s_{m}], m(\theta)= m\} \dd F_{\omega}(s,m) = 0
\end{equation}
Note that this holds regardless of $m' \in \mathcal M$ and whether $s_{m} \geq s_{m}'$ or $s_{m}' \geq s_{m}$. Therefore, 
\begin{equation}
    \begin{split}
              \bar{s}_{h}(\mu,\omega)&=\int_{\Theta}\mu(s,m)h(s)\dd F_{\omega}(s,m)\\
              &= \sum_{m \in \mathcal M} \int_{\Theta}  \mathbb I\{s(\theta) \geq s_{m}, m(\theta)= m\} h(s(\theta)) \dd F_{\omega}(s,m)\\
              &=\sum_{m \in \mathcal M} \int_{\Theta}  \mathbb I\{s(\theta) \geq s_{m}', m(\theta)= m\} h(s(\theta)) \dd F_{\omega}(s,m)\\
              &=\int_{\Theta}\mu(s,m)h(s)\dd F_{\omega}(s,m)\\
              &=\bar{s}_{h}(\mu',\omega)
    \end{split}
\end{equation}
where line equation holds from Equation \ref{zeromeasure} and all others are by definition. This finishes the proof of the claim.
\end{proof}

Therefore, if $\mu$ and $\mu'$ do not yield identical measures, then there are $m$ and $n$ such that $x_m > x'_m$ and $x'_n > x_{n}$. Since $x_m > x'_m$, it follows that $s_m' > s_m$. Likewise $x'_{n} > x_{n}$ implies that $s_{n} > s'_{n}$. Note that these imply: (i) $\mu'(s_n',n) = 1$ while  $\mu'(s_m',n) = 0$ and (ii) $\mu(s_m,m) = 1$ while $\mu(s_n',n) = 0$. Thus, the following inequalities hold:

\begin{equation}
        A_n(s_n',x_n') > A_m(s_m',x_m') \geq A_m(s_m,x_m) > A_n(s_n',x_n)\geq A_n(s_n',x_n')
\end{equation}

where the first inequality follows from (i), the second inequality follows from the fact that $x_m' < x_m$ and $A$ is monotone, the third inequality follows from (ii) and the fourth inequality follows from the fact that $x_n < x_n'$ and $A$ is monotone. This equation yields $A_n(s_n',x_n') > A_n(s_n',x_n')$, which is a contradiction. Therefore, all allocations implemented by $A$ yield the same $x$. Thus, from Lemma \ref{lem:cutoff}, if a monotone APM $A$ implements $\mu$ and $\mu'$, both allocations admit the highest-scoring measure $x_m$ agents from group $m$ and can differ (at most) on a measure $0$ set, proving that all allocations implemented by $A$ are essentially the same.
\end{proof}
\subsection{Proof of Theorem \ref{subsched}}
\label{subschedproof}
\begin{proof}
We characterize the optimal allocation for each $\omega\in\Omega$ and show that the claimed adaptive priority mechanism implements the same allocation. Fix an $\omega\in\Omega$ and suppress the dependence of $F_{\omega}$ and $f_{\omega}$ thereon, and define the utility index of a score as $\tilde s=h(s)$ with induced densities over $\tilde s$ given by $\tilde f_m$ for all $m\in\mathcal{M}$. Let the measure of agents from any group $m\in\mathcal{M}$ that is allocated the resource be $x_m \in [0,\bar{x}_m]$ where $\bar x_m = \int_{h(0)}^{h(1)} \tilde f_m(\tilde s) \dd \tilde s$. Observe that, conditional on fixing the measures of agents from each group that are allocated the resource $x = \{x_m\}_{m \in \mathcal M}$, there is a unique optimal allocation (\textit{i.e.,} $\xi$-maximal $\mu$ up to measure zero transformations). In particular, as $g$ and $h$ are continuous and strictly increasing, the optimal allocation conditional on $x$ satisfies $\mu^*(\tilde s,m;x)=1 \iff \tilde s\ge \underline{\tilde{s}}_m(x_m)$ for some thresholds $\{\underline{\tilde{s}}_m(x_m)\}_{m\in\mathcal{M}}$ that solve:
\begin{equation}\label{identity}
    \int_{\underline{\tilde s}_m(x_m)}^{h(1)} \tilde f_m(\tilde s)\dd\tilde s = x_m
\end{equation}
We can then express the problem of choosing the optimal $x = \{x_m\}_{m \in \mathcal M}$ as:
\begin{equation}
    \max_{x_m \in [0,\bar x_m], \, \forall m\in\mathcal{M}} \sum_{m \in \mathcal M} \int_{\underline{\tilde s}_m(x_m)}^{h(1)} \tilde s \tilde f_m(\tilde s)\dd\tilde s + \sum_{m \in \mathcal M} u_m(x_m) \quad \text{s.t.} \sum_{m \in \mathcal M} x_m \leq q
\end{equation}
where a solution exists by compactness of the constraint sets and continuity of the objective. We can derive necessary and sufficient conditions on the solution(s) to this problem by considering the Lagrangian:
\begin{equation}
\begin{split}
    \mathcal{L}(x,\lambda,\overline{\kappa},\underline{\kappa}) &= \sum_{m \in \mathcal M} \int_{\underline{\tilde s}_m(x_m)}^{h(1)} \tilde s \tilde f_m(\tilde s)\dd\tilde s + \sum_{m \in \mathcal M} u_m(x_m) \\
    \quad &+\lambda\left(q-\sum_{m \in \mathcal M} x_m\right) + \sum_{m\in\mathcal{M}}\overline{\kappa}_m(\bar{x}_m-x_m)+\sum_{m\in\mathcal{M}}\underline{\kappa}_m x_m
\end{split}
\end{equation}
The first-order necessary conditions to this program are given by:
\begin{equation}
\label{mainnec}
    \frac{\partial \mathcal{L}}{\partial x_m}=-\underline{\tilde s}_m'(x_m)\underline{\tilde s}_m(x_m)\tilde{f}_m(\underline{\tilde s}_m(x_m))+u_m'(x_m)-\lambda-\overline{\kappa}_m+\underline{\kappa}_m =0
\end{equation}
\begin{equation}
    \lambda\frac{\partial \mathcal{L}}{\partial \lambda} = \lambda\left(q-\sum_{m \in \mathcal M} x_m\right) =0
\end{equation}
\begin{equation}
\label{compslack}
    \overline{\kappa}_m\frac{\partial \mathcal{L}}{\partial \overline{\kappa}_m}=\overline{\kappa}_m(\bar{x}_m-x_m )=0
\end{equation}
\begin{equation}
\label{lastnec}
    \underline{\kappa}_m\frac{\partial \mathcal{L}}{\partial \underline{\kappa}_m}= \underline{\kappa}_mx_m =0
\end{equation}
for all $m\in\mathcal{M}$. By implicitly differentiating Equation \ref{identity}, we obtain that:
\begin{equation}
\label{stilde}
    -\underline{\tilde s}_m'(x_m)\tilde f_m(\underline{\tilde s}_m(x_m))=1
\end{equation}
Thus, we can simplify Equation \ref{mainnec} to:
\begin{equation}
\label{xfoc}
    \frac{\partial \mathcal{L}}{\partial x_m}=\underline{\tilde s}_m(x_m)+u_m'(x_m)-\lambda-\overline{\kappa}_m+\underline{\kappa}_m =0
\end{equation}

Observe that all constraints are linear. Thus, if the objective function is strictly concave, the first-order conditions are also sufficient. Observe by Equation \ref{stilde} that $\underline{\tilde s}_m(x_m)$ is a strictly decreasing function of $x_m$, and all cross-partial derivatives are zero. Therefore, the first summation is strictly concave. Moreover $u_m'$ is a decreasing function of $x_m$ by virtue of the assumption that $u_m$ is concave for all $m\in\mathcal{M}$. Then the second summation is concave. Thus, the objective function is strictly concave and the optimal allocation is unique.

Thus, to verify that our claimed adaptive priority mechanism is a first-best mechanism, it suffices to show that the allocation it implements satisfies Equations \ref{mainnec} to \ref{lastnec}. The adaptive priority mechanism $A_m(y_m,s)=h^{-1}\left(h(s)+u_m'(y_m))\right)$ in the transformed score space yields transformed scores $h\left(A_m(y_m, s)\right)=\tilde s+u_m'(y_m)$. Define $x_m$ as the admitted measure of agents from group $m$ under this mechanism. Agents in group $m\in\mathcal{M}$ are allocated the resource if and only if $\tilde s + u_m'(x_m) \ge s^C $  for some threshold $s^C$ that solves:
\begin{equation}
    \sum_{m\in\mathcal{M}}\int_{\max\{\min\{s^C-u_m'(x_m),h(1)\},h(0)\}}^{h(1)} \tilde f_m(\tilde s)\dd \tilde s = q
\end{equation}
We can therefore partition $\mathcal{M}$ into three sets that are uniquely defined: (i) interior $\mathcal{M}_{I}=\{m\in\mathcal{M}|s^C-u_m'(x_m)\in (h(0),h(1))\}$; (ii) no allocation $\mathcal{M}_{0}=\{m\in\mathcal{M}|s^C-u_m'(x_m)\ge h(1)\}$; (iii) full allocation $\mathcal{M}_{I}=\{m\in\mathcal{M}|s^C-u_m'(x_m)\leq h(0)\}$. For all $m\in\mathcal{M}_0$, we implement $x_m=0$. For all $m\in\mathcal{M}_I$, we implement $x_m=\bar{x}_m$. For all $m\in\mathcal{M}_I$, we implement $x_m\in(0,\bar{x}_m)$. For any $m\in\mathcal{M}_{I}$, the allocation threshold is $\underline{\tilde s}_m(x_m)=s^C-u_m'(x_m)$. For any $m\in\mathcal{M}_0$, the allocation threshold is $h(1)$. For any $m\in\mathcal{M}_I$, the allocation threshold is $h(0)$.

We now verify that this outcome satisfies the established necessary and sufficient conditions. For all $m\in\mathcal{M}_{I}$, by the complementary slackness conditions we have that $\underline{\kappa}_m=\overline{\kappa}_m=0$. Substituting the above into Equation \ref{mainnec} for all $m\in\mathcal{M}_{I}$ we obtain that:
\begin{equation}\label{lambdaeq}
    s^C-\lambda =0
\end{equation}
which is satisfied for $\lambda=s^C$. As $q=\sum_{m\in\mathcal{M}}x_m$, the complementary slackness condition for $\lambda$ is then satisfied. For all $m\in\mathcal{M}_{0}$, by complementary slackness we have that $\overline{\kappa}_m=0$ and Equation \ref{mainnec} is satisfied by:
\begin{equation}
    \underline{\kappa}_m=\lambda - h(1)-u_m'(0)
\end{equation}
For all $m\in\mathcal{M}_I$, by complementary slackness we have that $\underline{\kappa}_m=0$ and Equation \ref{mainnec} is satisfied by:
\begin{equation}
    \overline{\kappa}_m=h(0)+u_m'(\bar{x}_m)-\lambda
\end{equation}
This completes the proof of first-best optimality of $A^*$.  Moreover, as the optimal allocation is unique for all $\omega$, any allocation that differs from the allocation implemented by the optimal APM at any $\omega$ would not be first-best optimal. Therefore, any first-best-optimal mechanism must implement essentially the same allocation as $A^*$.
\end{proof}

\subsection{Proof of Theorem \ref{rationalization}}
\label{rationalizationproof}
\begin{proof}
First, we prove the if parts of the results. Part (i): When $u_m$ is linear, $u_m'$ is constant and the first-best optimal adaptive priority mechanism is a priority mechanism $P(s,m)=h^{-1}(h(s)+u_m')$. 
Part (ii): When $\tilde u_m'(x_m)\ge k$ for $x_m\leq x_m^{\text{tar}}$ and $\tilde u_m'(x_m)=0$ for $x_m>x_m^{\text{tar}}$  and $\sum_{m\in\mathcal{M}}x_m^{\text{tar}}<q$, observe that the optimal mechanism admits $x_m\ge x_m^{\text{tar}}$ for all $m\in\mathcal{M}$ in all states of the world, but conditional on $x_m\ge x_m^{\text{tar}}$ for all $m\in\mathcal{M}$ admits the highest-scoring set of agents. A quota $Q_m=x_m^{\text{tar}}$ and $Q_{R}=q-\sum_{m\in\mathcal{M}}x_m^{\text{tar}}$, with $D(R) = \vert \mathcal M \vert + 1$ implements this allocation and is first-best optimal for any authority that is extremely risk-averse.

Second, we prove the only if parts of the results. Part (i): Assume the utility functions are not linear and let $m$ denote a group where $u'_{m}$ is not constant in $[0,q]$. We say that a state $\omega$ has full support if $f_w$ has full support. A state $\omega$ has \textit{full support in $m$ and $n$} if $f_w(\cdot,m)>0$ and $f_w(\cdot,n)>0$ for some $m$ and $n$ and positive measures of only $m$ and $n$. Let $\omega$ be a state that has full support in $m$ and $n$. Moreover, assume both groups have a measure $q$ of agents. We first establish that in any optimal allocation, agents from both groups are allocated the resource.

\begin{claim}
\label{claim:int}
If preferences are non-trivial, then the optimal allocation has $x_n,x_m>0$.
\end{claim}
\begin{proof}
Toward a contradiction, suppose without loss of generality that $x_n=0$. This implies that $x_m=q$. By the necessary first-order condition from Theorem \ref{subsched} (combing Equations \ref{mainnec} and \ref{lastnec}), we have that:
\begin{equation}
\label{eq:intclaim}
\begin{split}
    u_m'(q)+h(0)&=u_n'(0)+h(1)+\underline{\kappa}_n \ge u_n'(0)+h(1) \\
\end{split}
\end{equation}
where the inequality follows as $\underline{\kappa}_n\ge0$. Thus, we have that:
\begin{equation}
     u_m'(q)-u_n'(0)\ge h(1)-h(0)>u_m'(q)-u_n'(0)
\end{equation}
where the first inequality follows by rearranging Equation \ref{eq:intclaim} and the second follows by the definition of non-triviality of preferences. This is a contradiction, thus $x_n,x_m>0$ in any optimal allocation.
\end{proof}

We now establish an equation relating $x_n$ and $x_m$ that will be useful in the steps to come.

\begin{claim}
\label{claim:characterization}
Let $\omega$ have full support in $m$ and $n$, $\mu$ denote a cutoff matching with cutoffs $s_m$ and $s_n$. Let $x_m$ and $x_n$ denote the measures of agents who are allocated the object at $\mu$. $\mu$ is optimal if and only if $u'_m(x_m) + h(s_m) = u'_n(x_n) + h(s_n)$ and $x_n+x_m=q$.
\end{claim}
\begin{proof}
By the necessary  and sufficient first-order conditions from Theorem \ref{subsched}, we again have that:
\begin{equation}
    u_m'(x_m)+h(s_m)-\overline{\kappa}_m+\underline{\kappa}_m=u_n'(x_n)+h(s_n)-\overline{\kappa}_n+\underline{\kappa}_n
\end{equation}
By Claim \ref{claim:int}, we have $x_m,x_n>0$. Thus, by the complementary slackness conditions (Equations \ref{compslack} and \ref{lastnec}), we have that $\overline{\kappa}_m=\underline{\kappa}_m=\overline{\kappa}_n=\underline{\kappa}_n=0$. Thus, we obtain:
\begin{equation}
\label{eq:FOC}
    u_m'(x_m)+h(s_m)=u_n'(x_n)+h(s_n)
\end{equation}
together with $x_n+x_m=q$, we have characterized the optimal allocation as claimed.
\end{proof}

Continue to let $x_{m}$ and $x_{n}$ denote the measures of group $m$ and $n$ agents at the optimal allocation under $\omega$, and $s_{m}$ and $s_{n}$ denote the cutoff scores for admission. There are now two cases to consider: (i) $u'_{m}(x_m)$ and $u'_{n}(x_{n})$ are locally constant. (ii) $u'_{m}(x_m)$ or $u'_{n}(x_{n})$ are not locally constant. If we are in case (i), we will construct an $\omega'$ with a unique optimal allocation $x_m'$ and $x_n'$ where $u'_{m}(x_m')$ or $u'_{n}(x_{n}')$ is not locally constant, and then show jointly how we arrive at a contradiction in both cases (i) and (ii).

To this end, suppose that we are in case (i). Let $x_m^*$ and $x_n^*$ denote the measures that are closest to $x_m$ and $x_n$ such that $u'_{m}(x_m)$ and $u'_{n}(x_{n})$ are not locally constant, \textit{i.e.,}:
\begin{equation}
\begin{split}
    x_k^* = \arg\min_{x_k'}\Bigg\{|x_k-x_k'| &\Big| u_k'(x_k)=u_k'(x_k') \text{ and for all } \varepsilon>0 \\ &\text{ either } u_k'(x_k'-\varepsilon)>u_k'(x_k) \text{ or } u_k'(x_k'+\varepsilon)>u_k'(x_k)\Bigg\}
\end{split}
\end{equation}
As $u_k'$ is continuous, this minimum is attained and $x_k^*$ is well-defined. Without loss of generality, assume $|x_m - x_m^*| \leq |x_n - x_n^*|$ and define both $\hat{x}_m=x_m^*$ and $\hat{x}_n=q-x_m^*$. We now construct a state $\omega'$ such that $\hat{x}$ is optimal:

\begin{claim}
\label{claim:omegaone}
Define $\omega'$ where $F_m(1)-F_m(s_m) = \hat{x}_m$, $F_n(1)-F_n(s_n) = \hat{x}_n$ and $\omega'$ has full support in $m$ and $n$. The allocation that admits the highest-scoring $\hat x_m$ group $m$ agents and the highest-scoring $\hat x_n$ group $n$ agents is the unique optimal allocation.
\end{claim}
\begin{proof}
By Claim \ref{claim:characterization}, as $\hat{x}_m+\hat{x}_n=q$ by construction, $\hat{x}$ is optimal if and only if Equation \ref{eq:FOC} holds. To this end, observe that if we admit $\hat{x}$, then the cutoff scores are the same as under $x$ as $F_m(1)-F_m(s_m) = \hat{x}_m$ and $F_n(1)-F_n(s_n) = \hat{x}_n$, by construction. Thus, we have that:
\begin{equation}
\begin{split}
    u_m'(\hat{x}_m)+h(s_m)&=u_m'(x_m)+h(s_m) = u_n'(x_n)+h(s_n)= u_n'(\hat{x}_n)+h(s_n)
\end{split}
\end{equation}
where the first equality holds by construction as $\hat{x}_m=x^*_m$ and $u_m'(x^*_m)=u_m'(x_m)$, the second equality holds by optimality of $x$, and the third equality holds as $|x_m - x_m^*| \leq |x_n - x_n^*|$, which implies that $u_n'(\hat{x}_n)=u_n'(x_n^*)$. Thus, Equation \ref{eq:FOC} holds, and $\hat{x}$ is optimal, as claimed.
\end{proof}

Observe that this construction also applies trivially in case (ii) with $x_m^*=x_m$. Thus, using this construction, we can now study cases (i) and (ii) together. In state $\omega'$, to implement this optimal allocation, we must have that $P(s,m)<P(s_n,n)$ for all but a measure zero set of $s$ such that $s<s_m$. We will now construct another state $\omega''$ such that any priority mechanism with this property is suboptimal.

First, suppose that $x_m^*\leq x_m$ and fix some $\varepsilon\in(0,x_m^*)$. Define $\tilde s_m$ as solving the following equation:
\begin{equation}
\label{eq:epseq}
    u'_m(\hat{x}_m - \varepsilon) + h(\tilde s_m) = u'_n(\hat{x}_n + \varepsilon) + h(s_n)
\end{equation}
We then have that:
\begin{equation}
\begin{split}
    \tilde s_m &= h^{-1}\left(h(s_n)+u'_n(\hat{x}_n + \varepsilon) -u'_m(\hat{x}_m - \varepsilon)\right) <h^{-1}\left(h(s_n)+u'_n(\hat{x}_n) -u'_m(\hat{x}_m)\right) = s_m
\end{split}
\end{equation}
where the first equality rearranges Equation \ref{eq:epseq} and the second inequality uses the facts that $u'_n(\hat{x}_n)\leq u'_n(\hat{x}_n + \varepsilon)$ and $u'_m(\hat{x}_m)<u'_m(\hat{x}_m - \varepsilon)$. We now construct a state $\omega''$ such that $(\hat{x}_m-\varepsilon,\hat{x}_n+\varepsilon)$ is optimal.

\begin{claim}
Define $\omega''$ where $1-F_m(\tilde s_m) = \hat{x}_m-\varepsilon$, $1-F_n(s_n) = \hat{x}_n+ \varepsilon$ with full support in $m$ and $n$. The allocation that admits the highest-scoring $(\hat{x}_m-\varepsilon,\hat{x}_n+\varepsilon)$ agents is the unique optimal allocation.
\end{claim}
\begin{proof}
Following the same steps as Claim \ref{claim:omegaone}, and the fact that Equation \ref{eq:epseq} holds by construction, we have that the claim holds.
\end{proof}

Observe that to implement this optimal allocation a priority mechanism must set $P(s,m)\ge P(s_n,n)$ for all but zero measure $s>\tilde{s}_m$. However, since $\tilde{s}_m<s_m$, this contradicts the optimality condition for state $\omega'$ that $P(s,m)<P(s_n,n)$. This is because for all but measure zero $s\in(\tilde{s}_m,s_m)$, which we have established is non-empty, we have that:
\begin{equation}
    P(s,m)\ge P(s_n,n) > P(s,m)
\end{equation}
which is a contradiction. To complete the proof, we need only consider the case that $x_m^*>x_m$. In this case, we can apply essentially the same steps and the result follows. Concretely, instead increasing $\hat{x}_m$ by $\varepsilon$ and following the same steps yields the required contradiction. 

We have now constructed three states $\omega,\omega',\omega''$ such that no priority mechanism can be optimal in each state when the authority is not risk-neural, completing the proof.

Part (ii): Assume that a quota policy is optimal, we now show that the authority's preferences must be extremely risk-averse. For each group $m \in \mathcal M$, let $c_m \in [0,1]$ and $c_m \neq c_n$ if $m \neq n$. Let $\omega$ be such that the scores of agents from each group $m$ are uniformly distributed between $[c_m,c_m+\epsilon]$, where $\epsilon$ is chosen to be small so that there is no overlap of these supports and each group has measure $q$ agents. Let $m_{\omega}$ denote the group with the highest $c_m$ at $\omega$. Now, compute the optimal allocation at $\omega$ and denote the measure of admitted agents from each group at the optimal allocation by $\{x_m^*(\omega)\}_{m \in \mathcal M}$. We first show that under any optimal quota policy, the level of the quotas must be set equal to the optimal allocation for all but the highest-scoring group:
\begin{claim}
\label{claim:Qpin}
If a quota $Q$ attains the optimal allocation, then for each $m \neq m_{\omega}$, $Q_m = x_m^*(\omega)$.
\end{claim}
\begin{proof}
If $Q_m>x_m^*(\omega)$, then we admit $x_m\ge Q_m>x_m^*(\omega)$, which is suboptimal as there is a unique optimal allocation by Theorem \ref{subsched}. If $Q_m<x_m^*(\omega)$ and $m\neq m_{\omega}$, then $x_m=Q_m$ as $c_{m_{\omega}}>c_m+\varepsilon$ and no agent from group $m$ can claim a merit slot. This is suboptimal. Thus, $Q_m = x_m^*(\omega)$ for all $m \neq m_{\omega}$.
\end{proof}

Next, create $\omega'$ by changing the highest-scoring group, \textit{i.e.,} $m_{\omega} \neq m_{\omega'}$. Let $x_{m_{\omega}}^*(\omega')$ denote the measure of admitted agents from group $m_{\omega}$ under $\omega'$. Applying Claim \ref{claim:Qpin}, If $Q$ attains the optimal allocation, then it must be that  $Q_{m_{\omega}} = x_{m_{\omega}}^*(\omega')$. Define $Q_m^*$ by $Q_m^* = x_m^*$ for all $m \in \mathcal M \setminus \{m_{\omega}\}$ and $Q_{m_{\omega}}^* = x_{m_{\omega}}^*(\omega')$.

Now, we have proved that if $Q$ is an optimal policy, then $Q_m = Q_m^*$ for all $m \in \mathcal M \setminus \{m_{\omega}\}$ and $Q_{m_\omega} = Q_{m_{\omega}}^*$. We now establish that merit slots must be processed after any positive measure quota slots if the merit slots are of positive measure:

\begin{claim}
If there is a quota policy that attains the first-best,  $Q$,  then $Q_m = Q_m^*$ and either $\sum_{m \in \mathcal M} Q_m^* = q$, \textit{i.e.,} there are no merit slots (merit slot processing does not matter), or $\sum_{m \in \mathcal M} Q_m^* < q$ and merit slots are processed after any positive measure quota slots.
\end{claim}
\begin{proof}
We have already proved  $Q_m = Q_m^*$. If $\sum_{m \in \mathcal M} Q_m^* = q$, there are no merit slots and any processing order yields the same result. If $\sum_{m \in \mathcal M} Q_m^* < q$, for a contradiction, assume merit slots are processed before quota slots for group $m$ and $Q_m^* >0$. There are two cases, $m \neq m_{\omega}$ and $m = m_{\omega}$. We start with the first case. Note that there is a cutoff $s_m$ for group $m$ with $s_m < c_m + \epsilon$ and all agents from group $m$ who score above $s_m$ are allocated the resource. Create $\omega''$ by taking measure $x_m/2$ of these agents who are allocated the resource and give them scores above $c_{m_{\omega}}$ (the highest-scoring group at $\omega$). The scores of the remaining $x_m/2$ agents are distributed uniformly at $[c_m, c_m+\epsilon]$. 

We now observe that the optimal allocations at $\omega$ and $\omega''$ are the same. This is because increasing the scores of already admitted agents does not change the preferences of the authority of whom to admit. Moreover, the optimal allocation at $\omega''$ cannot be attained if the quota slots for group $m$ are processed after the merit slots. This follows as, if merit slots are processed before quota slots for group $m$, a strictly positive measure of them would go to group $m$ agents at $\omega''$ since now they have a measure of agents with the highest scores, which violates optimality.

This proves the claim for $m \neq m_{\omega}$. To prove the result for $m=m_{\omega}$, replicate the above steps with $\omega'$ where $m_{\omega}$ is not the highest-scoring group.
\end{proof}

We now use these claims to establish that if quotas are first-best optimal, then $(u,h)$ must agree with $(\tilde u, \tilde h)$ on optimal allocations.
\begin{claim}
The quota first policy with $Q_m = x_m^{tar}$ maximizes the utility with respect to $\tilde u, \tilde h$.
\end{claim}
\begin{proof}
This is clear as for $\tilde u, \tilde h$, diversity utility dominates until $x_m^{tar}$ and has no effect after.
\end{proof}
This proves the result since if there exists a first-best optimal quota policy, then it is rationalized by $(\tilde u,\tilde h)$ with $x_m^{tar}=Q_m^*$. Hence, if there is a first-best quota mechanism, the authority is extremely risk-averse.
\end{proof}

\subsection{Proof of Proposition \ref{prop:nouncertainty}}
\begin{proof}
Let $x^*_m$ denote the measure of group $m$ agents in the optimal allocation, with $x^*=\{x^*_m\}_{m \in \mathcal M}$. A priority policy $P(s,m)=h^{-1}(h(s)+u_m'(x_m^*))=A_m(x_m^*,s)$ implements the same allocation as the optimal adaptive priority mechanism and by Theorem \ref{subsched}, is optimal. A quota mechanism with $(Q,D)$ where $Q_m = x^*_m$ implements $x^*$ for all $D$, and is therefore optimal.
\end{proof}

\subsection{Proof of Theorem \ref{prop:multischoolstable}}
\label{prop:multischoolstableproof}
\begin{proof}We first prove the following lemma.

\begin{lem}\label{lem:cutoffstructure}
Any stable matching is a cutoff matching.
\end{lem}

\begin{proof}
Assume that $\mu$ is a stable matching. Let  $S_{m,c} = \inf_{\theta} \{s_c(\theta): m(\theta) = m, \mu(\theta) = c\}$. Since $\mu$ satisfies within-group fairness, for all $m$ and $s' > S_{m,c}$, if $m(\theta) = m$ and $s_c(\theta) = s'$, $\mu(\theta) \succeq_{\theta} c$. Moreover, from part (iv) of the definition of matching, this extends to the case where $s' = S_{m,c}$. Concretely, suppose that $\mu(\theta)\neq c$, $c\succ_{\theta}\mu(\theta)$ and $s_c(\theta)=S_{m,c}$. Consider a sequence of types $\{\theta_k\}_{k\in\mathbb{N}}$ with common group $m$ and scores $\{s_c(\theta_k)\}_{k\in\mathbb{N}}$ such that $s_c(\theta_k)>S_{m,c}$ for all $k\in\mathbb{N}$ and $s_k(\theta)\rightarrow S_{m,c}$. Define the set $\Theta^{E}=\{\theta\in\Theta:c\succ\mu(\theta)\}$, which must be open by part (iv) of the definition of a matching. We have that $\theta_k\not\in\Theta^E$ for all $k\in\mathbb{N}$ but $\lim_{k\rightarrow\infty}\theta_k \in \Theta^E$, which contradicts that $\Theta^E$ is open. Thus, if $\mu$ is stable, then it is also a cutoff matching.
\end{proof}

Therefore, to characterize stable matchings, it is enough to characterize cutoffs that induce a stable matching, which we call \textit{stable cutoffs}.

\begin{defn}
A vector $S$ is a \textit{market-clearing cutoff} if it satisfies the following:
\begin{enumerate}
    \item $D_c(S) \leq q_c$ for all $c$.
    \item $D_c(S) = q_c$ if $S_{m,c} > 0$ for some $m \in \mathcal M$.
\end{enumerate}
\end{defn}
Since an authority can admit different measures of agents from different groups, there is a continuum of cutoffs that clear the market given $S_{-c}$, as long as $\{(0,\ldots,0)\}$ is not the only market-clearing cutoff.  Let $I(S_{-c})$ denote the set of market-clearing cutoffs. Let $I^*(S_{-c}) \subseteq I(S_{-c})$ denote the unique (by Lemma \ref{lem:cutoff}) cutoffs that implement the outcome under APM $A^*_c$ when the authority faces the induced type measure over the set $\tilde D_c(S_{-c})$. Define the map $T_c:[0,1]^{|\mathcal{M}|\times|\mathcal{C}|}\rightarrow[0,1]^{|\mathcal{M}|}$ as $T_c(S) = I^*_c(S_{-c})$ with $T:[0,1]^{|\mathcal{M}|\times|\mathcal{C}|}\rightarrow[0,1]^{|\mathcal{M}|\times|\mathcal{C}|}$ given by $T=\{T_c\}_{c\in\mathcal{C}}$. We first show that the set of fixed points of $T$ equals the set of stable cutoffs and that $T$ is increasing.

\begin{claim}\label{Tmap}
The set of fixed points of $T$ equals the set of stable cutoffs.
\end{claim}
\begin{proof}
If $S^*$, with corresponding matching $\mu^*$ (by Lemma \ref{lem:cutoffstructure}), is a fixed point of $T$, then each $c \neq c_0$ admits their most preferred measure $q_c$ agents in $\tilde D_c(S_{-c}^*)$ (by Theorem \ref{subsched}). Note that any $\hat \Theta$ that can block the matching must prefer $c$ to their allocation at $\mu^*$ and therefore  $\hat \Theta \subset \tilde D_c(S_{-c}^*)$. Then there cannot be a $\hat \Theta$ that blocks $\mu^*$ at $c$ since $c$ already attains the first-best utility under $\tilde D_c(S_{-c}^*)$ from the definition of $T_c(S)$ and Theorem \ref{subsched}. Conversely, if $S^*$, with corresponding matching $\mu^*$, is a not fixed point of $T$, then there exists $c$ such that  $T_c(S^*) \neq S^*_c$. Let $\hat \Theta$ denote the set of agents who are not matched to $c$ at $\mu^*$ but have scores greater than $T_c(S^*)$, and $\tilde \Theta$ denote the set of agents who are matched to $c$ at $\mu^*$ but have scores lower than $T_c(S^*)$. From optimality of $A^*_c$ (by Theorem \ref{subsched}), $\hat \Theta$ blocks $\mu^*$ at $c$ with $\tilde \Theta$.
\end{proof}

\begin{claim}
\label{claim:tinc}
$T$ is increasing.
\end{claim}
\begin{proof}
Fix an arbitrary $c\in\mathcal{C}$ and suppose that $S'_{-c}\ge S_{-c}$. Toward a contradiction suppose that there exists $m\in\mathcal{M}$ such that $t_{c,m}'=T_{c,m}(S')=I^*_c(S'_{-c})<I^*_c(S_{-c})=T_{c,m}(S)=t_{c,m}$, \textit{i.e.,} the admissions threshold for group $m$ at authority $c$ goes down. Let $f$ and $f'$ be the induced joint densities of agents over scores at $c$ and groups by the sets $\tilde D_c(S_{-c})$ and $\tilde D_c(S'_{-c})$, respectively. Let $\{x_{m,c}\}_{m \in \mathcal M}$ and $\{x_{m,c}'\}_{m \in \mathcal M}$ denote the measure agents who score above $t_{m,c}$ for their group (\textit{i.e.,} admitted under $A^*_c$) under  $\tilde D_c(S_{-c})$ and $\tilde D_c(S'_{-c})$, respectively. As $S'_{-c}\ge S_{-c}$, we have that $D^c(S_{-c},S_c)\subseteq D^c(S'_{-c},S_c)$ for all $S_c\in[0,1]^{|\mathcal{M}|}$. It follows that $f'(\theta_c)\ge f(\theta_c)>0$ for all $\theta_c=(s_c,m_c)\in[0,1]\times\mathcal{M}$. As $t_{c,m}'<t_{c,m}$, $f'$ has full support, and $f'\ge f$, we have that the measure of admitted group $m$ agents under increases $x_{c,m}'>x_{c,m}$. But as $\sum_{k\in\mathcal{M}}x_k'=\sum_{k\in\mathcal{M}}x_k=q$, we know that there exists an $m'\in\mathcal{M}$ such that $x_{c,m'}'<x_{c,m'}$. It follows that $t_{c,m'}'>t_{c,m'}$, otherwise, if $t_{c,m'}'\leq t_{c,m'}$, then $x_{c,m'}'\ge x_{c,m'}$. But now we have shown the following:
\begin{equation}
\begin{split}
    h_c(t_{c,m'}')+u_{m',c}'(x_{c,m'}')&> h_c(t_{c,m'})+u_{m',c}'(x_{c,m'}) \\
    &\ge h_c(t_{c,m})+u_{m,c}'(x_{c,m}) > h_c(t_{c,m}')+u_{m,c}'(x_{c,m}')
\end{split}
\end{equation}
where the first inequality follows by $t_{c,m'}< t_{c,m'}'$, $x_{c,m'}>x_{c,m'}'$, concavity of $u_m$ and strictly increasing $h_c$. The second inequality follows by optimality. This is because the facts that $t_{c,m}>0$ and $t_{c,m'}'<1$ imply that $\overline{\kappa}_{m}=\underline{\kappa}_{m'}=0$ and so Equation \ref{xfoc} implies that:
\begin{equation}
   h_c( t_{c,m'})+u_{m',c}'(x_{c,m'})-\overline{\kappa}_{m'}=h_c(t_{c,m})+u_{m,c}'(x_{c,m})+\underline{\kappa}_{m}
\end{equation}
with $\overline{\kappa}_{m'},\underline{\kappa}_{m}\ge 0$. The final inequality follows as $t_{c,m}'<t_{c,m}$ and $x_{c,m}'>x_{c,m}$. But this contradicts the optimality condition for APM (Theorem \ref{subsched}), which implies that $T_c\neq I^*_c$, which is a contradiction. Hence, for all $c$ and $m\in\mathcal{M}$, $T_{c,m}$ is an increasing function.
\end{proof}

As $T:[0,1]^{|\mathcal{M}|\times|\mathcal{C}|}\rightarrow [0,1]^{|\mathcal{M}|\times|\mathcal{C}|}$ is monotone and $[0,1]^{|\mathcal{M}|\times|\mathcal{C}|}$ is a lattice under the elementwise order $\ge$, Tarski's fixed point theorem implies that the set of stable matching cutoffs is a non-empty lattice.

Finally, we use the fact that the set of stable cutoffs is a complete lattice to argue that there is a unique cutoff consistent with stability.

\begin{claim}
\label{claim:cutoffunique}
The stable matching cutoffs are unique.
\end{claim}
\begin{proof}
Assume that there are multiple stable cutoffs. As the set of stable cutoffs is a lattice, there exists a largest ($S^+$) and smallest ($S^-$) stable cutoffs, where $S^+\ge S^{-}$, with strict inequality for some $m\in\mathcal{M}$, $c\in\mathcal{C}$ as $S^+ \neq S^-$. But then, as there is full support of agent types and authority $c$ fills the capacity under stable cutoffs $S^+$, it must exceed its capacity under $S^-$, which is a contradiction. Hence, we have shown that there exists a unique stable matching cutoff.
\end{proof}

The combination of Lemma \ref{lem:cutoffstructure} and Claim \ref{claim:cutoffunique} completes the proof.
\end{proof}

\subsection{Proof of Proposition \ref{thm:consistencycharacterization}}
\label{prop:cooperativestabilityproof}
\begin{proof}
If $\phi$ is equivalent to $A_c^*$, Claim \ref{Tmap} implies that $\phi$ is consistent with stability.

We prove consistency with stability implies that $\phi$ is equivalent to $A^*_c$ by the contrapositive. To this end, suppose that $\phi$ is not equivalent to $A^*_c$. It follows that there exists a full-support density $\{\tilde f(s_c,m)\}_{s_c\in[0,1],m\in\mathcal{M}}$ such that $\phi$ yields a different allocation than $A_c^*$ under $\tilde f$. The rest of the proof constructs a full-support measure $F$ with unique stable matching $\mu_F$ such that $\tilde f$ is the induced density of scores and groups of the agents who demand authority $c$ at $\mu_F$. Given such an $F$, we will have that $\phi$ cannot be consistent with stability as it yields a different allocation than $A_c^*$, which itself yields $\mu_F(c)$, the set of students $c$ is matched to in the unique stable matching.

We first define some notation. Given a density $f$, for any set of types $\check\Theta\subseteq\Theta$, we define the marginal density of agents with score $s_c\in[0,1]$ at authority $c$ in group $m\in\mathcal{M}$ as:
\begin{equation}
    f_{\text{marg}(\check\Theta)}(s_c,m)=\int_{\check\Theta}\mathbb{I}[s_c(\theta)=s_c,m(\theta)=m]\dd{F}(\theta)
\end{equation}

To construct such an $F$, we proceed in three steps. First, take a full-support density $f^0$ that satisfies the following two conditions: i) Define $\hat{S}_c\in[0,1]^{|\mathcal{M}|}$ as the cutoff vector that obtains by applying $A^*_c$ to $\tilde f$.\footnote{Which exists as any monotone APM admits a cutoff structure (Lemma \ref{lem:cutoffstructure}) and  the optimal APM is monotone (Theorem \ref{subsched}).} We assume that $f^0$ is such that authority $c$'s cutoff vector that is consistent with the unique stable matching, $\mu_{F_0}$, coincides with $\hat{S}_c$; ii) for all $m\in\mathcal{M}$ and $s_c<\hat{S}_{m,c}$, $f_{\text{marg}(\Theta)}^0(s,m)<\tilde f(s,m)$; and iii) all authorities have strictly positive cutoffs for all groups at the unique stable matching.
    
Second, transform $f^0$ into a new density $f^1$ that differs from $f^0$ on the set of types that is matched with $c$ under $\mu_{F_0}$, which we call $\Theta_c$.\footnote{Formally, $\Theta_c = \{ \theta: \theta \in D_c(\mu_{F^0}), s_c(\theta) \geq \hat S_{m(\theta),c)}\}$.} We define the scaling factor $\iota^1(s_c,m)$ as:
    \begin{equation}
        \iota^1(s_c,m)=\frac{\tilde f(s_c,m)}{f^0_{\text{marg}(\Theta_c)}(s_c,m)}
    \end{equation}
    Moreover, we define:
    \begin{equation}
        f^1(\theta)=
        \begin{cases}
            f^0(\theta)\iota^1(s_c(\theta),m(\theta)) & \text{if} \quad \theta\in\Theta_c, \\
                f^0(\theta) & \text{otherwise}.
        \end{cases}
    \end{equation}
    This changes the scores of the types who are allocated to $c$ under $\mu_{F^0}$ but does not change their total measure, their composition, or their scores at any other authority. Thus, the unique stable matching under $f^1$, $\mu_{F^1}$, coincides with $\mu_{F^0}$. Moreover, by assumption i) of step 1, we have that $f^1_{\text{marg}(\Theta_c)}(s_c,m)=\tilde f(s_c,m)$ for all $m$ and $s_c\ge \hat{S}_{m,c}$. 
    
Third, transform $f^1$ into a new density $f^2$ that differs on the set of unmatched agents under $f^0$ (and also therefore $f^1$ by step 2), $\tilde{\Theta}$, and define the set of types who strictly prefer $c$ to their assignment under $\mu_{F^0}$ (and also therefore $\mu_{F^1}$ by step 2), $\hat\Theta_c$.\footnote{Formally, $\tilde \Theta = \{ \theta: \theta \in D_c(\mu_{F^1}), s_c(\theta) < \hat S_{m(\theta),c}, s_{c'}(\theta) < S^{\mu_{F^1}}_{m(\theta),c'} \text{ for all } c' \neq c\}$, where $S^{\mu_{F^1}}_{m,c'}$ denotes the group $m$ cutoff at school $c'$ at the stable matching $\mu_{F^1}$, which is strictly positive by assumption iii) of step 2. Moreover, $\hat \Theta_c = \{ \theta: \theta \in D_c(\mu_{F^1}), s_c(\theta) < \hat S_{m(\theta),c} \}$.} We define a new scaling factor $\iota^2(s_c,m)$ as:
    \begin{equation}
        \iota^2(s_c,m)=\frac{\tilde f(s_c,m)-f_{\text{marg}(\hat\Theta_c)}(s_c,m)}{f_{\text{marg}(\tilde\Theta)}(s_c,m)}
    \end{equation}
    which is strictly positive by assumption ii) of step 1. We then define $f^2$ as:
    \begin{equation}
        f^2(\theta)=
        \begin{cases}
            f^1(\theta)(1+\iota^2(s_c(\theta),m(\theta))) & \text{if} \quad \theta\in\tilde\Theta, \\
            f^1(\theta) & \text{otherwise}.
        \end{cases}
    \end{equation}
    By construction, $f^2_{\text{marg}(\hat\Theta_c)}(s_c,m)=\tilde f(s_c,m)$ for all $m$ and $s_c<\hat{S}_{m,c}$. Moreover, $\mu_{F^2}=\mu_{F^1}=\mu_{F^0}$ as all $\theta\in\tilde\Theta$ remain unmatched.

We have now constructed a full-support density $f^2$ with unique stable matching $\mu_{F^2}$ (by Theorem \ref{prop:multischoolstable}) such that the density over $D_c(\mu_{F^2})$ coincides with $\tilde f$. Moreover, by Claim \ref{Tmap}, $A^*_c$ selects $\mu_{F^2}(c)$ from $D_c(\mu_{F^2})$. As $\phi$ selects a different allocation from $D_c(\mu_{F^2})$ (as it has density of types $\tilde f$), it is inconsistent with stability.
\end{proof}

\subsection{Proof of Theorem \ref{thm:eq}}
\label{thm:eqproof}
\begin{proof}
We prove that APM $A^*_c$ implements a dominant strategy by backward induction. Consider the terminal time $t=|\mathcal{C}|-1$. Some measure of agents $\lambda$ applies to the authority. Regardless of the measure $\lambda$, by Theorem \ref{subsched} we have that the APM $A^*_c$ is first-best optimal (to see this more concretely, simply index $\lambda$ by an arbitrary $\omega\in\Omega$ and apply Theorem \ref{subsched}). Thus, $A_c^*$ is dominant. Moreover, from Theorem \ref{subsched}, any strategy that differs from $A_c^*$ on a strictly positive measure set cannot be optimal. Thus any dominant strategy implements essentially the same allocation as $A^*_c$. Consider now any time $t<|\mathcal{C}|-1$, precisely the same argument applies and $A_c^*$ is (essentially uniquely) dominant.
\end{proof}

\subsection{Proof of Proposition \ref{prop:apmallocationstable}}\label{prop:apmallocationstableproof}
\begin{proof}
We first prove the following claim.
\begin{claim}
\label{claim:deterministic}
$\mu_{\Sigma^*}$ is (almost surely) a deterministic allocation that corresponds to a cutoff matching $\mu^*$.
\end{claim}
\begin{proof}
Since there is a continuum of agents, under any $\Sigma^*$, with probability $1$, any authority $c$ faces a given set of agents who apply $\Theta^{A,\Sigma^*}_c$ with induced measure $\lambda_c^{\Sigma^*}$. As $c$ uses APM $A^*_c$, with probability $1$, any agent $\theta$ is admitted to an authority if and only if $s_{c}(\theta) \geq S^{\Sigma^*}_{m,c}$, where $S^{\Sigma^*}_{m,c}$ denotes the cutoffs when APM $A^*_c$ is applied to agent measure $\lambda_c^{\Sigma^*}$. Since the agents have strict preferences, in any equilibrium, each agent applies to the $\succeq_{\theta}-$maximal authority in $\{c: s_{c}(\theta) \geq S^{\Sigma^*}_{m,c}\}$, and is admitted, which establishes that $\mu_{\Sigma^*}$ is (almost surely) deterministic allocation that corresponds to a cutoff matching with cutoffs $S^{\Sigma^*}_{m,c}$.
\end{proof}

We now establish that $\mu_{\Sigma^*}$ is the unique stable matching of the economy.

\begin{claim}
\label{claim:stableoutcome}
$\mu^*$ is the unique stable matching of this economy.
\end{claim}
\begin{proof}
For a contradiction, assume $\mu_{\Sigma^*}$ is not stable. Let $S$ denote the unique cutoffs associated with $\mu_{\Sigma^*}$. Since $\mu_{\Sigma^*}$ is not stable, by Claim \ref{Tmap}, $S$ is not a fixed point of $T$. Let $t_c = T_c(S)$. Since $S$ is not a fixed point of $T$, there exists $m\in\mathcal{M}$ and $c\in\mathcal{C}$ such that $t_{m,c} \neq S_{m,c}$. Moreover, let $\{x^t_{m,c}\}_{m \in \mathcal M}$ and  $\{x^s_{m,c}\}_{m \in \mathcal M}$ denote the measure of agents in $\tilde D_c(S_{-c})$ who are above the admission thresholds for authority $c$ under $t_c$ and $S_c$. As in Claim \ref{claim:tinc}, note that if there exists $m,c$ such that $t_{m,c} > S_{m,c}$, then from full support, we have that $x^s_{m,c} > x^t_{m,c}$. Since the authority fills its capacity in both cases, there must exists $m'$ such that $x^t_{m',c} > x^s_{m',c}$ which is only possible if $S_{m,c} > t_{m,c} $. By an identical argument, if there is $m,c$ such that $t_{m,c} < S_{m,c}$, then there exists $m'$ $S_{m',c} < t_{m',c}$. Therefore, whenever $t_{m,c} \neq S_{m,c}$, there exists $c$ and $m,m'$ such that   $t_{m,c} > S_{m,c}$ and $S_{m',c} > t_{m',c}$.  But now we have shown the following:
\begin{equation*}
    h_c(S_{c,m'})+u_{m'}(x^s_{c,m'})> h_c(t_{c,m'})+u_{m'}(x^t_{c,m'}) \ge h_c(t_{c,m})+u_{m}(x^t_{c,m}) > h_c(S_{c,m})+u_{m}(x^s_{c,m})
\end{equation*}
where the first inequality follows by $t_{c,m'}< S_{c,m'}$, $x^s_{c,m'} < x^t_{c,m'}$, and concavity of $u_m$. The second inequality follows by optimality. This is because the facts that $t_{c,m}>0$ and $t_{c,m'}'<1$ imply that the Lagrange multipliers in the proof of Theorem \ref{subsched} $ \overline{\kappa}_m=\underline{\kappa}_{m'}=0$ . The final inequality follows since $t_{m,c} > S_{m,c}$ and $x^t_{m,c} < x^s_{m,c}$. However, this is a contradiction since $h_c(S_{c,m'})+u_{m'}(x^s_{c,m'}) > h_c(S_{c,m})+u_{m}(x^s_{c,m})$ implies that there exists $\varepsilon>0$, an agent $\theta$ with score $s_c(\theta) = S_{c,m'} - \epsilon$ and type $m(\theta) = m$ has higher score under $A^*$ than the agent $\theta'$ with score $s_c(\theta') = S_{c,m}$ and type $m(\theta') = m$. Since $\theta'$ is admitted to $c$, $\theta$ would be if it applied to $c$. Moreover, from full support, there is such $\theta$ whose top choice is $c$ and the strategy of this agent is not a best response, which is a contradiction.
\end{proof}

The combination of Claims \ref{claim:deterministic} and \ref{claim:stableoutcome} completes the proof.
\end{proof}

\subsection{Proof of Proposition \ref{prop:ineff}}
\label{prop:ineffproof}
\begin{proof}
We prove the result by explicitly constructing an economy in which the optimal APMs lead to inefficiency. There are two authorities, $c$ and $c'$, both with capacity $1/2$ and two groups of agents, $m$ and $m'$. Both agent groups have a measure of $1$ and their scores are uniformly distributed in $[1/2,1]$. Authorities' utility functions are given by
\begin{equation}
 \xi_c\left(\bar{s}_{h},x\right)\equiv \bar{s}_{h}+ \frac{1}{4} \sqrt{x_{m}} + \frac{1}{8} \sqrt{x_{m'}}
 \end{equation}
 \begin{equation}
 \xi_{c'}\left(\bar{s}_{h},x\right)\equiv \bar{s}_{h}+ \frac{1}{4} \sqrt{x_{m'}} + \frac{1}{8} \sqrt{x_{m}}
 \end{equation}
with $h(x)\equiv x$ while all agents of type $m$ prefer authority $c'$ to $c$ while all agents of type $m'$ prefer authority $c$ to $c'$.\footnote{This assumption on the preferences and the distribution of scores violate our full support assumption, but adding an arbitrarily small full support density to all types makes arbitrarily small changes in the utility under the stable matching and optimal allocation but complicates the calculation, so we omit it for expositional clarity.}

We will now derive the stable outcome of this economy, which is (up to measure zero transformation) the unique outcome implemented when the authorities use the optimal APM. Let $x_m^{c}$ denote the measure of type $m$ agents at authority $c$. First, note that higher-scoring agents from the same group go to the more preferred authority. To see why this is true, note that if $m(\theta)=m(\theta') = m$, $s(\theta)>s(\theta')$ and $\mu(\theta)=c$ while $\mu(\theta')=c'$, $c$ and $\theta$ would violate within group fairness since $\theta$ has higher priority at $c$ than $\theta'$ regardless of the allocation. As a result, in any stable allocation $\mu$, the highest-scoring $x_{m'}^c$ type $m'$ agents are assigned to $c$ and the next highest-scoring $x_{m'}^{c'}$ agents are assigned to authority $c'$, while rest of the type $m'$ agents are not assigned to any authority. The allocation for type $m$ agents is analogous. Moreover, since $q=1/2$ for both authorities, $x_{m'}^{c'}$=1/2-$x_{m'}^c$ and $x_{m}^{c}$=1/2-$x_{m}^{c'}$ and the allocation is completely determined by the measures $x_{m'}^{c}$ and $x_{m}^{c'}$.

Next, note that at $\mu$, the adaptive priority of the lowest-scoring type $m$ and $m'$ agents must be equal at both authorities. To see why this is true, take authority $c$ without loss of generality. Let $s_{m'}^c = 1-x_{m'}^c$ and $s_{m}^c = 1-x_{m'}^c - x_{m}^c$ denote the scores of the lowest-scoring type $m$ and $m'$ agents and $A_m$ denote the optimal APM. For a contradiction, assume $A_m(x_m^{c},s_m^c) > A_{m'}(x_{m'}^{c},s_{m'}^c)$. Since agents of type $m'$ with scores lower than $s_m^c$ are unassigned at $\mu$, for small enough $\epsilon$, a type $m$ agent with score $s_m^c-\epsilon$ and authority $c$ blocks the matching. Similarly, assume $A_m(x_m^{c},s_m^c) < A_{m'}(x_{m'}^{c},s_{m'}^c)$. Since agents of type $m'$ with scores lower than $s_m^{c'}$ are assigned to authority $c$ or unmatched at $\mu$, a type $m'$ agent with score $s_{m'}^c-\epsilon$ and authority $c$ blocks the matching $\mu$. Thus, the following equations must be satisfied:
\begin{equation}
\begin{split}
    A_m(x_m^{c},s_m^c) = A_{m'}(x_{m'}^{c},s_{m'}^c) \text{  and  } A_m(x_m^{c'},s_m^{c'}) = A_{m'}(x_{m'}^{c'},s_{m'}^{c'})
\end{split}
\end{equation}
As the optimal APM in this setting is given by:
\begin{equation}
    A^*_{\hat m,\hat c}(y_{\hat m},s)\equiv s+u_{\hat m,\hat c}'(y_{\hat m})
\end{equation}
for all $\hat m\in\{m,m'\}$ and $\hat c\in\{c,c'\}$, we have that:
\begin{equation}
    1-x_{m'}^{c} + \frac{1}{8} \frac{1}{\sqrt{x_{m'}^c}} =  1-x_{m'}^{c} - x_{m}^{c} +   \frac{1}{4}\frac{1}{\sqrt{1/2 - x_{m'}^c}}
\end{equation}
and:
\begin{equation}
    1-x_{m}^{c'} + \frac{1}{8} \frac{1}{\sqrt{x_{m}^{c'}}} =  1-x_{m}^{c'} - x_{m'}^{c
    '} +   \frac{1}{4} \frac{1}{\sqrt{1/2 - x_{m}^{c'}}}
\end{equation}

These equations are identical up to relabelling and so $x_{m'}^{c} = x_{m'}^{c'} = x^*$ for some $x^*$. Thus, we need to find the solution to the following single equation to characterize the allocation:
\begin{equation}
    1-x^* + \frac{1}{8} \frac{1}{\sqrt{x^*}} =  \frac{1}{2} + \frac{1}{4}  \frac{1}{\sqrt{1/2 - x^*}}
\end{equation}
Observe that this equation can be rewritten as the fixed point equation:
\begin{equation}
x^*=\frac{1}{2}+\frac{1}{8}\frac{1}{\sqrt{x^*}}-\frac{1}{4}  \frac{1}{\sqrt{1/2 - x^*}}
\end{equation}
We observe that the RHS satisfies the following properties: (i) $\lim_{x^*\rightarrow 0}\text{RHS}(x^*)=\infty$, (ii) $\lim_{x^*\rightarrow \frac{1}{2}}\text{RHS}(x^*)=-\infty$, and (iii) $\text{RHS}'(x^*)<0$ for all $x^*\in(0,\frac{1}{2})$. Thus, there exists a unique solution. Moreover, we can guess-and-verify that this solution is $x^*=\frac{1}{4}$.

In summary, if both authorities use the optimal APM, then the outcome is
\begin{equation}
    \mu(\theta) =\begin{cases}
    c & \text{ if } m(\theta) = m, s(\theta) \in [1/2,3/4) \text{ or } m(\theta) = m', s(\theta) \in [3/4,1]\\
    c'& \text{ if }  m(\theta) = m', s(\theta) \in [1/2,3/4) \text{ or } m(\theta) = m, s(\theta) \in [3/4,1]\\
    \theta &\text{otherwise}
    \end{cases}
\end{equation}
In this outcome, both authorities have an average score of $3/4$ and admit measure $1/4$ agents from both groups, giving them a utility of $15/16$. Thus, total utilitarian welfare is $15/8$ under the decentralized outcome.

We now show that this does not attain the efficient benchmark. A necessary condition for the (utilitarian) efficient outcome is that for $c$:
\begin{equation}
    \frac{1}{4} \frac{1}{\sqrt{x_m^{c}}} = \frac{1}{8} \frac{1}{\sqrt{1/2 - x_m^{c}}} 
\end{equation}
and for $c'$:
\begin{equation}
    \frac{1}{4} \frac{1}{\sqrt{x_{m'}^{c'}}} = \frac{1}{8} \frac{1}{\sqrt{1/2 - x_{m'}^{c'}}}
\end{equation}
This implies that $x_m^c=x_{m'}^{c'}=4/10$ and $x_{m'}^c=x_m^{c'}=1/10$. In this case, the same set of agents is admitted overall, so the score contribution to utility remains 3/4 on average across the authorities. Total utilitarian welfare is now:
\begin{equation}
    3/2+1/2\times\sqrt{4/10}+1/4\times\sqrt{1/10}\approx 1.895>1.875=15/16
\end{equation}
Completing the proof.
\end{proof}

\subsection{Proof of Proposition \ref{thm:apmq}}
\label{thm:apmqproof}
\begin{proof}
First, we define a fictitious \textit{composite authority} with utility function defined over vectors of total scores $\bar{s}_h=\{\bar{s}_h^c\}_{c\in\mathcal{C}}$, and aggregate allocation to each group $x=\{x_m\}_{m\in\mathcal{M}}$. To do this, we define:
\begin{equation}
\begin{split}
    &\tilde u(\{ x_m\}_{m\in\mathcal{M}}) = \max_{\{x_{m,c}\}_{c\in\mathcal{C}}}\sum_{c\in\mathcal{C}}\sum_{m\in\mathcal{M}}u_{m,c}(x_{m,c})\\
    & \quad \text{s.t.}\, \sum_{c\in\mathcal{C}}x_{m,c}\leq x_m, \, \sum_{m\in\mathcal{M}}x_{m,c}\leq q_c, \, \forall m\in\mathcal{M}, c\in\mathcal{C}
\end{split}
\end{equation}
and $\tilde{\bar{s}}_h=\sum_{c \in \mathcal C}\bar{s}_h^c$. We write the utility function of this composite authority as
\begin{equation}
    \tilde\xi\left(\tilde{\bar{s}}_h,x\right) = \tilde{\bar{s}}_h + \tilde u(x)
\end{equation}

We first establish that $\tilde u$ satisfies the properties necessary to invoke Proposition \ref{nonsepsubsched}, which establishes the optimality of the claimed APM for the fictitious authority.
\begin{claim}
The function $\tilde u$ is concave and partially differentiable in each argument.
\end{claim}
\begin{proof}

First, we establish concavity. That is, for all $\lambda\in[0,1]$ and $x,x'\in\mathbb{R}_{+}^{|\mathcal{M}|}$, we have that $\tilde u(\lambda x'+(1-\lambda)x)\ge \lambda\tilde u(x')+(1-\lambda)\tilde u(x)$. Let $\{x_{m,c}^*\}_{m \in \mathcal M, c \in \mathcal C}$ and $\{x_{m,c}^{*'}\}_{m \in \mathcal M, c \in \mathcal C}$ correspond to optimal values under $x$ and $x'$. Under $\tilde x=\lambda x'+(1-\lambda)x$, we have that $\lambda x_{m,c}^{*'}+(1-\lambda)x_{m,c}^{*}$ is feasible for all $m \in \mathcal M$ and $c \in \mathcal C$. Thus, we have that:
\begin{equation}
\begin{split}
    \tilde u(\tilde x)&\ge \sum_{m\in\mathcal{M}}\sum_{c\in\mathcal{C}}u_{m,c}(\lambda x_{m,c}^{*'}+(1-\lambda)x_{m,c}^*) \\
    &\ge \sum_{m\in\mathcal{M}}\sum_{c\in\mathcal{C}}\lambda u_{m,c}( x_{m,c}^{*'})+(1-\lambda)u_{m,c}(x_{m,c}^*) \\
    &=\lambda\tilde u(x')+(1-\lambda)\tilde u(x)
\end{split}
\end{equation}
where the second inequality is by concavity of $u_{m,c}$ for all $m\in\mathcal{M},c\in\mathcal{C}$.

Second, we establish partial differentiability in each argument. That is, for all $x\in\mathbb{R}_{++}^{|\mathcal{M}|}$, $\frac{\partial}{\partial x_m}\tilde u(x)=\tilde u^{(m)}(x)$ exists. This follows by Corollary 5 in \citeAp{milgrom2002envelope}. Concretely, the domain of optimization can be taken to be a compact and convex subset of a normed vector space -- a sufficiently large cube in $\mathbb{R}_{+}^{|\mathcal{M}|\times|\mathcal{C}|}$ equipped with the Euclidean norm, for example. The objective function does not depend on $x$, and constraints are linear in $x$ (and therefore both continuous and continuously differentiable). Moreover, as $x\gg 0$, there exists a $\{x_{m,c}\}$ that satisfies all constraints with strict inequality.

\end{proof}

It follows that the objective function of the composite authority satisfies Assumption \ref{nonsep}, and so Proposition \ref{nonsepsubsched} implies that the non-separable APM $\tilde A_m(y,s)=h^{-1}\left(h(s)+\tilde u^{(m)}(y)\right)$ uniquely implements the first-best optimal allocation for the composite authority.

It remains to establish that the quota functions implement the optimal allocation $\{x_{m,c}\}$ conditional on $\{x_m\}$. Let $\lambda_m$ be the Lagrange multiplier on the $x_m$ constraint, $\gamma_c$ be the Lagrange multiplier on the $q_c$ constraint and $\underline{\kappa}_{m,c}$ be the Lagrange multiplier on the positivity constraint. Under our maintained Inada condition, we have that $\underline{\kappa}_{m,c}=0$. Moreover, by Corollary 5 in \citeAp{milgrom2002envelope}, we have that $\tilde u^{(m)}(x)=\lambda_m$, $\tilde u_{q_c}(x)=\gamma_c$, and $u_{m,c}'(x_{m,c}^*)=\lambda_m+\gamma_c-\underline{\kappa}_{m,c}$. Hence, we obtain that:
\begin{equation}
    x_{m,c}^*=\left(u_{m,c}^{'}\right)^{-1}\left(\tilde u^{(m)}(x)+\tilde u_{q_c}(x)\right)
\end{equation}
Thus, the following profile of quota functions implements the optimal cross-sectional allocation:
\begin{equation}
    Q_{m,c}(x)=\left(u_{m,c}^{'}\right)^{-1}\left(\tilde u^{(m)}(x)+\tilde u_{q_c}(x)\right)
\end{equation}
Completing the proof.
\end{proof}
\clearpage
\section{Additional Results for the Example (Section \ref{weitzmansection})}

\subsection{Formal Equivalence Between Prices \textit{vs.} Quantities and Priorities \textit{vs.} Quotas}
\label{weitzmancomp}
The structure of the comparative advantage of priorities over quotas from Section \ref{weitzmansection} hints at a more formal relationship between our analysis of affirmative action policies and Weitzman's analysis of price and quantity regulation. In Weitzman's model, there is a single firm producing a quantity of a single good $x\in\mathbb{R}$ with production costs $C(x,\zeta)$ and benefits $B(x,\zeta')$:
\begin{equation}
    \begin{split}
      C(x,\zeta)&=a_0(\zeta)+(C'+a_1(\zeta))(x-\hat x)+\frac{C''}{2}(x-\hat x)^2 \\
        B(x,\zeta')&=b_0(\zeta')+(B'+b_1(\zeta'))(x-\hat x)+\frac{B''}{2}(x-\hat x)^2
    \end{split}
\end{equation}
where $B',C',C''>0$, $B''<0$, and $\zeta$ and $\zeta'$ are random variables. The regulator can either set a price that the firm must charge (after which the firm chooses its optimal production quantity) or mandate the production of a given quantity. The comparative advantage of prices over quantities $\Delta^{\text{Weitzman}}$ is then defined as the difference between expected benefits net of costs under the optimal price regime minus the corresponding net benefits under the optimal quantity regime. This comparative advantage is given by:
\begin{equation}
    \Delta^{\text{Weitzman}}=\frac{C''^{-1}}{2}\left(1+C''^{-1}B''\right)\text{\normalfont Var}[a_1(\zeta)]
\end{equation}
The intuition for this formula is that when benefits are more curved than costs $|B''|>C''$, reducing variability in production is more valuable than the gain of having producers minimize costs. Thus, quantities are preferred. On the other hand, when costs are more curved than benefits, prices are preferred as there is greater production when producers have the lowest marginal costs of production.

These trade-offs are, in a certain sense, formally analogous to those that we have highlighted between priorities and quotas. In particular, under the mapping $C''^{-1}\mapsto \kappa$, $B''\mapsto-\gamma\beta$, $\text{\normalfont Var}[\omega]\mapsto\text{\normalfont Var}[a_1(\zeta)]$, we have that $\Delta^{\text{Weitzman}}=\Delta$. The intuition for this is that $C''^{-1}$ in the Weitzman framework determines how sensitive production is to changes in marginal cost, while $\kappa$ in our framework determines how sensitive the admitted measure of minority students is to the relative scores. Moreover, $B''$ corresponds to curvature in the benefits of production while $\gamma\beta$ corresponds to curvature in the benefits of admitting more minority students. Finally, $\text{\normalfont Var}[a_1(\zeta)]$ corresponds to the authority's uncertainty in the level of marginal costs of production while $\text{\normalfont Var}[\omega]$ corresponds to the authority's uncertainty regarding the marginal cost of admitting more minority students in terms of lost total score. Thus, the positive selection effect whereby priorities admit more minority students in the states of the world where they score highest is directly analogous to the effect that price regulation gives rise to the greatest production in states where the firm's marginal cost is lowest. Moreover, the guarantee effect whereby quotas prevent variation in the measure of admitted minority students across states of the world is analogous to the ability of quantity regulation to stabilize the level of production. Importantly, our results therefore allow one to apply established price-theoretic intuition for the benefits of price \textit{vs} quantity choice to matching markets without an explicit price mechanism.

\subsection{Beyond Affirmative Action: Medical Resource Allocation}\label{sec:Medical}

The lessons of this paper apply not only to affirmative action in academic admissions, but also more broadly to settings in which centralized authorities must allocate resources to various groups. One prominent such context is the allocation of medical resources during the Covid-19 pandemic. An important issue faced by hospitals is how to prioritize health workers (doctors, nurses and other staff) in the receipt of scarce medical resources: hospitals wish to both treat patients according to clinical need and ensure the health of the frontline workers needed to fight the pandemic. To map this setting to our example, suppose that the score $s$ is an index of clinical need for a scarce medical resource available in amount $q$, the measure of frontline health workers is $\kappa$, and $\omega$ indexes the level of clinical need in the patients currently (or soon to be) treated by the hospital, which is unknown. The risk aversion of the authority $\gamma\beta$ corresponds to both a fear of not treating sufficiently many frontline workers and excluding too many clinically needy members of the general population.

In practice, both priority systems and quota policies have been used, as detailed extensively by \citeAp{pathak2020leaving}.\footnote{Some other papers that study the allocation of scarce medical resources are \citeAp{akbarpour2021economic}, \citeAp{grigoryan2021effective} and \citeAp{dur2022allocating}.} The primary concern that has been voiced is that if a priority system is used, some groups (or characteristics) may be completely shut out of allocation of the scarce resource and that this is unethical, so quotas should be preferred. Our framework can be used to understand this argument: if there is an unusually high draw of $\omega$, a priority system would lead to the allocation of very few resources to frontline workers, and vice-versa. Our Proposition \ref{mechcomp} implies that if the authority is very averse to such outcomes ($\gamma\beta $ is high), quotas will be preferred and for exactly the reasons suggested. However, we also highlight a fundamental benefit of priority systems in inducing positive selection in allocation: when $\omega$ is high, it is beneficial that fewer resources go to the less sick medical workers and more to the relatively sicker general population. More generally, we argue that an adaptive priority mechanism that awards frontline workers a score subsidy that depends on the number of more clinically needy frontline workers could further improve outcomes.

An important additional consideration in this context arises if the hospital or authority must select a regime (priorities or quotas) before it understands the clinical need of its frontline workers $\kappa$, after which it can decide exactly how to prioritize these workers or set quotas, but before ultimate demand for medical resources $\omega$ is known. It follows from Proposition \ref{mechcomp} that the comparative advantage of priorities over quotas is:
\begin{equation}
    \mathbb{E}[\Delta]=\frac{1}{2}\left(\mathbb{E}[\kappa]-(\text{\normalfont Var}[\kappa]+\mathbb{E}^2[\kappa])\gamma\beta \right)\text{\normalfont Var}[\omega]
\end{equation}
Thus, an increase in uncertainty $\text{\normalfont Var}[\kappa]$ regarding the need of frontline workers leads to a greater preference for quotas. This highlights a further advantage of quotas in settings where a clinical framework must be adopted in the face of uncertainty regarding the clinical needs of frontline workers, as was the case at the onset of the Covid-19 pandemic.

\subsection{Optimal Precedence Orders}\label{precedencesection}

Thus far we have modelled quotas by first allocating minority students to quota slots and then allocating all remaining students according to the underlying score. However, we could have instead allocated $q-Q$ places to all agents according to the underlying score and then allocated the remaining $Q$ places to minority students. The order in which quotas are processed is called the \textit{precedence order} in the matching literature and their importance for driving outcomes has been the subject of a growing literature \citepAp[see \textit{e.g.,}][]{dur2018reserve,dur2020explicit,pathak2020immigration}. Our framework can be used to understand which precedence order is optimal, a question that has not yet been addressed.

In this example, the same factors that determine whether one should prefer priorities or quotas determine whether one should prefer processing quotas second or first. By virtue of uniformity of scores, it can be shown in the relevant parameter range that a priority subsidy of $\alpha$ is equivalent to a quota policy of $\kappa \alpha$ when quotas are processed second. Thus, the comparative advantage of priorities over quotas is exactly equal to the comparative advantage of processing quotas second over first. The intuition is analogous: processing quotas second allows for positive selection while processing quotas first fixes the number of admitted minority students. Thus, on the one hand, when the authority is more risk-averse, they should process quota slots first to reduce the variability in the admitted measure of minority students. On the other hand, when they are less risk-averse, they should process quotas second to take advantage of the positive selection effect such policies induce.
\begin{cor}
\label{prec}
The optimal quota-second policy achieves the same value as the optimal priority policy; quota-second policies are preferred to quota-first policies if and only if $\frac{1}{\kappa}\ge\gamma\beta $.
\end{cor}
\begin{proof}
We show that a quota-second policy $Q$ is equivalent to a priority subsidy of $\alpha(Q)=\frac{Q}{\kappa}$. A quota-second policy admits the highest-scoring  $x=\kappa(1-\omega)+Q$ minority students, floored by zero and capped by $\min\{\kappa,q\}$. A priority policy $\alpha(Q)=\frac{Q}{\kappa}$ admits the highest-scoring  $x=\kappa(1+\alpha(Q)-\omega)=\kappa(1-\omega)+Q$ minority students, floored by zero and capped by $\min\{\kappa,q\}$. Thus, state-by-state, quota-second policy $Q$ and priority subsidy $\alpha(Q)=\frac{Q}{\kappa}$ yield the same allocation. The claims then follow from Proposition \ref{mechcomp}.
\end{proof}

We emphasize that this equivalence is a result of the uniform distribution of scores and merely illustrates the similarity between priority policies and processing quotas second. This result does not hold in the more general model we study in the remainder of the paper. Indeed, in Theorem \ref{rationalization}, we show that for any quota policy to be optimal in the presence of uncertainty, it must process quotas first.

\clearpage

\section{Additional Quantitative Results}

In this Appendix, we describe both the methodology and results of the two robustness exercises that are not discussed in full detail in the main text. First, we estimate the gains from APM when we assume that CPS sets one tier size for all tiers rather than separately optimizing the sizes of the four tiers. Second, we estimate the gains from APM under alternative utility functions that differentially penalize underrepresentation and overrepresentation.

\subsection{Estimation with Homogeneous Reserves}\label{sec:homogeneousreserves}

As we have motivated, in this section we estimate an alternative model, where CPS chooses a single reserve size, $r$, instead of separate reserve sizes for all tiers. Formally, we replace the vector of reserve sizes of the four socioeconomic tiers, $r = (r_1,r_2,r_3,r_4)$ by $r=(r,r,r,r)$. In this setting, we define the marginal benefit of increasing reserve size as

\begin{equation}
    G(r,\Lambda;\beta,\gamma) = \frac{\partial}{\partial r}\Xi(r,\Lambda;\beta,\gamma)
\end{equation}

As in the general model, any (interior) reserve policy $r^*$ must satisfy  $ G(\hat{r}^*,\hat{\Lambda};\beta,\gamma) = 0$. This first-order condition yields one moment, and so we can estimate one parameter. To this end, we fix $\gamma$, and for each $\gamma \in [1,10]$, and we estimate $\beta^*(\gamma)$ as the exact solution to the following empirical moment condition:
\begin{equation}
   G(\hat{r}^*,\hat{\Lambda};\beta^*(\gamma),\gamma) = 0
\end{equation}

\begin{figure}
    \centering
    \caption{Estimated Slope of Utility Under Homogeneous Reserves}
    \includegraphics[width=0.7\textwidth]{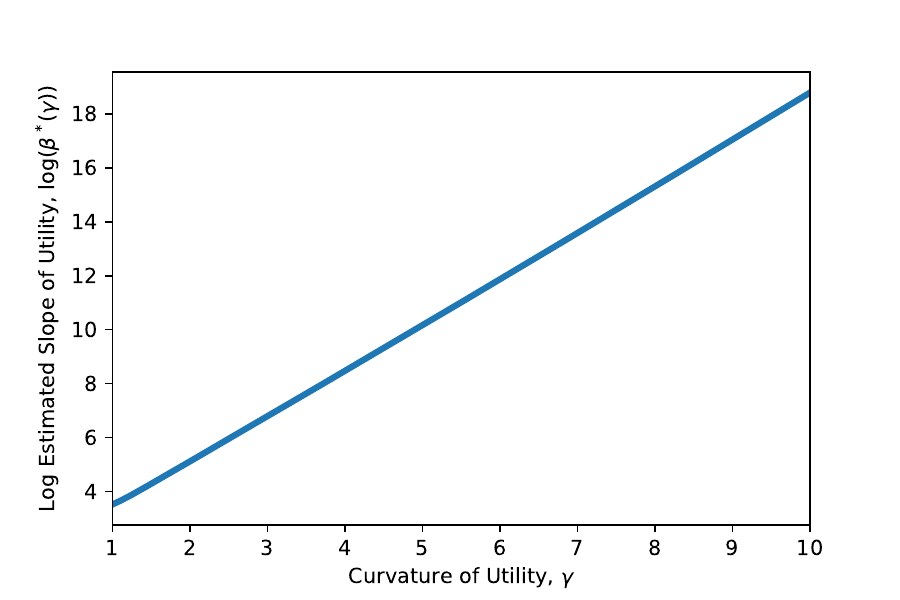}
    \fnote{This graph plots the estimated logarithm of the slope of utility $\log \beta^*(\gamma)$ in the homogeneous reserve case as we vary the curvature of utility $\gamma \in [1,10]$.}
    \label{fig:betaestimate}
\end{figure}

\begin{figure}
    \centering
    \caption{Payoff Gains from APM Under Homogeneous Reserves}
    \includegraphics[width=0.7\textwidth]{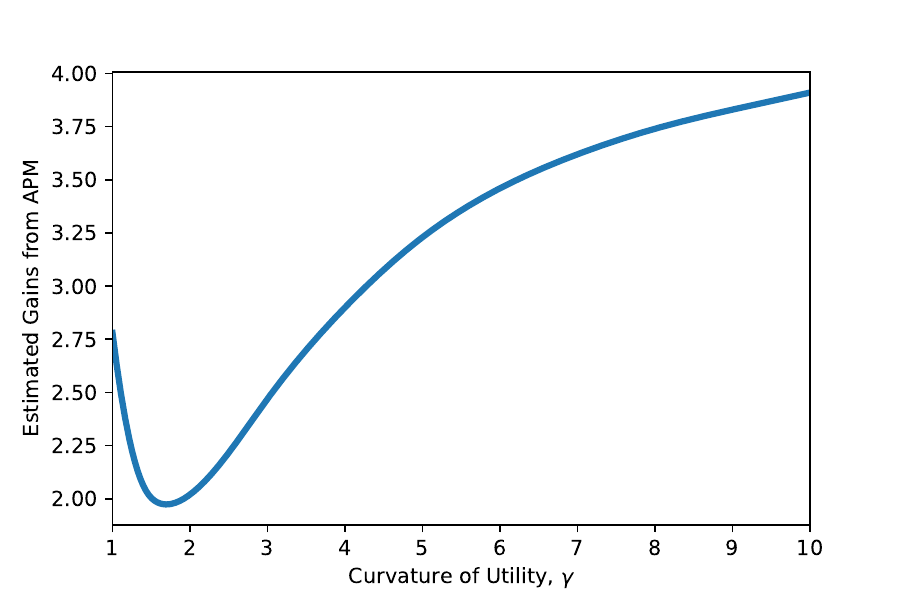}
    \fnote{This graph plots the estimated difference in payoffs in the homogeneous reserves case between the optimal APM and the CPS policy as we vary the curvature of utility $\gamma \in [1,10]$.}
    \label{fig:onemomentgains}
\end{figure}

Figure \ref{fig:betaestimate} plots the logarithm of the estimated $\beta^*(\gamma)$. The estimated $\beta^*(\gamma)$ is increasing in $\gamma$. As the loss term $|x_t-0.25|$ is in $(0,1)$, $\beta^*(\gamma)$ is increasing and convex in $\gamma$, where $\beta^*(1) = 34$ and $\beta^*(10) = 1.436 \times 10^8$. In Figure \ref{fig:onemomentgains}, we plot the gains as a function of $\gamma$, which shows that even though the estimated value for $\beta$ moves quite a lot, the empirical gains range from $2$ to $4$ points. This also shows that the estimated gain from APM of $2.1$ under our benchmark specification is close to the lower bound of the estimated gains under the alternative specification with homogeneous reserves.

\begin{figure}
    \centering
    \caption{The Gains from APM as a Fraction of the Loss From Underrepresentation Under Homogeneous Reserves}
    \includegraphics[width=0.7\textwidth]{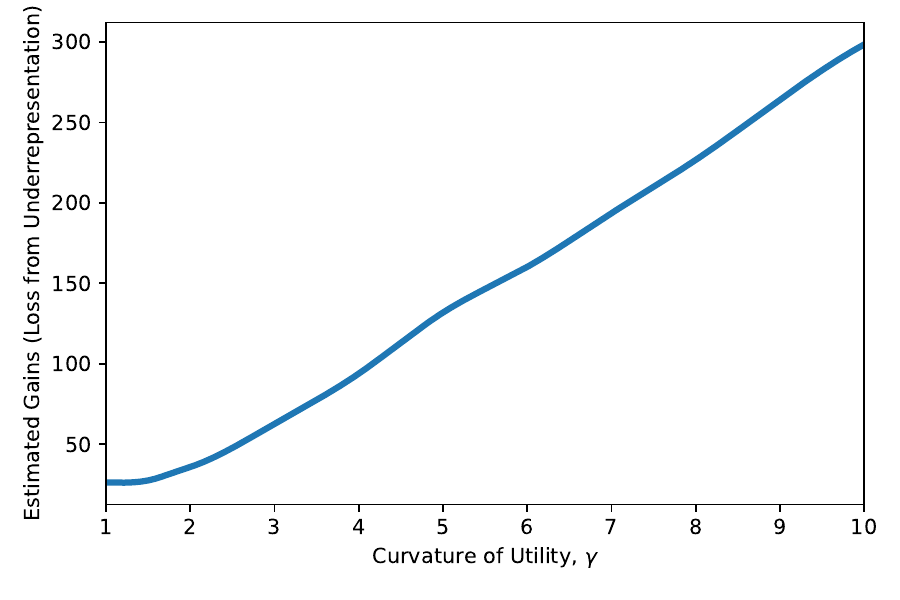}
    \fnote{This graph plots the estimated difference between the payoffs under the optimal APM under homogeneous reserves as a fraction of loss from underrepresentation as we vary the curvature of utility $\gamma\in[1,10]$.}
    \label{fig:onemomentpercentagegains}
\end{figure}

Finally, we benchmark these gains as a fraction of loss from underrepresentation under the CPS policy, where the loss of underrepresentation is calculated under the estimated parameter values. In Figure \ref{fig:onemomentpercentagegains}, we plot the gains under APM as a percentage of diversity loss under the CPS policy. These range from $26\%$ to $300\%$. Our baseline percentage gain estimate of $37.5\%$ is again close to the lower bound that we estimate under the alternative specification with homogeneous reserves.

\subsection{Gains from APM Under Different Utility Functions}\label{sec:alternativeutilityfunctions}
In this section, as we have motivated, we estimate alternative objective functions to investigate the robustness of our findings.

First, we analyze a setting that includes a loss term only for underrepresented tiers (and does not penalize overrepresentation of any tier). To this end, we replace the term $|0.25 - x_t|$ with $\min\{0,(0.25-x_t)\}$ and perform the same estimation with the following parametric utility function:

\begin{equation}\label{CPSutility1}
  \xi(\bar{s},x;\beta,\gamma) = \bar{s} +  \beta \sum_{t=1}^4 (\min\{0,(0.25-x_t)\})^{\gamma}
\end{equation}

The estimated parameter values are $\beta^* = -52058$ and $\gamma^* = 3.87467$. We compute the difference between the empirical payoffs under APM and the CPS reserve policy to be $0.262$, which is significantly lower than our estimate of $2.1$. However, the reason for this is that the diversity domain is estimated to be less important under this specification, and the diversity loss under the CPS policy is $2.71$. Thus, improvements from APM correspond to $9.6 \%$ of the loss from underrepresentation, which is attenuated relative to our baseline specification, but remains non-negligible.

Second, we allow CPS to care differentially
about underrepresentation and overrepresentation by considering a utility function with separate coefficients for underrepresented and overrepresented tiers. To this end, we define the following loss function:

\begin{equation}
    f(x_t,\beta_l,\beta_h,\gamma) = \begin{cases}
            \beta_{l} (0.25 - x_t)^{\gamma} & \text{ if } x_t \leq 0.25\\
            \beta_{h} (x_t - 0.25)^{\gamma} & \text{ if } x_t > 0.25
    \end{cases}
\end{equation}
where $\beta_l$ indexes the loss from underrepresentation of a tier, while $\beta_h$ indexes the loss from overrepresentation. We then perform the same estimation with the following parametric utility function:

\begin{equation}\label{CPSutility2}
  \xi(\bar{s},x;\beta,\gamma) = \bar{s} + \sum_{t=1}^4 f(x_t,\beta_l,\beta_h,\gamma) 
\end{equation}

This yields the following estimated values: $\beta_l^* = -1362270$, $\beta_h^* = -12278$, $\gamma^* = 5.28021$. We compute the difference between the empirical payoffs under APM and the CPS reserve policy to be $0.195$ and the loss from underrepresentation under the CPS policy to be $2.24$. Thus, we conclude that improvement from APM corresponds to $8.7 \%$ of loss from underrepresentation under the CPS policy, which is similar to what we obtain under the specification in which there is no loss from overrepresentation.

\clearpage

\section{Extension to More General Authority Preferences}
\label{ap:preferences}
In this Appendix, we relax Assumption \ref{sep} to allow for (i) non-separable diversity preferences, (ii) non-separable score and diversity preferences, (iii) non-differentiable preferences, and (iv) non-concave diversity preferences. We show how these changes in assumptions lead to certain modified APM mechanisms becoming first-best optimal.

\subsection{Non-Separable Diversity Preferences}
First, we relax Assumption \ref{sep} and instead suppose that the authority's preferences satisfy the following assumption:

\begin{assumption}
\label{nonsep}
The authority's utility function can be represented as:
\begin{equation}
    \xi\left(\bar{s}_{h},x\right)\equiv g\left(\bar{s}_{h}+u(x)\right)
\end{equation}
for some continuous, strictly increasing function $g:\mathbb{R}\rightarrow\mathbb{R}$ and a concave, partially differentiable $u$ in each argument.
\end{assumption}

In this environment, we define a \textit{non-separable APM} $\tilde A=\{\tilde A_m\}_{m\in\mathcal{M}}$ where $\tilde A_m:\mathbb{R}^{|\mathcal{M}|}\times[0,1]\rightarrow\mathbb{R}$. This implements allocation $\mu$ in state $\omega$ as per Definition \ref{def1} (under the modification of point 1 in Definition \ref{def1} to allow $A_m$ to depend on $x$ rather than just $x_m$).

We generalize Theorem \ref{subsched} to show that the following non-separable APM uniquely implements the first-best optimal allocation:

\begin{prop}
\label{nonsepsubsched}
The non-separable APM $\tilde{A}_m^*(y,s)\equiv h^{-1}(h(s)+u^{(m)}(y))$ and uniquely implements the first-best optimal allocation.\footnote{Where we define $u^{(m)}(y)=\frac{\partial}{\partial y_m}u(y)$.}
\end{prop}
\begin{proof}
Follow every step in the proof of Theorem \ref{subsched} with $\sum_{m\in\mathcal{M}}u_m(x_m)$ replaced by $u(x)$ and $u_m'(x_m)$ replaced by $u^{(m)}(x)$.
\end{proof}

Thus, allowing for non-separable diversity preferences does not substantially change the analysis of adaptive priority mechanisms. One must simply adapt the APM to be non-separable to allow cross-group diversity concerns to shape the marginal benefits of admitting agents from various groups. The main difference is that this a non-separable APM does not necessarily allow the greedy implementation of Algorithm \ref{alg1}. This is because, in the presence of cross-group adaptive priorities, it is no longer enough to rank agents within their own group. A small adaptation to this algorithm that dynamically admits agents, starting from the highest-scoring agents in each group, would naturally implement the unique first-best optimal allocation.

\subsection{Non-Separable Score and Diversity Preferences}
Second, we relax Assumption \ref{sep} and instead suppose that the authority's preferences are represented by:

\begin{assumption}
\label{fullnonsep}
The authority's Bernoulli utility function can be represented as:
\begin{equation}
    \xi\left(\bar{s}_{h},x\right)
\end{equation}
where $\xi$ is monotone, differentiable, and concave.
\end{assumption}

We define a state-dependent APM $\hat A=\{\hat A_m\}_{m\in\mathcal{M}}$ where $\hat{A}_m:\mathbb{R}^{|\mathcal{M}|}\times[0,1]\times\Omega\rightarrow\mathbb{R}$. This implements allocation $\mu$ in state $\omega$ as per Definition \ref{def1} (where point 1 in Definition \ref{def1} is modified to allow $A_m$ to depend on both $x$ and $\omega$).

In this more general setting, we now find a state-dependent APM that implements the optimal allocation.

\begin{prop}
The following state-dependent APM implements a first-best optimal allocation:
\begin{equation}
    A_m(y,s,\omega)\equiv h^{-1}\left(h(s)+\frac{{\xi}_{x_m}\left(\bar{s}_{h}(y,\omega),y\right)}{{\xi}_{\bar{s}_{h}}\left(\bar{s}_{h}(y,\omega),y\right)}\right)
\end{equation}
where $\bar{s}_h(y,\omega)$ is the score index in state $\omega$ when the highest-scoring $y=\{y_m\}_{m\in\mathcal{M}}$ agents of each attribute are allocated.
\end{prop}
\begin{proof}
Follow every step in Theorem \ref{subsched} with $\sum_{m \in \mathcal M} \int_{\underline{\tilde s}_m(x_m)}^{h(1)} \tilde s \tilde f_m(\tilde s)\dd\tilde s + \sum_{m \in \mathcal M} u_m(x_m)$ replaced with $\xi\left(\bar{s}_h(y,\omega),x\right)$ where $\bar{s}_h(y,\omega)=\sum_{m \in \mathcal M} \int_{\underline{\tilde s}_m(x_m)}^{h(1)} \tilde s \tilde f_{m,\omega}(\tilde s)\dd\tilde s$.
\end{proof}

There are two substantial differences in this optimal policy from our baseline APM. First, the policy depends on the joint distribution of agents in the population. Thus, specifying it \textit{ex ante} is likely to be extremely challenging in any practical setting. This is necessary because the marginal rate of substitution between diversity and scores depends on the level of scores, which depends on the distribution of agents. Second, without assumptions on the shape of the distribution of agents, there is no guarantee that this policy is monotone and thus no guarantee that it implements a unique policy.

Thus, while Proposition \ref{nonsepsubsched} showed that cross-group separability is largely inessential for our main conclusions, separability between score and diversity preferences is key to the power of APM.

\subsection{Non-Differentiable Preferences}

In this section, we retain the majority of Assumption \ref{sep}, where we instead suppose that the authority's diversity preferences $\{u_m\}_{m\in\mathcal{M}}$ are potentially non-differentiable at finitely many points.

As $u_m$ is concave, the left and right derivatives of $u_m$, $u^{-}_m$ and $u^{+}_m$, exist. The definition of our first-best APM is not applicable to this case since $u_m'$ might not exist. Therefore, we define the following generalized optimal APM $A_m^*(y_m,s)\equiv h^{-1}(h(s)+u_m^{-}(y_m))$, which simply replaces $u_m'$ with $u_m^{-}$ in the definition. By concavity of $u_m$, $u_m^{-}$ is monotone decreasing. Thus, this generalized optimal APM (as it is a montone APM) implements a unique allocation by Proposition \ref{lem:APMprop}. Moreover, the unique allocation that it implements is an optimal allocation:

\begin{prop}
Let $\mu^*$ denote the allocation implemented by the generalized optimal APM. $\mu^*$ is an optimal allocation. 
\end{prop}
\begin{proof}
We first prove a claim. An allocation in this setting is a cutoff allocation if there exists cutoffs $\{s_m\}_{m \in \mathcal M}$ such that an agent $\theta$ is assigned the resource if and only if $s(\theta) \geq s_m$ and $m(\theta) = m$.

\begin{claim}
There exists a unique optimal allocation $\mu'$ in the sense that all other allocations that attain the optimal payoff differ from $\mu'$ on at most a measure zero set of types. Moreover, there exists an optimal allocation that is a cutoff allocation.
\end{claim}
\begin{proof}
In the setting of Theorem \ref{subsched}, observe that $\underline{\tilde s}_m(x_m)$  is strictly decreasing in $x_m$. This, together with the concavity of $u$ implies that the objective is strictly concave and constraints are linear. Therefore an optimal allocation exists and is unique up to measure zero transformations. Given this allocation $\mu'$ (with measures $x_m$), an optimal cutoff allocation is obtained by the cutoff scores $s_m^*$ that satisfy
\begin{equation}
 s_m' = \sup\left\{s_m\in[0,1]: \int_{s_m}^{1} \tilde f_m(\tilde s)\dd\tilde s = x_m \right\}
\end{equation}
\end{proof}

Using this claim, toward a contradiction, assume there exists another allocation $\mu'$, which gives the authority a strictly higher utility. Moreover, take $\mu'$ to be an optimal cutoff allocation (which must exist by the claim). As $\mu'$ differs from $\mu^*$ and both are cutoff allocations, we have that there exist two groups $m,n\in\mathcal{M}$ such that: (i) $s_m'>s_m^*$ and $x_m'<x_m^*$ and (ii) $s_n'<s_n^*$ and $x_n'>x_m^*$. We have that:
\begin{equation}
    A_m^*(x,s)\ge A_m^*(x_m^*,s)>A_m^*(x_m^*,s_m^*)\ge A_n^*(x_n^*,\hat s) \ge A_n^*(\hat{x},\hat{s})
\end{equation}
for all $s\in(s_m^*,s_m')$, $\hat{s}\in(s_n',s_n^*)$, $x\leq x_m^*$, $\hat{x}\ge x_n^*$. The first inequality follows by concavity of $u_m$, the second follows by the fact that $h$ is strictly increasing, the third follows by the definition of APM and the fact that $\mu^*(s_m^*,m)=1$ and $\mu^*(\hat{s},n)=0$, and the fourth follows from concavity of $u_n$. Thus, we have that, for all $s\in(s_m^*,s_m')$, $\hat{s}\in(s_n',s_n^*)$, $x\leq x_m^*$, $\hat{x}\ge x_n^*$:
\begin{equation}
    u_m^{-}(x)+h(s)>u_n^{-}(\hat{x})+h(\hat{s})
\end{equation}
Thus, the total marginal utility obtained by replacing any positive measure type $m$ students with scores $s\in(s_m^*,s_m')$ with an identical measure of type $n$ students with scores $\hat{s}\in(s_n',s_n^*)$ is positive. But this contradicts the optimality of $\mu'$. Thus, if $\tilde\mu$ is optimal, then $\tilde\mu=\mu^*$ (up to a measure zero set).
\end{proof}

\subsection{Non-Concave Preferences}

In this section, we relax the assumption that the $u_m$ are concave.

\begin{prop}\label{prop:nonconcaveresult}
If $\mu$ is an optimal allocation, then $\mu$ is implemented by $A^*$.
\end{prop}
\begin{proof}
Without concavity, the optimal allocation characterized in the proof of Theorem \ref{subsched} is no longer unique. However, the Lagrangian conditions we have derived are still necessary for any optimal allocation $x = \{x_m\}_{m \in \mathcal M}$. Thus, any optimal allocation is implemented by $A^*$.
\end{proof}

This result shows that any optimal allocation is implemented by the optimal APM. However, when $\{u_m\}_{m \in \mathcal M}$ are not concave, $A^*$ is not necessarily monotone. Therefore, $A^*$ does not necessarily implement a unique allocation. Indeed, it is possible that $A^*$ implements suboptimal allocations, as it will implement any locally optimal allocation. Therefore, a mechanism defined by an arbitrary selection from the allocations implemented by $A^*$ would not be first-best optimal. However, $A^*$ may still help decision-making in this setting as it implements any optimal allocation.

\clearpage
 \section{Extension of the Main Results to Discrete Economies}

\label{discapp}
In this Appendix, we extend the results in the main text to discrete economies and thereby establish that the core of our analysis generalizes from the continuum framework. Concretely, we show that appropriate analogs of Theorems \ref{subsched}, \ref{rationalization}, and \ref{thm:eq} carry over to discrete economies. Together, these establish the optimality of APM, characterize the (sub)-optimality of priorities and quotas, and demonstrate the dominance of APM in discrete economies. As discrete economies do not necessarily admit a unique stable matching (as is well known), the first part of Theorem \ref{prop:multischoolstable} does not hold (uniqueness), but the second part does (stable matchings are cutoff matchings).

 \subsection{Primitives}
An authority has $q$ resources to allocate. At each state $\omega$, the economy the authority faces corresponds to agents  $\Theta^{\omega}  = \{\theta_1 ,\ldots,\theta_{N(\omega)} \}$ where $q\leq|N(\omega)|$.  As in the continuum case, $\theta \in[0,1]\times\mathcal{M}$ denotes the type of an agent  who has score $s$ and belongs to group $m$. We denote the score and group of any type $\theta$ by $s(\theta)$ and $m(\theta)$, respectively. For simplicity, we assume that no two agents have the same score at any $\omega$, formally, if $\{\theta,\theta'\}\subseteq \Theta^{\omega}$, then $s(\theta) \neq s(\theta')$.

An allocation $\mu:\Theta\rightarrow\{0,1\}$ specifies for any type $\theta\in\Theta$ whether they are assigned to the resource. The set of possible allocations is $\mathcal{U}$ and $\Omega$ is the set of all possible economies. An allocation is feasible if it allocates no more than measure $q$ of the resource. A mechanism is a function $\phi:\Omega\rightarrow \mathcal{U}$ that returns a feasible allocation for any possible $\Theta^{\omega}$.

The authority believes $\omega$ has distribution $\Lambda \in \Delta(\Omega)$. $x(\mu,\omega) = \{x_m(\mu,\omega)\}_{m \in \mathcal M}$ denotes the number of agents of each group allocated the resource at matching $\mu$, while $\bar{s}_{h}(\mu,\omega) = \sum_{\theta \in \Theta^{\omega}} \mu(\theta) h(s(\theta))$ denotes the utility the authority derives from scores at $\mu$. The preferences of the authority are given by $ \xi: \mathbb{R}^{|\mathcal{M}|+1} \to \mathbb R$:
\begin{equation}
    \xi\left(\bar{s}_{h},x\right)\equiv \bar{s}_{h}+\sum_{m\in\mathcal{M}}u_{m}(x_m)
\end{equation}
where $h$ is continuous and strictly increasing and $u_m$ is concave for all $m\in\mathcal{M}$.

\subsection{Optimal Mechanisms in Discrete Economies}
We adapt our definition of the Adaptive Priority Mechanisms to the discrete setting. An \textit{adaptive priority policy} $A=\{A_m\}_{m\in\mathcal{M}}$, where $A_m:\mathbb{R}\times[0,1]\rightarrow\mathbb{R}$. The adaptive priority policy assigns priority $A_m(y_m,s)$ to an agent with score $s$ in group $m$ when $y_m$ of agents of the same group is allocated the object. Given an adaptive priority policy, an APM implements allocations in the following way: 

\begin{defn}[Adaptive Priority Mechanism]\label{defn:discreteapm}
An adaptive priority mechanism, induced by an adaptive priority $A$, implements an allocation $\mu$ in state $\omega$ if the following are satisfied:
\begin{enumerate}
    \item Allocations are in order of priorities: $\mu(\theta) = 1$  if and only if 
    \begin{itemize}
    \item[(i)] for all $\theta'$ with $m(\theta') \neq m(\theta)$ and $\mu(\theta') = 0$,
    \begin{equation}\label{discreteapmequation}
        A_{m(\theta)}(x_{m(\theta)}(\mu,\omega),s(\theta)) \geq A_{m(\theta')}(x_{m(\theta')}(\mu,\omega)+ 1,s(\theta'))
    \end{equation}
    \item[(ii)]  for all $\theta'$ with $m(\theta') = m(\theta)$ and $\mu(\theta')=0$, $s(\theta) > s(\theta')$
    \end{itemize}
    \item The resource is fully allocated:
    \begin{equation}
        \sum_{m \in \mathcal M} x_m(\mu,\omega) = q
    \end{equation}
\end{enumerate}
\end{defn}

Definition \ref{defn:discreteapm} makes two modifications relative to the continuum model. First, the measures of agents from each group are replaced by the number of agents from each group. Second, when $m(\theta) \neq m(\theta')$, the adaptive priority of $\theta'$ is now evaluated in the case where an extra agent from $m(\theta')$ is assigned the resource.\footnote{This was not the case in the continuum model since all types of agents have measure $0$ and therefore replacing $\theta$ with $\theta'$ has no effect the evaluation of diversity.} Unlike the continuum case, it is possible for a monotone APM to implement two different allocations, since it can assign the same priority to two different agents, which could happen only for a zero-measure set of agents in the continuum model.

Define  $A^*_m(y_m,s)\equiv h(s)+u_m(y_m)-u_m(y_m-1)$, which will turn out to be the optimal APM. We first show that $A^*$ is monotone, and all allocations that $A^*$ implements give the authority the same utility.

\begin{lem}\label{discretepayoffequiv}
$A^*$ is monotone. Moreover, if $A^*$ implements $\mu$ and $\mu'\neq\mu$ in state $\omega$, then $\xi(\mu,\omega) = \xi(\mu',\omega)$. 
\end{lem}
\begin{proof}
Monotonicity is immediate from the definition of $A^*$ and concavity of $u_m$. Assume that $A^*$ implements two different allocations, $\mu$ and $\mu'$ at $\omega$. Let $x_l$ and $x'_l$ denote the number of group $l \in \mathcal M$ agents assigned the resource at $\mu$ and $\mu'$. Since $A^*$ is monotone and $\mu \neq \mu'$, there are $m$ and $n$ such that $x_m > x'_m$ and $x'_{n} > x_{n}$. Let $\tilde \theta_{l}$ and $\tilde \theta'_{l}$ denote the lowest-scoring  type $l$ agent assigned the resource at $\mu$ and $\mu'$, respectively. Similarly, let $\hat \theta_{l}$ and $\hat \theta'_{l}$ denote the highest-scoring type $l$ agents who is not assigned the resource at $\mu$ and $\mu'$, respectively. Let $\tilde \mu$ denote the matching given by: $\tilde \mu(\theta) = \mu(\theta)$ if $\theta \not \in \{\tilde \theta_{m},\hat \theta'_{n}\}$, $\tilde \mu(\tilde \theta_{m}) = 0$ while $\mu(\hat \theta_{n}) = 1$. $\tilde \mu$ starts with $\mu$, takes the resource away from the lowest-scoring group $m$ agent who has it, $\tilde \theta_{m}$, and allocates it to the highest-scoring group $n$ agent who does not have it, $\hat \theta_{n}$. Note that since $A^*$ is monotone, from $x_m > x'_m$ and $x'_{n} > x_{n}$, under $\mu'$,  $\hat \theta_{n}$ is already allocated the resource while $\tilde \theta_{m}$ is not.
\begin{claim}\label{onestep}
$\tilde \mu$ is implemented under $A^*$ in state $\omega$ and $\xi(\mu,\omega) = \xi(\tilde \mu,\omega)$.
\end{claim}
\begin{proof}
Since $A^*$ implements $\mu$ and $\mu(\hat \theta_{n}) = 0$, we have that $A_m^*(s(\tilde \theta_{m}),x_m) \geq A_n^*(s(\hat \theta_{n}),x_{n}+1)$. Conversely, since $A^*$ also implements $\mu'$ and $\mu'(\hat \theta_{m}') = 0$, we have that $A_n^*(s(\tilde \theta_{n}'),x_n') \geq A_m^*(s(\hat \theta_{m}'),x_{m}'+1)$. Moreover, since $x_m > x'_m$ and $x'_{n} > x_{n}$, we have that $s(\hat \theta_m') \ge s(\tilde \theta_m)$ and  $s(\hat \theta_{n}) \ge s(\tilde \theta_n')$ . From this, it follows that:
\begin{equation}
\begin{split}
    A_n^*(s(\hat\theta_n),x_n+1)&\ge A_n^*(s(\tilde\theta_n'),x_n+1)\ge A_n^*(s(\tilde \theta_{n}'),x_n')\\
    &\geq A_m^*(s(\hat \theta_{m}'),x_{m}'+1)\ge A_m^*(s(\hat\theta_m'),x_m)\ge A_m^*(s(\tilde\theta_m),x_m)
\end{split}
\end{equation}
where the first inequality holds as  $s(\hat \theta_{n}) \ge s(\tilde \theta_n')$, the second inequality holds as $x_n' > x_n$ (which implies $x_n' \ge x_n + 1$) and $A^*_n$ is decreasing in its second argument, the third inequality holds as $A^*$ also implements $\mu'$ (as stated above), the fourth inequality holds as $x_m' < x_m$ (which implies $x_m' + 1 \le x_m $) and $A^*_n$ is decreasing in its second argument, and the fifth inequality holds as $s(\hat \theta_m') \ge s(\tilde \theta_m)$. Thus, $A_m^*(s(\tilde\theta_m),x_m) \leq A_n^*(s(\hat \theta_{n}),x_{n}+1)$. This shows that  $A_m^*(s(\tilde \theta_{m}),x_m) = A_n^*(s(\hat \theta_{n}),x_{n}+1)$, which implies that $\tilde \mu$ is implemented under $A^*$ and  $\xi(\mu,\omega) = \xi(\tilde \mu,\omega)$.
\end{proof}
Note that Claim \ref{onestep} shows that starting from a matching $\mu$ which is implemented by $A^*$, taking away the object from a particular agent who does not have it in $\mu'$ and allocating it to a particular agent who has it in $\mu'$, we arrive at another matching $\tilde \mu$ that is implemented under $A^*$ and gives the authority the same payoff. Therefore, starting from any $\mu$ that is implemented by $A^*$ and repeating this construction (by replacing $\mu$ at step $i$ with $\tilde \mu$ at step $i-1$) where at each step we take the resource from an agent who is not allocated the resource at $\mu'$ and assign it to an agent who is, in finitely many steps we arrive at $\mu'$. Since the payoff stays the same at each step, $\mu'$ gives the authority the same payoff as $\mu$.
\end{proof}

\begin{thm}\label{discreteoptimality}
If $\mu$ is implemented by $A^*$, then $\mu$ is an optimal matching.
\end{thm}
\begin{proof}
First, note that an optimal matching exists since the economy (and therefore the set of matchings) is finite. We first show the following lemma.
\begin{lem}\label{improvement}
If $\mu$  is not implemented by $A^*$, then there exists $\mu'$ that gives the authority a strictly higher payoff.
\end{lem}
\begin{proof}
If $\mu$  is not implemented by $A^*$, then there exists  $\theta$ and $\theta'$ such that $\mu(\theta) = 0$, $\mu(\theta')=1$ and either $m(\theta) = m(\theta')$ and $s(\theta) > s(\theta')$ or $m(\theta) \neq m(\theta')$ and
\begin{equation}
\begin{split}
        h(s(\theta)) + u_{m(\theta)}(x_{m(\theta)}(\mu)+1) -  u_{m(\theta)}(x_{m(\theta)}(\mu)) > \\ h(s(\theta')) + u_{m(\theta')}(x_{m(\theta')}(\mu)) -  u_{m(\theta')}(x_{m(\theta')}(\mu)-1)
\end{split}
\end{equation}
However, in both cases, a $\mu'$ that allocates the resource to $\theta$ instead of $\theta'$ (while not changing any other agent's matching) strictly improves the utility of the authority.
\end{proof}
Lemma \ref{improvement} proves that the optimal matching cannot be a matching that is not implemented by $A^*$. Since the optimal matching exists, then it is implemented by $A^*$. From Lemma \ref{discretepayoffequiv}, all matchings implemented by $A^*$ give the authority the same payoff, proving the result.
\end{proof}

Note that Lemma \ref{discretepayoffequiv} and Theorem \ref{discreteoptimality} imply that any mechanism that is defined by an arbitrary singleton selection from the set of matchings that $A^*$ implements would achieve the optimal matching under any $\omega$ and therefore would be first-best optimal.

 \subsection{Priorities \textit{vs.} Quotas in Discrete Economies}

Now, we define Priority and Quota Mechanisms in the discrete model and extend our (sub)optimality results to discrete economies.

A \textit{priority policy} $P:\Theta \to [0,1]$ awards an agent of type $\theta\in\Theta$ a priority $P(\theta)$.

\begin{defn}[Priority Mechanisms]
A priority mechanism, induced by a priority policy $P$, allocates the resource in order of priorities until measure $q$ has been allocated, with ties broken uniformly and at random.
\end{defn}

A \textit{quota policy} is given by $(Q,D)$, where $Q=\{Q_m\}_{m \in \mathcal M}$ and $D:\mathcal{M}\cup\{R\}\rightarrow \{1,2,\ldots,|\mathcal{M}|+1\}$ is a bijection. The vector $Q$ reserves $Q_m$ objects for agents in group $m$, with residual capacity $Q_R=q-\sum_{m\in\mathcal{M}}Q_m$ open to agents of all types. The bijection $D$ (often called the precedence order) determines the order in which the groups are processed.

\begin{defn}[Quota Mechanisms]
\label{quotadefdisc}
A quota mechanism, induced by a quota policy $(Q,D)$, proceeds by allocating $Q_{D^{-1}(k)}$ objects to agents from group $D^{-1}(k)$ (if there are sufficient agents from this group) to the resource in ascending order of $k$, and in descending order of score within each $k$. If there are insufficiently many agents of any group to fill the quota, the residual capacity is allocated to a final round in which all agents are eligible.
\end{defn}

We also extend the definitions of risk-neutrality and high risk aversion to the discrete setting. Authority preferences are \textit{non-trivial} if for all $m,n\in\mathcal{M}$:
    \begin{equation}
      h(1) + (u_{n}(1) - u_{n}(0)) >  h(0)  + (u_{m}(q) - u_{m}(q-1)) 
    \end{equation}
    
The authority is \textit{risk-neutral} if for all $m \in \mathcal M$, $u_m(x) = c_m x$ for some $c_m \ge 0$ and all $x \in \{0,1,\ldots,q\}$. Define $\tilde u$ and $\tilde h$ as follows: there exists $x_m^{\text{tar}}$ such that $\tilde u_m(x_m+1) - \tilde u_m(x_m)= 0$ for all $x_m \geq x_m^{\text{tar}}$ and $\tilde u_m(x_m+1) - \tilde u_m(x_m) \geq h(1) - h(0)$ for $x_m < x_m^{\text{tar}}$ and  where $\sum_{m\in\mathcal{M}}x_m^{\text{tar}}\leq q$. Let $\tilde \xi$ denote the preferences of the authority under $\tilde u$ and $\tilde h$. The authority with preferences $\xi$ is \textit{extremely risk-averse} if the set of optimal allocations under $\xi$ and $\tilde \xi$ coincide for all $\omega$.

\begin{thm}
The following statements are true:
\begin{enumerate}
    \item If there is no uncertainty, then there exist first-best priority and quota mechanisms.
    \item Suppose that the authority has non-trivial preferences. There exists a first-best priority mechanism if and only if the authority is risk-neutral. This mechanism is given by $P(s,m) = s+u_m(1)-u_m(0)$. 
    \item Suppose that the authority has non-trivial preferences. There exists a first-best quota mechanism if and only if the authority is extremely risk-averse. This mechanism is given by $Q_m=x_{m}^{\text{tar}}$ and $D(R) = \vert \mathcal M \vert + 1$.
\end{enumerate}
\end{thm}
\begin{proof}
Part (1):
\begin{claim}
Let $\mu$ denote an optimal allocation at $\omega$. Then $\mu$ is a cutoff matching.
\end{claim}
\begin{proof}
If $\mu$ is not a cutoff matching, then there exists $(s,m)$ and $(s',m)$ where $\mu(s,m) = 1$, $\mu(s',m) = 0$ and $s' > s$. Define $\mu'$ by setting: $\mu'(s,m) = 0$, $\mu(s',m) = 1$ and $\mu(\tilde s, \tilde m) = \mu(\tilde s, \tilde m)$ for all $(\tilde s, \tilde m)$ such that $(\tilde s, \tilde m) \not \in \{(s,m),(s',m)\}$. Observe that, $\xi(\mu',\omega) - \xi(\mu,\omega) = s' - s > 0$. Therefore, $\mu$ is not an optimal allocation, which is a contradiction.
\end{proof}
Let $\mu$ denote an optimal allocation under $\omega$,  $\{\hat{s}_m(\mu,\omega)\}_{m \in \mathcal M}$ denote the cutoff scores at $\mu$ and $s^*$ denote an arbitrary number. Any priority policy that assigns $P(\hat{s}_m(\omega),m) = s^*$ for all $m \in \mathcal M$ and is strictly increasing in the first argument allocates the resource to any agent who has a higher score than the cutoff for their group and implements the optimal allocation.

Let $x_m$ denote the number of group $m$ agents who are allocated the resource at an optimal allocation under $\omega$. Then a quota policy that sets $Q_m = x_m$ allocates the resource to any agent who has a higher score than the cutoff for their group and implements the optimal allocation.

Part (2): The if part of the result follows from observing the priority policy  $P(s,m) = s+u_m(1)-u_m(0)$ is equivalent to the optimal APM  $A^*$ under risk neutrality since $u_m(1)-u_m(0) = u_m(y_m+1)-u_m(y_m)$ for all $m$, $y_m$. Thus, by Theorem \ref{discreteoptimality}, $P(s,m) = s+u_m(1)-u_m(0)$ is first-best optimal.

To prove the only if part, assume risk neutrality does not hold and let $m$ denote a group such that $u_m$ does not satisfy risk neutrality. For a contradiction, assume that $P$ is an optimal priority policy. First, we observe that $P(s,m)$ must be strictly increasing in $s$ for all $m$. To see why, assume $P(s,m) = P(s',m)$ where $s>s'$ and just consider an $\omega$ where there are $q-1$ group $m$ agents with scores strictly higher than $s$, and no other agents. Clearly, the optimal allocation would be to allocate the resource to all agents but $(s',m)$, while $P$ allocates the resource to $(s',m)$ with at least probability $1/2$.

Second, let $m$ denote a group such that $u_m$ does not satisfy risk neutrality. Take another arbitrary group $n$. We have the following:

\begin{claim}
Either (i) there exists $t < q$, $s_m$, $s_n$ such that
\begin{equation}\label{eq:discreteequality}
    u_m(t+1) - u_m(t) + h(s_m) =  u_n(q-t) - u_n(q-t-1) + h(s_n)
\end{equation}
or (ii) there exists $t<q$ such that
\begin{equation}\label{firsteq}
     u_m(t+1) - u_m(t) + h(1) < u_n(q-t) - u_n(q-t-1)  + h(0)
\end{equation}
\begin{equation}\label{secondeq}
     u_m(t) - u_m(t-1) + h(0) > u_n(q-t+1) - u_n(q-t) + h(1)
\end{equation}
\end{claim}
\begin{proof}
From non-triviality, we know that $u_m(1) - u_m(0) + h(1) >  u_n(q) - u_n(q-1) + h(0)$ and $u_n(1) - u_n(0) + h(1) > u_m(q) - u_m(q-1)  + h(0)$. The result then follows from the fact that $h$ is continuous and strictly increasing and  $u_m$ and $u_n$ are concave.
\end{proof}

We first prove the result under case (ii). Fix two agents with scores $s_m \in (0,1)$, who belong to group $m$ and $s_n \in (0,1)$, who belong to group $n$. Assume that there are $t-1$ group $m$ agents and $q-t$ group $n$ agents with higher scores than $\max\{s_n,s_m\}$, so a total of $t$ group $m$ agents and $q-t+1$ group $n$ agents. Note that in this case, only one agent will not be allocated the resource in the optimal allocation, and that would be either $(s_m,m)$ or $(s_n,n)$. From equation \ref{secondeq}, $(s_m,m)$  is more preferred than $(s_n,n)$ and therefore it must be that $ P(s_n,n) < P(s_m,m) $, as otherwise $P$ would not be optimal. Next, assume that there are $t$ group $m$ agents and $q-t-1$ group $n$ agents with higher scores than $\max\{s_n,s_m\}$.  From equation \ref{firsteq},  $(s_n,n)$ is more preferred than $(s_m,m)$ and therefore it must be that $ P(s_m,m) < P(s_n,n)$, which is a contradiction.

We now prove the result under case (i).
\begin{claim}
\label{claim:Pstuff}
In case (i), any optimal priority policy $P$ must satisfy $P(s_m+\epsilon,m) > P(s_n,n)$ for all $\epsilon > 0$ and  $P(s_m-\epsilon,m) < P(s_n,n)$ for all $\epsilon > 0$
\end{claim}
\begin{proof}
From Equation \ref{eq:discreteequality}, we see that when there are $t$ group $m$ agents and $q-t-1$ group $n$ agents with higher scores, $(s_m+\epsilon,m)$ is strictly preferred to $(s_n,n)$, which is strictly preferred to $(s_m-\epsilon,m)$.
\end{proof}
Since $u_m$ is not linear, there exists an $l$ such that $u_m(l+1) - u_m(l) < u_m(l) - u_m(l-1)$. There are two possibilities: $l \leq t$ or $l > t$. First, suppose that  $l \leq t$. We have that:
\begin{equation}
    u_m(l) - u_m(l-1) + h(s_m)> u_m(l+1) - u_m(l)+h(s_m) \ge u_n(q-l) - u_n(q-l+1) + h(s_n)
\end{equation}
where the first inequality follows from $u_m(l+1) - u_m(l) < u_m(l) - u_m(l-1)$ and the second inequality follows as $u_m(t+1) - u_m(t) + h(s_m) =  u_n(q-t) - u_n(q-t-1) + h(s_n)$, $u_m$ and $u_n$ are concave, and $l\leq t$. Thus, for sufficiently small $\epsilon>0$, we have that:
\begin{equation}
    u_m(l) - u_m(l-1) + h(s_m-\epsilon) > u_n(q-l) - u_n(q-l+1) + h(s_n)
\end{equation}
Given this inequality, we see that when there are $l-1$ group $m$ agents and $q-l$ group $n$ agents with higher scores, $(s_m-\epsilon,m)$ is strictly preferred to $(s_n,n)$. Thus, to implement the optimal allocation, it must be that $P(s_m-\epsilon,m) \geq P(s_n,n)$, which is a contradiction to Claim \ref{claim:Pstuff}.

Second, suppose that $l > t$. We know that:
\begin{equation}
    u_m(t+1) - u_m(t) + h(s_m) =  u_n(q-t) - u_n(q-t-1) + h(s_n)
\end{equation}
As $l > t$, from concavity of $u_m$ and $u_n$,
\begin{equation}
    u_m(l) - u_m(l-1) + h(s_m) \le  u_n(q-l+1) - u_n(q-l) + h(s_n)
\end{equation}
From concavity of $u_n$ and $u_m$:
\begin{equation}
    u_m(l+1) - u_m(l) + h(s_m) <  u_n(q-l) - u_n(q-l-1) + h(s_n)
\end{equation}
Thus, for sufficiently small $\epsilon>0$, we have that:
\begin{equation}
    u_m(l+1) - u_m(l) + h(s_m+\epsilon) <  u_n(q-l) - u_n(q-l-1) + h(s_n)
\end{equation}
Given this inequality, we see that when there are  $l$ group $m$ agents and $q-l-1$ group $n$ agents with higher scores, $(s_n,n)$  is strictly preferred to $(s_m+\epsilon,m)$. Thus, to implement the optimal allocation, it must be that $P(s_m+\epsilon,m) \leq P(s_n,n)$, which is a contradiction to Claim \ref{claim:Pstuff}.

Part (3):  To prove the if part, fix an $\omega$ and let $\mu^*$ denote the optimal allocation under $\omega$. Let  $x_m^*$ denote the number of group $m$ agents allocated the resource at $\mu^*$ and $x_m(\omega)$ denote the total number of group $m$ agents under $\omega$. 

\begin{claim}
If the authority is extremely risk-averse, then $x_m^* \geq \min\{x_m(\omega),x_m^{tar}\}$
\end{claim}
\begin{proof}
Assume for a contradiction this is not the case. Then $x_m^* < x_m(\omega)$ and $x_m^* < x_m^{tar}$. Since $\sum_{m \in \mathcal M} x_m^{tar} \leq q$ and $x_m^* < x_m^{tar}$, there exists $n \in \mathcal M$ such that $x_n^* > x_n^{tar}$. Let $s_n$ denote the score of the lowest-scoring group $n$ agent who is allocated the resource, and let $s_m$ denote the score of any group $m$ agent who is not allocated the resource, which exists as $x_m^* < x_m(\omega)$. Since the authority is extremely risk-averse, we have the following:
\begin{equation}
    h(s_m) + u_m(x_m^*+1) - u_m(x_m^*) > h(s_n) - u_n(x_n^*) + u_m(x_n^*-1)
\end{equation}
However, this contradicts the optimality of $\mu^*$ and proves the claim.
\end{proof}

\begin{claim}
If the authority is extremely risk-averse, $x_m^* > x_m^{tar}$ and $x_n^* > x_n^{tar}$, $\mu^*(s,m) = 0$ and $\mu^*(s',n) = 1$, then $s' > s$.
\end{claim}
\begin{proof}
Assume for a contradiction that $s > s'$.\footnote{Remember that $s'=s$ was ruled out by assumption.} The difference in the utility of the authority when allocating the resource to $(s,m)$ rather than $(s',n)$ is given by
\begin{equation}
    h(s) + u_m(x_m^*+1) - u_m(x_m^*) - (h(s') - u_n(x_n^*) + u_m(x_n^*-1)) = h(s) - h(s') > 0
\end{equation}
which is a contradiction to optimality of $\mu^*$.
\end{proof}

The previous two claims show that under any $\omega$, the optimal allocation admits (i) the highest-scoring $x_m^{tar}$ agents from each group (provided that they exist) and (ii) highest-scoring agents who are not in (i), until the capacity is exhausted. Clearly, the quota policy $Q_m=x_{m}^{\text{tar}}$ and $D(R) = \vert \mathcal M \vert + 1$ implements this outcome at every $\omega$.

To prove the only if part, assume that $\{Q_m\}_{m \in \mathcal M}$ is part of an optimal quota policy.

\begin{claim}\label{beforequota}
For and each $m,n \in \mathcal M$ and any $t,l$ such that $t\leq Q_m$, $Q_m>0$ and $l\geq Q_n$, we have that:
\begin{equation}
    u_m(t) - u_m(t-1) + h(0) \geq u_n(l+1) - u_n(l) + h(1)
\end{equation}
\end{claim}
\begin{proof}
Assume that at $\omega$, there are $t$ group $m$ agents, one of which one has score $0$ and $l+1$ group $n$ agents with scores higher than $1-\epsilon_1$ and $q$ agents from other groups who have scores higher than $1-\epsilon_2$, where $\epsilon_1 > \epsilon_2>0$. As $t\leq Q_m$ and $Q_n<l+1$, $t$ group $m$ agents and $Q_n < l+1$ group $n$ agents are admitted under $Q$. Since $Q$ is optimal for all $\epsilon_1$, we must have that:
\begin{equation}
    u_m(t) - u_m(t-1) + h(0) \geq u_n(l+1) - u_n(l) + h(1-\epsilon_1)
\end{equation}
The statement then follows from continuity of $h$ by taking the limit $\epsilon_1 \to 0$.
\end{proof}
\begin{claim}\label{meritlastdiscrete}
Merit slots are processed last at the optimal quota policy.
\end{claim}
\begin{proof}
For a contradiction, assume there is a merit slot that is processed before a quota slot. Let $l$ denote the last merit slot that precedes a quota slot. Let $m$ denote a group that has a quota slot after $l$. We consider a state in which: (i) there are $q$ group $n$ agents with scores $\hat s - \epsilon_i$, where $\epsilon_i>0$ for all $i \in \{1,\ldots,q\}$ (let $\underline{\hat s}$ denote the score of the highest-scoring agent from this group), (ii) there are $Q_m$ group $m$ agents with scores $\hat s + \epsilon_j$ for $j \in \{1,\ldots,Q_m\}$ (let $\overline{\hat s}$ denote the score of lowest-scoring agent from this group) and one with score $\hat s/2$, and (iii) $q$ agents from other groups with scores in $(\underline{\hat s},\overline{\hat s})$. A group $m$ agent with score $\hat s + \epsilon_k$ for some $k$ is matched to $l$, thus $(\hat s/2,m)$ is matched to a later quota slot, while some agents with type $(\hat s - \epsilon_j,n)$ are rejected for some $j$. Let $\hat s - \epsilon_{j'}$ be the score of the highest-scoring such agent. From the optimality of the quota policy we have that
\begin{equation}
\label{eq:claimhigh}
    u_m(Q_m+1) - u_m(Q_m) + h(\hat s/2) \geq u_n(Q_n+1) - u_n(Q_n) + h(\hat s - \epsilon_{j'})
\end{equation}
Let $s^*$ be the score of the lowest-scoring group $n$ agent (\textit{i.e.,} $s^* = \min_{i \in \{1,\ldots,q\}} \hat s - \epsilon_i$). Next, consider the modified version of the above state, all group $n$ agents are the same,  but all of the other $Q_m$ group $m$  agents as well as $q$ agents from other groups now have scores in $(s^*-\hat \epsilon,s^*)$ and the group $m$ agent who had a score of $\hat s/2$ now has a score of $\hat s/2 + \hat \epsilon$ for $\hat \epsilon>0$. Note that now the group $n$ agent with score $\hat s - \epsilon_{j'}$  is allocated the slot $l$ or an earlier slot, while the agent $(\hat s/2 + \hat \epsilon,m)$ is not allocated to any slot. Thus
\begin{equation}
 u_m(Q_m+1) - u_m(Q_m) + h(\hat s/2 + \hat \epsilon) \leq u_n(Q_n+1) - u_n(Q_n) + h(\hat s - \epsilon_{j'})
\end{equation}
which, since $h$ is strictly increasing, implies that $u_m(Q_m+1) - u_m(Q_m) + h(\hat s/2) < u_n(Q_n+1) - u_n(Q_n) + h(\hat s - \epsilon_{j'})$. This contradicts Equation \ref{eq:claimhigh}, proving the claim.
\end{proof}

Given the previous two claims, the following claim proves the result.

\begin{claim}\label{afterquota}
If merit slots are processed last, then for all $l \geq Q_m$ and $j \geq Q_n$
\begin{equation}
    u_m(l+1) - u_m(l) = u_n(j+1) - u_n(j)
\end{equation}
\end{claim}
\begin{proof}
Assume for a contradiction this does not hold. Without loss of generality, assume $u_m(l+1) - u_m(l) > u_n(j+1) - u_n(j)$ and define $\delta$ as

\begin{equation}
    \delta = \left(u_m(l+1) - u_m(l)\right) - \left( u_n(j+1) - u_n(j) \right)
\end{equation}

Consider a state with $q-1$ agents with scores higher than $s^*$, of which exactly $Q_m$ are group $m$ agents and $Q_n$ are group $n$ agents. Moreover, there is one more group $m$ agent with score $s'<s^*$ (denote this agent by $\theta_m$) and one more group $n$ agent with score $s'' \in (s',s^*)$ where $h(s'') - h(s') < \delta$ (denote this agent by $\theta_n$). Note that all agents apart from $\theta_m$ and $\theta_n$ are allocated the resource before the final merit slot. Moreover, since $\theta_n$ has a higher score, she obtains the final merit slot. However, this is a contradiction to the optimality of $Q$ as $h(s'') - h(s') < \delta$ and allocating that resource to $\theta_m$ gives the authority higher utility. This proves the claim.
\end{proof}
Taken together, claims \ref{beforequota} and \ref{afterquota} prove that a fictitious authority that is extremely risk-averse with $x_m^{\text{tar}}=Q_m$ agrees with the authority on the optimal allocation, for all $\omega$. To see this, observe that claim \ref{beforequota} implies that diversity preferences dominate any concern for scores when a group is allocated less than $Q_m$. Moreover, conditional on being allocated at least $Q_m$, it is as if there is no residual diversity preference, by claim \ref{afterquota}. This proves the only if part of (3), which finishes the proof of the result.
\end{proof}

\subsection{Characterization of Stable Allocations and Dominance of APM in Discrete Economies}

In this section, we extend our discrete model to the multiple authority case. We show that the cutoff structure of stable matchings studied in Section \ref{sec:characterizationofstability} and the dominance of the optimal APM in the decentralized admissions setting studied in Section \ref{section:decentralized} can be extended to the discrete setting. Let $\Theta_0$ denote the set of agents. $\mathcal C = \{c_0,c_1,\ldots,c_{|\mathcal C|-1}\}$ denote the set of authorities. $q_c$ denotes the capacity of authority $c$ and $q_{c_0} \geq |\Theta_0|$. $\theta=(s,m,\succ)\in[0,1]^{|\mathcal C|}\times\mathcal{M}\times\mathcal{R}=\Theta$, where $\mathcal{R}$ is set of all complete, transitive, and strict preference relations over $\mathcal{C}$ such that $c_0$ is less preferred than all $c\in\mathcal{C}$. For each type $\theta$, $s_c(\theta)$ denotes the score of $\theta$ at authority $c$ and $m(\theta)$ denotes the group of $\theta$.

A matching in this environment is a function $\mu: \mathcal C \cup \Theta \to 2^{\Theta} \cup \mathcal C$ where $\mu(\theta) \in \mathcal C$ is the authority any type $\theta$ is assigned and $\mu(c) \subseteq \Theta$ is the set of agents assigned to authority $c$, which satisfies $|\mu(c)| \leq q_c$ for all $c$. $x_c(\mu) = \{x_{m,c}(\mu)\}_{m \in \mathcal M}$ denotes the number of agents of each group assigned to school $c$ at $\mu$ while $\bar{s}_{h_c}(\mu) = \sum_{\theta \in \mu(c)}  h(s(\theta))$ denotes the score utility the authority derives from $\mu$. The preferences of the authority are given by:
    \begin{equation}
        \xi_{c}(\bar{s}_{h_c},x_c)=\bar{s}_{h_c}+\sum_{m\in\mathcal{M}}u_{m,c}(x_{m,c})
    \end{equation}
where $h_c$ is continuous and strictly increasing and $u_{m,c}:\mathbb{R}\rightarrow\mathbb{R}$ is concave for all $m\in\mathcal{M}$ and $c\in\bar{\mathcal{C}}$.

We first extend the dominance of APM to discrete economies. As in Section \ref{section:decentralized}, agents apply to the authorities sequentially, who decide which agents to admit. We index the stage of the game by $t\in \mathcal{T}=\{1,\ldots,|\mathcal{C}|-1\}$. Each stage corresponds to an authority $I(t)$, where $I:\mathcal{T}\rightarrow \mathcal{T}$. At each stage $t$, any unmatched agents choose whether apply to authority $I(t)$. Given the set of applicants, authority $I(t)$ chooses to admit a subset of these agents. Given this, histories are indexed by the path of the remaining of agents who have not yet matched, $h^{t-1}=(\Theta_0,\Theta_1,\ldots,\Theta_{t-1})\in\mathcal{H}^{t-1}$. Given each history $h^{t-1}$ and set of applicants $\Theta^{A}_c\subseteq\Theta$, a strategy for an authority returns a set of agents $\Theta^{G}_c\subseteq \Theta$ whom they will admit such that $\Theta^G_c\subseteq\Theta^{A}_c$ and $|\Theta^G_c|\leq q_c$ for each time at which they could move $t\in\mathcal{T}$, $a_{c,t}:\mathcal{H}^{t-1}\times\mathcal{P}(\Theta)\rightarrow\mathcal{P}(\Theta)$, where $\mathcal{P}(\Theta)$ is the power set over $\Theta$. A strategy for an agent returns a choice of whether to apply to authorities at each history and time for all agent types $\theta\in\Theta$, $\sigma_{\theta,t}:\mathcal{H}^{t-1}\rightarrow[0,1]$. We moreover say that a strategy $a_{\tilde c,t}$ for an authority $\tilde c$ at time $t$ is \textit{dominant} if it maximizes authority utility regardless of $\{\{a_{c,t}\}_{c\in\mathcal{C}/\{\tilde c\}},\{\sigma_{\theta,t}\}_{\theta\in\Theta}\}_{t\in\mathcal{T}}$ and $I$.

\begin{thm}
\label{prop:discretedominance}
The APM $A^*_c$ is a dominant strategy for all authorities.
\end{thm}
\begin{proof}
We prove that APM $A^*_c$ implements a dominant strategy for all authorities in all stages by backward induction. Consider the terminal time $t=|\mathcal{C}|-1$. Some set of agents $\hat \Theta \subseteq \Theta$ applies to the authority. Regardless of $\hat \Theta$, by Theorem \ref{discreteoptimality} we have that the set of agents chosen under any selection from APM $A^*_c$ is first-best optimal. Thus, $A_c^*$ is dominant. Consider now any time $t<|\mathcal{C}|-1$, precisely the same argument applies and $A_c^*$ is dominant.
\end{proof}

In Theorem \ref{prop:multischoolstable}, we showed that in the continuum model, there is stable matching and this matching is a cutoff matching. It is well known that in discrete models there may be multiple stable matchings, so the first part of the result does not hold. However, we can extend the second part of Theorem  \ref{prop:multischoolstable}. Recall that in the discrete setting, a matching $\mu$ is stable if there are no blocking pairs, that is, there does not exist an agent $\theta$ and an authority $c \in \mathcal C$ (which includes the dummmy authority) such that $c \succ_{\theta} \mu(\theta)$ and either (i) $c$ does not fill its capacity or (ii) there exists $\theta' \in c$ such that $\xi_{c}(\bar{s}_{h_c}(\mu'),x_c(\mu')) > \xi_{c}(\bar{s}_{h_c}(\mu),x_c(\mu))$, where $\mu'(c) = \mu(c) \setminus \theta' \cup \theta$.

\begin{prop}
    If $\mu$ is a stable matching, then it is a cutoff matching.
\end{prop}
\begin{proof}
    If $\mu$ is not a cutoff matching, then there exist $\theta = (s,m)$, $\theta' = (s',m)$ and $c \in \overline{\mathcal C}$ such that $\mu(\theta') = c$, $c >_{\theta} \mu(\theta)$ and $s> s'$. Define $\mu'$ as follows. $\mu'(c) = \mu(c) \setminus \theta' \cup \theta$ and $\mu'(c') = \mu(c')$ for all $c' \neq c$. As $s > s'$ and $h_c$ is strictly increasing,  $\xi_{c}(\bar{s}_{h_c}(\mu'),x_c(\mu')) > \xi_{c}(\bar{s}_{h_c}(\mu),x_c(\mu))$, which contradicts the stability of $\mu$.
\end{proof}

 \subsection{Discrete Model: Example under Imperfect Information}\label{examplesection}
We develop a simple example to show how the qualitative trade-offs between priorities and quotas we have identified are those present in discrete matching markets. There are 4 students, $\Theta =\{\theta_1,\theta_2,\theta_3,\theta_4\}$ and one authority $c$ with capacity two. Students $\theta_3$ and $\theta_4$ belong to an underrepresented minority. The scores of minority students are distributed independently and uniformly on $[0,1]$, so that $s_3,s_4 \sim U[0,1]$. For simplicity, we assume there is no uncertainty over the scores of other students: $s_1 = s_2 = 1$.\footnote{The qualitative result here does not change as long as non-minority students draw their scores from a distribution that FOSD $U[0,1]$.} We further specify that the authority has the following utility function:\footnote{$\mu$ denotes a matching, where $\mu(c) \subset \Theta$ and $|\mu(c)|=2$. We assume $\mu$ is stable, which uniquely determines the allocations.}
\begin{equation}
    W(\beta,\mu) = \beta \mathbb{I}\{\mu(c) \cap \{\theta_3,\theta_4\}\neq\emptyset\} + \sum_{i: \mu(\theta_i) = c} s_i
\end{equation}
This function embodies the main trade-off we have studied: the trade-off between scores and diversity. The first term indicates that whenever the authority admits at least one minority student, the utility of authority increases by $\beta$, which denotes the strength of affirmative action or diversity preferences. The second term simply indicates that the authority cares about scores and wants to admit the highest-scoring students they can. An alternative interpretation in this context, where allocating to agents with low scores is perceived as unfair, is that the authority wants to ensure outcomes that are fair in this sense.

The authority implements a stable matching and has two different policies at their disposal to influence the outcome of the matching mechanism. The first is a priority subsidy, denoted by $\alpha \in [0,1]$. A subsidy $\alpha$ simply increases the scores of minority students by $\alpha$ and moves the distribution of scores of minority students to $U[\alpha,1+\alpha]$. The second is a minority quota $Q \in \{0,1,2\}$ that reserves $Q$ seats for minority students.

We start by characterizing the first-best where the authority can choose the matching they most prefer in each state of the world. Intuitively, if the score of the highest-scoring minority student is sufficiently high, then the designer prefers to admit her and one of the non-minority students. Otherwise, it is optimal for the designer to admit the two highest-scoring students, that is the two non-minority students. The first-best matching $\mu^*$ is therefore given by:

\begin{equation}
 \mu^*(c) =
\begin{cases}
        \{\theta_1,\theta_2\} \quad, \ \text{if} \quad \max_{i \in \{3,4\}} s_i < 1-\beta,\\
        \{\theta_1,\theta_k\} \quad, \ \text{if} \quad \max_{i \in \{3,4\}} s_i > 1-\beta \text{ and } s_k = \max_{i \in \{3,4\}} s_i.
\end{cases}
\end{equation}

See that if the authority had perfect information and knew $s_3$ and $s_4$ that both a priority subsidy and a quota can implement the first-best.\footnote{Formally speaking, this is only true outside of the knife-edge case where $s_3=s_4$, which is probability zero. In this case, there is no subsidy that can implement the first-best.} In particular, if $\max_{i\in\{3,4\}}s_i>1-\beta$, then both a quota $Q=1$ that reserves one seat for minority students and a priority subsidy for minority students of $\alpha \in (1-\min_{i\in\{3,4\}} s_i ,1-\max_{i\in\{3,4\}}s_i)$ implement the first-best. When $\max_{i\in\{3,4\}}s_i\leq1-\beta$, then both a quota of $Q=0$ and a subsidy $\alpha=0$ implement the first-best. Thus, with perfect information, both policies yield the first-best and there is no trade-off for the authority.

We now consider the second best where an authority is constrained to implement a priority or a quota before the realization of uncertainty. Note that implementing the first-best is impossible with both quotas and priorities as neither can be adapted to the underlying realized scores of the minority students. We now solve for the optimal quota and priority designs and compare their values. In order to characterize the optimal reserve policy, one first notes that reserving both seats for minority students is always strictly dominated by reserving only one. Thus the designer only needs to compare a policy with a quota of one against a policy with no quotas. With no quota, no minority student is admitted and the utility of the designer $W_{nr}(\beta) = 2$ with probability one. On the other hand, if a quota of one is used, only the highest-scoring minority student is admitted. Thus, the expected utility of the designer is:
\begin{equation}
    \mathbb{E}[ W(\beta) ]= \beta + 1 + \mathbb{E} \left[\max\{s_3,s_4\}\right] = \frac{5}{3} + \beta
\end{equation}
The optimal quota policy is therefore to reserve one seat for minority students if $\beta > 1/3$ and reserve no seats otherwise. Moreover, the utility of the designer under the optimal quota policy is:
\begin{equation}
    V_Q(\beta) =
\begin{cases}
        2\quad &, \text{if} \quad \beta \leq \frac{1}{3},\\
        \dfrac{5}{3} + \beta \quad &, \text{if} \quad \beta >\frac{1}{3}.\\
\end{cases}
\end{equation}

We now compute the optimal priority design and the authority's value thereof. To this end, we first calculate the utility of the designer under subsidy $\alpha$. We start by calculating the matching conditional on the realized scores. There are three cases to consider. If both minority students have scores above $1-\alpha$, then both of them are admitted to the authority. If both minority students score below $1-\alpha$, then neither of them is admitted. Lastly, if one of them scores above $1-\alpha$ while the other scores below $1-\alpha$, then only one minority student is admitted. The following equation gives the utility of the designer as a function of $\beta$ and $\alpha$:
\begin{equation}
\begin{split}
        \mathbb{E}[W(\beta,\alpha)] =& \int_{1-\alpha}^1 \int_{1-\alpha}^1 (s_3 + s_4 + \beta) ds_3 ds_4+ 2 \int_{1-\alpha}^1 \int_0^{1-\alpha} (1 + s_4 + \beta) d s_3 d s_4\\
        &+ \int_0^{1-\alpha} \int_0^{1-\alpha} 2 d s_3 d s_4\\
        =& -(1+\beta) \alpha^2 + 2 \beta \alpha + 2
\end{split}
\end{equation}
A quick calculation shows that the optimal subsidy is always interior and $\alpha^* = \frac{\beta}{1+\beta}$. Plugging the optimal subsidy policy into the authority's payoff function, we obtain:
\begin{equation}
    V_P(\beta) = 1 + \beta + \frac{1}{1+\beta}
\end{equation}
\begin{figure}[t]
    \centering
        \caption{Comparative Statics for the Preference Between Priorities and Quotas}
    \includegraphics[width=0.6\linewidth]{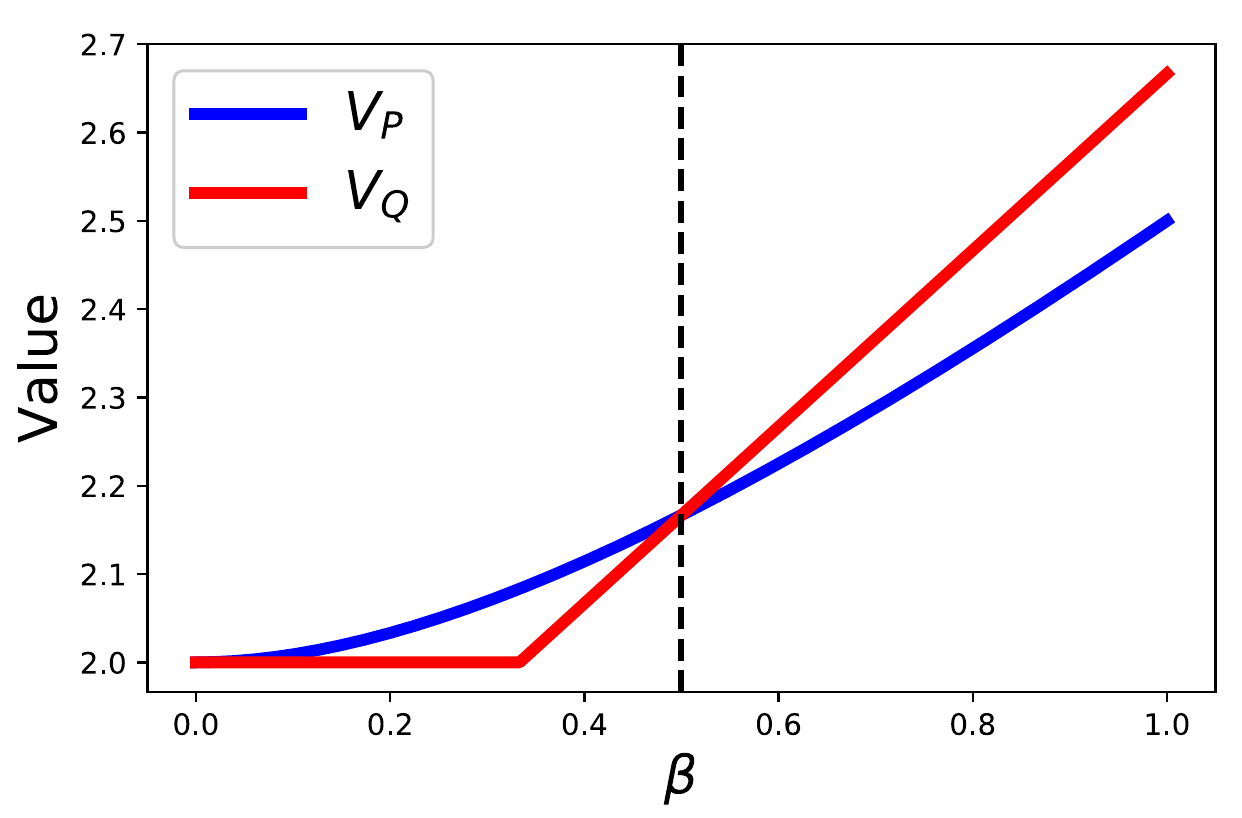}
\fnote{Values of the optimal priority policy, $V_P$, and optimal quota policy, $V_Q$, as a function of the strength of the diversity preference $\beta$. The dashed black line corresponds to $\beta=1/2$ and is the point at which both policies yield the same value.}
    \label{examplefig}
\end{figure}
We now compare the value of the optimal quota and priority designs as the strength of the affirmative action motive changes. Comparing $V_P(\beta)$ and  $V_Q(\beta)$ shows that the optimal policy depends on the strength of affirmative action preferences of the authority. Figure \ref{examplefig} plots these two values as a function of the affirmative action motive with the dotted line giving the value $\beta=\frac{1}{2}$ at which the two value functions cross. Importantly, we see that a quota policy is optimal whenever $\beta > 1/2$ and a priority subsidy policy is optimal whenever $\beta < 1/2$. 

This example highlights the main differences between priorities and quotas under uncertainty and suggests when we might expect to prefer one over the other. When the preference for diversity is low, the authority only wants to admit a minority student if her score is high enough. In this case, a subsidy is a better policy as its outcome can depend on the relative scores of the students. In particular, it only admits minority students if they obtain sufficiently high scores while a quota admits minority students equally across states of the world. Consequently, priority designs generate a desirable positive selection of minority students which tends to improve scores. However, the drawback of a subsidy policy is that it applies to all students and can therefore cause either the admission of a second minority student with a lower average score or fail to admit any minority students. On the other hand, if the preference for diversity is sufficiently high, then the authority wants to admit one minority student for sure, regardless of her score. In this case, the subsidy policy is undesirable as even under the optimal subsidy, there are many realizations where neither or both minority students are admitted, while the reserve policy ensures that one minority student is admitted in all states of the world.

\clearpage

 \section{Implementation, Precedence Orders, and an Illustration from H1-B Visa Allocation}
\label{ap:implementation}
In this appendix, we show that (with no uncertainty) priority and quota policies can implement the same set of allocations. We apply this insight to study the effect of precedence orders in US H1-B visa allocation.

 \subsection{Equivalence of Priorities and Quotas for Implementation}
In Proposition \ref{prop:nouncertainty}, we showed that if there is no uncertainty, both priorities and quotas can achieve the optimal allocation.  We say that a priority policy $P$ is monotone if $P(s,m)$ is strictly increasing in $s$. Note that since the authority prefers higher-scoring agents to lower-scoring ones, monotone policies perform better non-monotone policies. We will now show that, in the setting of Section \ref{sec:single}, for a given $\omega$ (which we suppress for the rest of this section), these quota and monotone priority policies are equivalent in the sense that any allocation that is achieved by one can also be achieved by the other.

\begin{prop}\label{equivalence}
$\mu$ is implemented by a quota policy if and only if it is also implemented by a monotone priority policy.
\end{prop}
\begin{proof}
Assume that $\mu$ is implemented by a quota policy. Then $\mu$ is a cutoff allocation since the resource is allocated in descending order of score. Let $s_m$ denote the lowest-scoring agent from group $m$ who is allocated the object at $\mu$ for $m \in \mathcal M$.  Let $\bar s = \max_{m \in \mathcal M} s_m$. Define the priority policy as $P(s,m) = s + (s_m - \bar s)$. Note that if $\mu(s,m) = 1$ and $\mu(s',m') = 0$, then $s + s_m - \bar s > s' + s_{m'} - \bar s$ and therefore $P(s,m) > P(s',m')$. As $P$ allocates the resource to measure $q$ highest-scoring agents under $P$ and measure $q$ of agents who are allocated the resource under the quota policy has higher priorities than those who are not, $P$ implements the same allocation as the quota policy.

Conversely, assume that $\mu$ is implemented by a monotone priority policy. Let $x_m$ denote the measure of agents from group $m$ allocated the object at $\mu$  for $m \in \mathcal M$. Let $Q$ denote a quota policy where $Q_m = x_m$. Under any processing order, $Q$ implements the same allocation and allocates the resource to the highest-scoring measure $x_m$ agents from group $m$, for all $m$. This is the allocation under $P(s,m)$ since $P$ is a monotone priority policy and  allocates the resource to the highest-scoring measure $x_m$ agents from group $m$, which proves the result.
\end{proof}

In the next section, we use this result to provide a diagnostic test for evaluating quota policies with different precedence orders by the strength of the equivalent priority policy.

 \subsection{Application of a Diagnostic Test for the Effect of Precedence Orders to US H1-B Visa Allocation}
\label{h1bapp}
In this Appendix, we argue that in light of Proposition \ref{equivalence}, priorities can be used as a diagnostic test for the effect of precedence orders in the context of US H1-B Visa allocation, which has had historical issues in implementation arising from the choice of precedence order. The American H1-B visa program enables American companies to temporarily employ educated foreign workers in high-skill occupations.\footnote{See \citeAp{pathak2020immigration} for a detailed account of H-1B policies and reforms.} The statutory law enacted by the U.S. Congress mandates the total number of visas to be granted and The U.S. Customs and Immigration Service (USCIS) implements this mandate. The visa allocation is governed by the H-1B Visa Reform Act of 2004 that established an annual system in which 65,000 visas were made available for all eligible applicants and an additional 20,000 visas were reserved for applicants with advanced degrees. Until 2009, USCIS used the arrival time of the application to determine priorities. Since then, the priorities are determined according to a uniform lottery.

As we have emphasized, under quota policies, specifying the processing order is critical. Between 2009 and 2019, USCIS used a Reserve-Initiated processing rule. In 2020, in accordance with the 2017 {\it Buy American and Hire American Executive Order}, USCIS switched to a Unreserved-Initiated rule, in order to award visas to the most-skilled workers. \citeAp{pathak2020immigration} document this switch and give a detailed account of the consequences. In particular, they calculate the effect of this change on visa allocation between 2013-2017. They find that from (on average) 55,900 applicants with advanced degrees, 33,495 of them obtain a visa under the Reserve-Initiated rule, while 38,843 of them obtain a visa under the Unreserved-Initiated rule. This fact underscores how  the complexity of quota policies can lead to issues in implementation, even in the simplest case with two groups.

We now use the structure of Proposition \ref{equivalence} to provide a diagnostic test that authorities can use to see the degree of effective affirmative action when employing a quota policy and apply it to the H1-B and Boston Public Schools settings.\footnote{We note that the H-1B lottery is a setting where the perfect information assumption is justified. First, the only object that is allocated is the visa and all applicants prefer obtaining the visa to not obtaining it. This removes any uncertainty over the preferences of the individuals. Second, the priorities are determined according to a uniform random lottery and the market is large. In particular, between 2013 and 2017, each year, 85,000 visas are allocated to an average of 137,017 reserve ineligible and 55,900 reserve eligible applicants. Thus, although there is uncertainty at the individual level, the distribution of lottery numbers conditional on reserve eligibility is essentially fixed.} Concretely, when an authority is considering designing its precedence order, it can simply compute the implied priority subsidy being afforded to each group. In the context of H1-B allocation, we assume the uniform random lottery of USCIS is implemented by drawing a number uniformly from the interval $[0,100]$. In the counterfactual priority mechanism, there are no quotas for reserve category applicants, but they get a score subsidy of $\alpha$, i.e. their random numbers are distributed uniformly on $[\alpha,100+\alpha]$. Computing the implied $\alpha$ under both processing orders to compare the policies, we obtain Table \ref{subsidycalc}. Note that even though both quota policies correspond to 20,000 visas being reserved for applicants with advanced degrees, there is an important difference in the number of visas allocated to advanced degree applicants and therefore in the subsidy required to achieve that allocation. In particular, the Unreserved-Initiated order leads to a 12-point subsidy increase relative to the Reserve-Initiated benchmark. 


\begin{table}[t]
\begin{center}
\caption{Equivalent Priorities for Different Precedence Orders}
\label{subsidycalc}
\begin{tabular}{l|ll|ll|}
                              & \multicolumn{2}{c|}{\# Applicants} & \multicolumn{2}{l|}{\# Reserve-eligible Visas} \\
                              & General     & Reserve-eligible     & R-I Rule              & NR-I Rule              \\ \hline
5-yr Average (2013-2017)      & 137,017     & 55,900               & 33,495                & 38,834                 \\
Equivalent Subsidy ($\alpha$) &             &                      & 23                    & 35                    
\end{tabular}
\fnote{Allocation of H-1B visas under Reserve-Initiated and Nonreserve-Initiated rules along with the equivalent subsidies that would induce these allocations.}
\end{center}
\end{table}
\bibliographystyleAp{econometrica.bst}
\bibliographyAp{bib-matching}

\end{appendices}
\end{document}